\documentclass[linktocpage,
 aps,
 prx,
 twocolumn,
 floats,
 epsf,
 amsmath,amssymb,
 superscriptaddress,
]{revtex4-2}

\usepackage[utf8]{inputenc}
\usepackage[english]{babel} % Spracheinstellung
\usepackage[T1]{fontenc}
\usepackage{amsmath}
\usepackage{bm} % dick und kursiv
\usepackage{amsfonts}
\usepackage{amssymb}
\usepackage{dsfont}
\usepackage{graphicx}

\usepackage{commath}
\usepackage{orcidlink}
\usepackage[normalem]{ulem}

\usepackage{empheq} %boxed equations

\usepackage{enumitem} %enumerate items with reference

\usepackage{xcolor} % Vektorgrafiken
\usepackage{transparent}

\usepackage{comment}

\usepackage{tabularx} %vertically centered table

\usepackage{soul}
\usepackage{hyperref} % Links
\usepackage{url} % URL-Links

\newcommand{\ee}{\mathrm{e}}
\newcommand{\ii}{\mathrm{i}}

\newcommand{\upup}{{\uparrow\uparrow}}
\newcommand{\updnx}{{\hat{\uparrow\downarrow}}}

\newcommand{\updn}{{\uparrow\downarrow}}
\newcommand{\dnup}{{\downarrow\uparrow}}

\newcommand{\ph}{{ph}}
\newcommand{\pp}{{pp}}
\newcommand{\phx}{{\overline{ph}}}

\newcommand{\sgn}{\mathrm{sgn}}

\newcommand{\todo}[1]{{\color{red}[\textbf{ToDo:} #1]}}

\newcommand\SRC[1]{\textcolor{orange}{({\bf SR}: #1)}}

\newcommand{\MGC}[1]{{\color{blue}({\bf MG}: #1)}}

\definecolor{darkgrass}{RGB}{0,128,0}
\newcommand{\changed}[1]{{\color{black}{#1}}}

\newcommand{\TUVienna}
{\affiliation{Institute of Solid State Physics, TU Wien, 1040 Vienna, Austria}}

\begin{document}

\title{Instabilities in self-consistent diagrammatic approaches and how to cure them}

\author{Marcel Gievers \orcidlink{0000-0002-6951-7003}}
\thanks{Both authors contributed equally.}
\TUVienna
\author{Herbert Eßl\orcidlink{0009-0005-9883-8104}}
\thanks{Both authors contributed equally.}
\TUVienna
\author{Stefan Rohshap \orcidlink{0009-0007-2953-8831}}
\TUVienna
\author{Anna Kauch \orcidlink{0000-0002-7669-0090}}
\TUVienna
\author{Alessandro Toschi \orcidlink{0000-0001-5669-3377}}
\TUVienna

\date{\today}

\begin{abstract}
While self-consistent diagrammatic approaches are widely used to compute the physical properties of correlated quantum materials, their applicability may get severely hindered precisely in the parameter regions where the most exciting physics is observed.   One of the major issues, referred to as \emph{``misleading convergence''}, is the tendency of iterative schemes  to converge to \emph{unphysical} fixed points for intermediate-to-strong electronic interactions, \emph{regardless} of the numerical accuracy of the computation. Here, we explicitly verify that the origin of this problem in several established self-consistent many-electron approaches, defined in the general diagrammatic framework of the boson-exchange formalism, resides \emph{exclusively} in the stability condition of the respective iteration schemes, and \emph{not} in an intrinsic breakdown of their self-consistent diagrammatic description. This insight enables a simple and general remedy, as recently proposed in Phys.~Rev.~Lett.~\textbf{137}, 016502 (2026):
The redefinition of the iterative procedure, by inverting the unstable eigendirections of the Jacobian associated with the fixed point of the self-consistent algorithm. We illustrate the successful outcome of this procedure by means of systematic calculations performed on exactly solvable testbed models. Our results demonstrate that the physical fixed point of the diagrammatic schemes we considered can be stabilized, {\sl de facto}, across the entire parameter range, including the most challenging nonperturbative/strong-coupling regimes.

\end{abstract}

\maketitle

\section{Introduction}

One of the hardest challenges of contemporary condensed-matter theory is represented, indisputably, by the simulation of quantum materials in the presence of significant electronic correlations. In fact, on the one hand, correlations are believed to drive a plethora of fascinating phenomena in solid-state physics and in cold atoms trapped in optical lattices (e.g., high-temperature superconductivity \cite{Lee2006Doping,Scalapino2012Common,Keimer2015Quantum}, metal-to-insulator \cite{Georges1998Dynamical,Imada1998Metal,Kotliar2004Strongly} and quantum phase transitions \cite{Loehneysen2007,Sachdev_2011,Brando2016,Fus2024arXiv}, unconventional magnetic patterns \cite{Fradkin2015Colloquium} etc.). On the other hand, their inclusion poses important hurdles to several of the most advanced theoretical approaches. First of all, the (thermodynamically) large number of electronic degrees of freedom to be considered calls for a description based on quantum field theory, and hence, partly or totally, on Feynman diagrammatics. Second,  a theoretical treatment restricted to one-particle Green's functions (which experimentally correspond to ARPES measurements) does \emph{not} suffice to grasp the underlying physics: In the presence of significant electronic correlations, the comparison of one- and two-particle spectroscopy is mandatory to fully unveil the correct physical picture. Third, the values of the effective electronic interaction to be considered are often comparable or even larger than the electronic bandwidth and temperature.

A quite versatile class of schemes, which formally fulfill these requirements, is represented by self-consistent diagrammatic approaches derived within the boson-exchange (BE).
The BE formulation of diagrammatic approximations, in fact, is known to be quite advantageous both for its numerical efficiency and for the transparency of its physical interpretation \cite{Krien2019Single,Krien2019Parquetlike,Krien2020Boson,Krien2020Tiling,Krien2022Plain,Krien2022Explaining,Bonetti2022Single,Gievers2022Multiloop,Fraboulet2022Single,Adler2024Nonperturbative,Kiese2024Embedded,meixner2025,Patricolo2025Single,Gievers2025Subleading,Meixner2026Nonperturbative,meixner2026}.
In this broad context, depending on the level of approximation (or, more generally, on the iterated quantities that are explicitly calculated in the self-consistent procedure), one can identify different classes of methods, associated with an increasing level of numerical complexity. We start by (i)  the less demanding schemes, where only the self-energy and the physical response functions are iterated. This class corresponds to the TRILEX-based approaches \cite{Ayral2015,Vucicevic2017,Stepanov2019Consistent}, with the famous GW technique representing a subclass of it \cite{Hedin1965new,Aryasetiawan1998GW,Vucicevic2017,Golze2019GW}. Then, a step higher, we have (ii) the algorithms, where, in addition to the self-energy and the response functions, the three-leg fermion-boson vertex (the so-called ``Yukawa coupling'' or ``Hedin vertex'') is also iterated. These  schemes are classified as \emph{``single-boson exchange''} (SBE) approaches \cite{Krien2019Single,Bonetti2022Single,Kiese2024Embedded} since all microscopic processes entailing the exchange of a single bosonic collective mode are explicitly included in the respective self-consistent procedures. Eventually, (iii) the most advanced schemes are considered, where even the four-leg (two-particle) vertex functions, describing the exchange of more than one bosonic mode, are iterated. Henceforth, this class of schemes, to which also methods based on parquet equations generally belong \cite{Bickers2004Selfconsistent,Essl2026Stabilizing}, is often associated with the name of \emph{``multi-boson exchange''} (MBE) \cite{Krien2020Tiling,Krien2022Plain,Fraboulet2022Single,Gievers2022Multiloop,Fraboulet2026Multiloop,AlEryani2026Functional}. \changed{We note that, in the literature, the labels SBE and MBE are not always used consistently. They either refer to the BE decomposition used to classify the vertices or to the vertices that are determined self-consistently \footnote{For instance, the ``embedded MBE'' approach of Ref.~\cite{Kiese2024Embedded} takes the $U$-irreducible vertex $\Lambda^U$ of a self-consistently embedded cluster as fixed input and iterates only the SBE vertices, which places it in the SBE class in our terminology. Conversely, the multiloop SBE fRG of Refs.~\cite{Gievers2022Multiloop,Fraboulet2026Multiloop,AlEryani2026Functional} includes the flow of the MBE vertices $M_r$, which, at loop convergence, reproduces the self-consistent parquet solution and thus corresponds to the MBE class in our terminology. In this paper, we call the method MBE when the MBE vertices are self-consistently included and SBE when MBE vertices are only given as an input or neglected.}.}

Evidently, all classes of methods mentioned above satisfy the first two requirements, in that, through the Feynman diagrammatics, they are explicitly  based on quantum field theory, while, at the same time,  they allow for a self-consistent evaluation of one-particle (Green's function, ARPES spectra) and two-particle properties (dynamical response functions in all sectors) of the many-electron problem considered. However, their applicability often faces serious troubles in the most interesting cases of intermediate-to-strong electronic correlations. Specifically, by increasing values of the electronic interaction $U$ and/or inverse temperature $\beta = 1/T$ (with $k_\mathrm{B} \equiv 1$), the iterative procedures of these approaches display a general tendency to (abruptly) converge to \emph{unphysical} fixed-point solutions, i.e., to solutions that can violate intrinsic physical properties of the problem considered, \emph{irrespective} of the numerical accuracy of the calculation. The origin of such \emph{``misleading convergences''}, for their apparent similarity to those observed in self-consistent perturbative expansions of the Luttinger--Ward functional (LWF) \cite{Kozik2015Nonexistence,Stan2015Unphysical,Rossi2015Skeleton,Tarantino2017Selfconsistent}, was generally ascribed to the intrinsic multivaluedness of this functional and to the associated  divergences \cite{Gunnarsson2017Breakdown} of important classes of scattering diagrams (namely the two-particle-irreducible vertex functions) \cite{Schaefer2013Divergent,Janis2014Critical,Schaefer2016Nonperturbative,Chalupa2018Divergences,Vucicevic2018Practical,Springer2020Interplay}. However, as recently noted in Ref.~\cite{Essl2026Origin},  for diagrammatic approaches that are \emph{not} directly derivable from the LWF (which is the case for TRILEX, SBE, and MBE), this interpretation appears
rather questionable. Notably, a closely related instability of the physical solution of the parquet equations was independently identified via Monte Carlo simulations, in the critical region of the two-dimensional Ising universality class \cite{Pollet2026Two}.

In our work, we explicitly address this question, by  demonstrating that the misleading convergence observed in all classes of diagrammatic methods listed above is \emph{not} due to an intrinsic breakdown of the many-electron functional theory, but \emph{exclusively} to the  stability condition of the iterative algorithms used in the calculations. This indeed represents a fundamental piece of information, which paves the way towards a general remedy, applicable to all these diagrammatic methods. In fact, while the
convergence issues directly linked to the multivaluedness of the LWF \cite{Kozik2015Nonexistence,Kim2020,Kim2022}, still represent an \emph{unsolved} problem, a clear strategy can be exploited to stabilize the fixed points of all the iterative schemes considered. In fact, as discussed in Refs.~\cite{Essl2026Origin,Essl2026Stabilizing}, as well as, more generally in Refs.~\cite{guckenheimer1983,strogatz2024nonlinear, chen2024nonlinear}, the stability condition of the fixed point of a given iterative scheme is encoded in the eigenspectrum of its Jacobian, with the fixed point becoming \emph{unstable}, when the absolute value of one of its (complex) eigenvalues exceeds unity. If this happens, however, the fixed point can be made locally stable again by changing the iteration scheme.
In this work, we do so by inverting the sign of the corresponding eigenvalue, i.e., by stabilizing the associated eigendirection in the computational space.

To achieve these goals, we apply self-consistent algorithms belonging to the three classes listed before (TRILEX, SBE, MBE) to exactly solvable models (i.e., the Hubbard model in the atomic limit and the zero-point model) and compare the  numerical results obtained for different one- and two-particle quantities against the values of the respective exact solutions.
In this framework, we are then able (i) to perform a rigorous study of the stability/instability regions within the conventional iteration scheme, and (ii) to demonstrate that the modified iteration procedure, proposed in Ref.~\cite{Essl2026Origin} and tested on the parquet equations in Ref.~\cite{Essl2026Stabilizing}, \emph{does fully suffice} to achieve, for all the methods considered, a convergence to the \emph{physical fixed point} in the parameter regions where it was unstable for the conventional iteration algorithm. We successfully apply the stabilization scheme in the different diagrammatic methods both with perturbative and nonperturbative input. In particular, we underline how the physical fixed point is stabilized, in this way, even in the nonperturbative (intermediate-to-strong) coupling regimes, i.e., also \emph{after} the crossing of one or more vertex divergences.

This achievement is quite remarkable both from a fundamental as well as from a practical perspective. On the one hand, it demonstrates, indisputably, the complete distinction of ``misleading convergence'' problems in BE approaches from the intrinsic convergence problem plaguing the self-consistent diagrammatic expansion of the LWF. On the other hand, it identifies a working procedure to stabilize self-consistent algorithms in the strong-coupling regimes. This practical aspect is highly relevant, e.g., for diagrammatic extensions of the dynamical mean-field theory (DMFT) \cite{Rohringer2018Diagrammatic}, which are often expressed in the BE formalism \cite{Kiese2024Discrete,Kiese2024Embedded}.

Eventually, we must underline that, in the course of our study, we have also discovered the occurrence of a \emph{different kind} of two-particle vertex divergences compared to those mentioned above. In fact, while the latter, extensively discussed in the current literature \cite{Schaefer2013Divergent,Janis2014Critical,Schaefer2016Nonperturbative,Gunnarsson2016Parquet,Gunnarsson2017Breakdown,Chalupa2018Divergences,Vucicevic2018Practical,Melnick2020Fermi,Reitner2020Attractive,Chalupa2021Fingerprints,Mazitov2022Local,Pelz2023Highly,Adler2024Nonperturbative,Essl2024General,Moghadas2026Effective}, affect the two-particle-irreducible vertex diagrams only, our analysis demonstrates that divergences can also appear, \changed{in contrast to what was generally expected}, in the SBE diagrams, with a direct impact on the stability of the self-consistent algorithms.

The paper is organized as follows. In Sec.~\ref{sec:Methods_and_models}, we introduce the Hubbard atom and zero-point model and then define the different diagrammatic schemes (TRILEX, SBE, MBE) as well as the stabilization method. In Sec.~\ref{sec:Stability}, we analyze the stability of the various methods with respect to the appearance of vertex divergences and unstable eigendirections. In Sec.~\ref{sec:Results}, we show the successful application of the stabilization procedure to all the diagrammatic schemes, with different (nonperturbative/perturbative) inputs. We conclude the paper with some remarks on more realistic implications in Secs.~\ref{sec:Outlook} and \ref{sec:Conclusion}. In the appendices, we provide more details on analytical expressions and numerical algorithms as well as additional results in other parameter regimes and for the zero-point model.

\section{Models and methods}
\label{sec:Methods_and_models}

In this section, we introduce the Hubbard atom, the workhorse for our proof of principle, and its single-frequency limit, the zero-point model. Furthermore, we define the different self-consistent schemes (TRILEX, SBE, MBE) used for our calculations. Finally, we derive the specific expressions of the stabilization method, introduced in Ref.~\cite{Essl2026Origin}, for the three classes of diagrammatic schemes considered.

\subsection{Models}

\subsubsection{Hubbard atom (HA)}

In the zero-hopping limit (or atomic limit) of the Hubbard model, the lattice sites decouple completely, resulting in a lattice of isolated interacting sites. The action of one of these (equivalent) sites, referred to as \emph{Hubbard atom} (HA), is given by
\begin{align}
    \nonumber S_\text{HA} &= -\tfrac{1}{\beta}\sum_\nu (\ii\nu+\mu)(\bar c_\uparrow^\nu c_\uparrow^\nu+\bar c_\downarrow^\nu c_\downarrow^\nu)\\
    &\phantom{=}~+ \tfrac{1}{\beta^3}\sum_{\nu,\nu',\omega}U\bar c_\uparrow^{\nu'+\omega} \bar c_\downarrow^{\nu} c_\downarrow^{\nu+\omega} c_\uparrow^{\nu'}.
    \label{eq:Hubbard_atom}
\end{align}
Here, the fermionic Grassmann fields $\bar c_\sigma^\nu, c_\sigma^\nu$ depend on spins $\sigma=\,\uparrow,\downarrow$ and Matsubara frequencies $\nu$. We use $\nu, \nu' =(2n+1)\pi/\beta$ for fermionic and $\omega=2n\pi/\beta$ for bosonic Matsubara frequencies, where $n\in\mathbb{Z}$. $U$ denotes the Hubbard interaction and $\mu$ the chemical potential. The system is exactly solvable, since its entire Hilbert space comprises four eigenstates (no occupation $|0\rangle$ with $E_0=0$, single occupation $|\!\uparrow\rangle, |\!\downarrow\rangle$ with $E_{\uparrow}=E_\downarrow=-\mu$, double occupation $|\updn\rangle$ with $E_\updn=U-2\mu$).  We further introduce the parameter $\delta\mu = \mu - U/2$, which indicates how far away the system is from half-filling \footnote{At $\delta\mu = 0$, we have $\mu = U/2$ and $n_\sigma=\langle c^\dagger_\sigma c_\sigma\rangle = 1/2$. Thus the system is at half-filling.}. Throughout this paper, we express $U$ and $\delta\mu$ in units of temperature $T=1/\beta$ so the parameter space is two-dimensional. 

As the HA allows for a formally exact solution, it is frequently used as a benchmark for diagrammatic methods in Hubbard-model systems \cite{Pairault2000Strong,Thunstroem2018Analytical,wallerberger2021,Essl2024General,Rohshap2025Two,AlEryani2026Functional,Grosso2026Adaptive}. From our perspective, it is important to underline that, despite the simplicity of the underlying Hamiltonian, the two-particle correlations of the HA are highly nontrivial and encode many nonperturbative features of more realistic models \cite{Chalupa2021Fingerprints,Adler2024Nonperturbative}. The full exact solution of the model is presented in App.~\ref{sec:HA-vertices}.

\subsubsection{Zero-point model (ZPM)}

An even more drastic simplification is the \emph{zero-point model (ZPM)} \cite{Rossi2015Skeleton}, which has the spin $\sigma=\,\uparrow,\downarrow$ of the fermionic fields $\bar c_\sigma$ and $c_\sigma$ as \emph{only} degree of freedom:
\begin{align}
    S_\mathrm{ZP} = -G_0^{-1} \left(\bar c_\uparrow c_\uparrow + \bar c_\downarrow c_\downarrow\right) + U \bar c_\uparrow \bar c_\downarrow c_\downarrow c_\uparrow .
    \label{eq:ZPM}
\end{align}
In the spirit of Ref.~\cite{Kim2020}, we set the Green's function to $G_0^{-1}=\delta\mu$ with $\delta\mu\in\mathbb{C}$ for a better connection to the HA. In this way, the imaginary part plays the role of a single (fully decoupled) Matsubara frequency which we set to $\mathrm{Im}\,\delta\mu=\ii\pi$ while the real part $\mathrm{Re}\,\delta\mu$ refers to the doping away from particle-hole symmetry (for more details see App.~\ref{sec:ZP_BE} and Sec.~A.3 in Ref.~\cite{Essl2026Stabilizing}). This model radically simplifies the system, which even allows a fully analytical expression of its Luttinger--Ward functional (LWF), making it a model that has often been studied in the past in the context of multivaluedness of the LWF and misleading convergence \cite{Rossi2015Skeleton,Kim2020,vanhoucke2024,Essl2026Origin,Essl2026Stabilizing}.
Physically and from a functional dependence, the ZPM resembles the coherent potential approximation (CPA) solution of the binary mixture model in infinite dimensions \cite{Essl2026Origin}, which can capture a metal-insulator transition by increasing interaction at half-filling \cite{Schaefer2016Nonperturbative}. Further the ZPM is, from a functional point of view, similar to the mean-field spin-density wave model introduced in Ref.~\cite{Altshuler1998} to represent a nearly antiferromagnetic Fermi liquid. 

In the context of this work, it is quite remarkable that, in the framework of stability analyses, one observes qualitative similarities with the results obtained for the (more complex) HA for both conventional parquet \cite{Essl2026Stabilizing} and the investigation in this work.

\subsection{Boson-exchange (BE) formalism}

\subsubsection{General BE equations}
\label{sec:MBE_formalism}

The boson-exchange (BE) decomposition of the full four-point vertex $F$ provides a very efficient and physically transparent way of handling the complicated frequency dependence of two-particle vertices, allowing to derive  diagrammatic schemes with different degrees of complexity for solving self-consistently the many-electron problem. 
Introduced originally for the fermionic Hubbard model \cite{Krien2019Single}, it has been later extended to generic energy-conserving fermionic models with quartic interactions \cite{Gievers2022Multiloop,Gievers2025PhD,AlEryani2026Functional}.

We start by recalling that the four-point vertex $F$, which  describes, physically, the full scattering amplitude between two interacting fermions, is formally defined by the connected two-particle correlation function and the fermionic Green's function $G$,
\begin{subequations}
    \begin{align}
        \langle c_{\sigma_1}^{\nu_1}c_{\sigma_2}^{\nu_2}\bar c_{\sigma_{2'}}^{\nu_{2'}}\bar c_{\sigma_{1'}}^{\nu_{1'}}\rangle_\mathrm{con} &= G_{\sigma_1}^{\nu_1}G_{\sigma_2}^{\nu_2} F^{\nu_{1'}\nu_{2'}\nu_1\nu_2}_{\sigma_{1'}\sigma_{2'}\sigma_1\sigma_2} G_{\sigma_{2'}}^{\nu_{2'}}G_{\sigma_{1'}}^{\nu_{1'}},
        \label{eq:F}\displaybreak[2]\\
        \mathrm{with}~G_\sigma^\nu &= - \langle c_\sigma^\nu\bar c_\sigma^\nu\rangle.
    \end{align}
\end{subequations}
Energy conservation reduces $F$ to a function depending on three frequencies. Furthermore, due to SU(2) symmetry of the considered models, the spin arguments $\sigma_i$ of $F$ can be combined to the four physical channels $r$ ($c$harge, $m$agnetic, $s$inglet, $t$riplet) \cite{Bickers2004Selfconsistent,Rohringer2012Local,Rohringer2018Diagrammatic}. Hence, the four-point vertex is expressed as $F_r^{\omega\nu\nu'}$, where $\omega$ is a bosonic Matsubara frequency and $\nu$, $\nu'$ are fermionic Matsubara frequencies (for details see App.~\ref{sec:Frequencies}).

In the BE formalism, all diagrammatic contributions to $F$ are classified according to the criteria of their $U$ reducibility (a Feynman diagram for $F$ is defined $U$ reducible if it can be split in two diagrams by cutting one interaction ($U$) line) as well as of their two-particle reducibility (a Feynman diagram for $F$ is defined two-particle-reducible (2PR) if it can be split in two diagrams by cutting two fermionic propagators lines).
In this way, $F$ gets naturally decomposed into $U$-reducible vertices, called \emph{single-boson exchange} (SBE) vertices $\nabla_r^{\omega\nu\nu'}$, 2PR (but $U$-irreducible) \emph{multi-boson exchange} (MBE) vertices $M_r^{\omega\nu\nu'}$, and the fully two-particle-irreducible (2PI) rest function $\Lambda_r^{\omega\nu\nu'}$ [cf.\ Eq.~\eqref{eq:Diagrammatic_decomposition}]. The SBE vertices $\nabla_r^{\omega\nu\nu'}$ comprise a \emph{bosonic propagator} $W_r^{\omega}$ coupled to two fermion-boson vertices $\lambda_r^{\omega\nu}$, known as \emph{Hedin} vertices: 
\begin{align}
    \nabla_r^{\omega\nu\nu'} = \lambda_r^{\omega\nu}W_r^\omega\lambda_r^{\omega\nu'}.
    \label{eq:SBE-vertex}
\end{align}
A major technical advantage of the BE decomposition is that the more involved frequency dependence of the full two-particle vertex $F_r^{\omega\nu\nu'}$ is rearranged in vertices with a simpler frequency structure, i.e., $W_r^\omega$ depends on a single bosonic frequency and $\lambda_r^{\omega\nu}$ on one bosonic and one fermionic frequency.
Further, the BE decomposition, based on the exchange of one (or more) bosonic excitations, is directly  related to the physical (linear) response of the system in the respective channels. Hence, the BE formalism naturally links the results of different kinds of one- and two-particle spectroscopy in a transparent way. For this reason, the BE decomposition is not only exploited to render self-consistent diagrammatic algorithms more performing, but also to interpret, in a rigorous way, the physical information encoded in their results \cite{Krien2019Single,Krien2022Plain,Adler2024Nonperturbative}.

For practical applications, the bosonic propagators $W_r^\omega$ can be computed through the corresponding (bosonic) Dyson equation [see also Eq.~\eqref{eq:W_r_recursive}],
\begin{align}
    W^\omega_r = \frac{U_r}{1-U_r P_r^\omega},
    \label{eq:W_r}
\end{align}
where bare vertices $U_r$ are defined in the four physical channels:
\begin{align}
    U_c = -U, \quad U_m = U, \quad U_s = -2U, \quad U_t = 0,
    \label{eq:U_r}
\end{align}
taking the role of the corresponding ``bare bosonic propagator''.

The bosonic self-energies (or polarizations) $P_r^\omega$ are obtained by a Schwinger--Dyson equation, 
\begin{align}
    P^\omega_r = \tfrac{1}{\beta}\sum_{\nu''}\lambda_r^{\omega\nu''}\Pi_r^{\omega\nu''},
    \label{eq:P_r}
\end{align}
where the bubbles $\Pi_r^{\omega\nu}$ are defined as channel-specific products of Green's functions $G^\nu$:
\begin{subequations}
    \label{eq:Pi_r}
    \begin{alignat}{3}
        &\Pi^{\omega\nu}_{r=c,m} &&= \Pi^{\omega\nu}_\ph &&= -G^\nu G^{\omega+\nu},\displaybreak[2]\\
        &\Pi^{\omega\nu}_{r=s,t} &&= \Pi^{\omega\nu}_\pp &&= \tfrac{1}{2}G^{-\nu}G^{\omega+\nu}.
    \end{alignat}
\end{subequations}
The Hedin vertices are computed by another Schwinger--Dyson equation,
\begin{align}
    \lambda^{\omega\nu}_r &= 1 + \tfrac{1}{\beta}\sum_{\nu''}\Pi_r^{\omega\nu''} T_r^{\omega\nu''\nu}, 
    \label{eq:lambda_r}
\end{align}
where $T_r^{\omega\nu\nu'}$ are the $U$-irreducible contributions of the full vertices $F_r^{\omega\nu\nu'}$ with respect to the channel $r$:
\begin{align}
    T_r^{\omega\nu\nu'} &= F_r^{\omega\nu\nu'}-\lambda^{\omega\nu}_r W_r^\omega \lambda_r^{\omega\nu'}.
    \label{eq:T_r_from_F_r}
\end{align}
The MBE vertices $M_r^{\omega\nu\nu'}$ fulfill the Bethe--Salpeter equations [see also Eq.~\eqref{eq:M_r_symmetrized}]:
\begin{align}
    M^{\omega\nu\nu'}_r &= \tfrac{1}{\beta}\sum_{\nu''}(T_r^{\omega\nu\nu''}-M_r^{\omega\nu\nu''})\Pi_r^{\omega\nu''}T_r^{\omega\nu''\nu'}.
    \label{eq:M_r}
\end{align}
Due to Pauli's exclusion principle and the purely on-site nature of the interaction considered, both the bare four-point vertex and the SBE vertices vanish in the triplet channel, i.e., $U_t=0=W_t$ and $\lambda_t^{\omega\nu}=1$. In contrast, MBE vertices in the triplet channel do not vanish by symmetry arguments, i.e., $M_t^{\omega\nu\nu'}\neq 0$ in general.

Finally, for computing the self-energy $\tilde \Sigma$ beyond its Hartree term,
\begin{align}
    \Sigma_\mathrm{H}=U\tfrac{1}{\beta}\sum_{\nu''}G^{\nu''},
    \label{eq:Sigma_Hartree}
\end{align}
we use the Schwinger--Dyson equation written in SBE vertices:
\begin{subequations}
    \label{eq:Sigma}
    \begin{align}
        \tilde\Sigma^\nu &= \tfrac{1}{\beta}\sum_{\nu''}W_c^{\nu''-\nu}\lambda_c^{\nu''-\nu,\nu}G^{\nu''}-\tfrac{1}{\beta}\sum_{\nu''}U_cG^{\nu''}
        \label{eq:Sigma_c}\\
		&= \tfrac{1}{\beta}\sum_{\nu''}W_m^{\nu''-\nu}\lambda_m^{\nu''-\nu,\nu}G^{\nu''}-\tfrac{1}{\beta}\sum_{\nu''}U_mG^{\nu''}
        \label{eq:Sigma_m}\\
		&=-\tfrac{1}{2\beta}\sum_{\nu''}W_s^{\nu''+\nu}\lambda_s^{\nu''+\nu,-\nu}G^{\nu''}+\tfrac{1}{2\beta}\sum_{\nu''}U_sG^{\nu''}
        \label{eq:Sigma_s}\\
        &=\tfrac{1}{2\beta}\sum_{\nu''}\left[ W_c^{\nu''-\nu}\lambda_c^{\nu''-\nu,\nu}+W_m^{\nu''-\nu}\lambda_m^{\nu''-\nu,\nu}\right]G^{\nu''}.
        \label{eq:Sigma_symmetrized}
    \end{align}
\end{subequations}
Evidently, while for the exact solution, the representations of $\tilde\Sigma^\nu$ in different channels are fully equivalent \cite{Schäfer2021How,Dong2022Mechanism}, in perturbative approaches, they can lead to different results \cite{Patricolo2025Single}.
For our numerical analysis, we take Eq.~\eqref{eq:Sigma_symmetrized}, an average over the representation in the $c$ and $m$ channel, as this turns out as numerically robust with respect to frequency asymptotics (cf.\ App.~\ref{sec:change_it_maps}).  
Equations~\eqref{eq:Sigma} only consider the frequency-dependent part of the self-energy excluding the Hartree term $\Sigma_\mathrm{H}$, Eq.~\eqref{eq:Sigma_Hartree}, so the total self-energy yields $\Sigma^\nu = \tilde\Sigma^\nu + \Sigma_\mathrm{H}$. The fermionic Green's function is then obtained from the fermionic Dyson equation,
\begin{align}
    G^\nu = \frac{1}{[G_0^{-1}]^\nu-\Sigma^\nu},
    \label{eq:G}
\end{align}
where $G_0^\nu=1/(\ii\nu+\mu)$ is the noninteracting Green's function.

\begin{figure*}
    \includegraphics[width=1.0\textwidth]{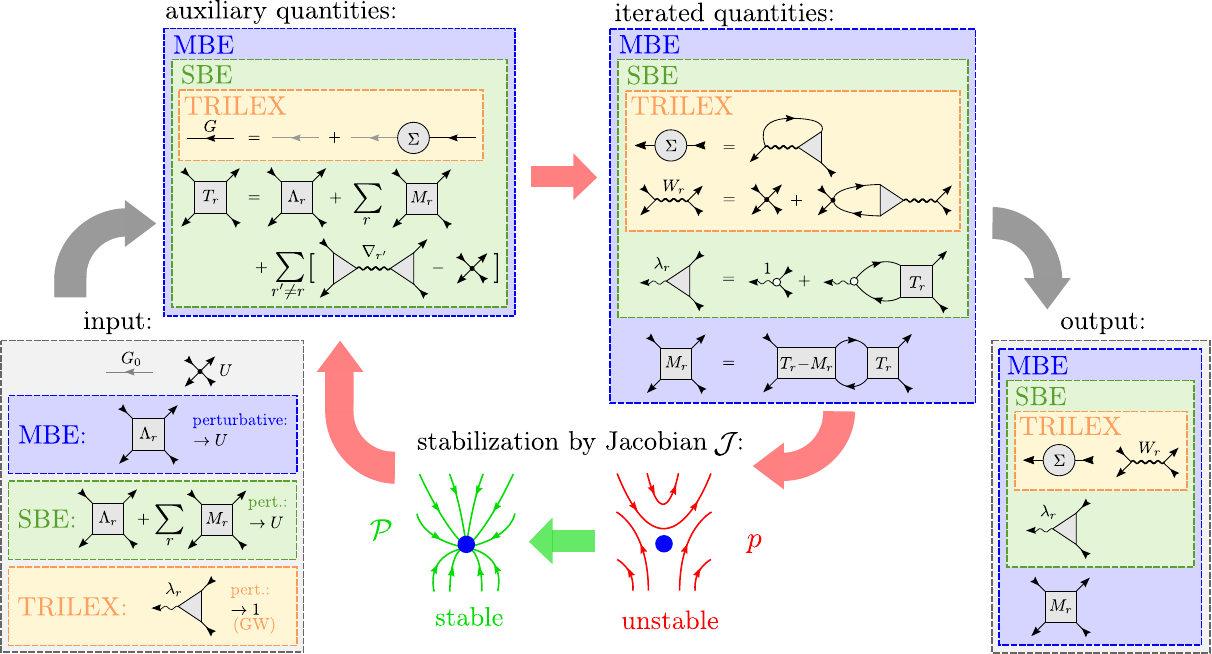}
    \caption{Overview of the different iteration schemes. The boson-exchange (BE) equations~\eqref{eq:SBE-vertex}--\eqref{eq:G}, here illustrated diagrammatically, are solved iteratively until convergence (red arrows mark the self-consistency loop).  TRILEX iterates over $\Sigma$, $W_r$ with input of $\lambda_r$, SBE iterates over $\Sigma$, $W_r$, $\lambda_r$ with input of $\Lambda_r+\sum_r M_r$, Eq.~\eqref{eq:Lambda^U}, and MBE iterates over $\Sigma$, $W_r$, $\lambda_r$, $M_r$ with input of $\Lambda_r$. In our analysis of nonperturbative approaches, the input is taken exactly, while in the perturbative approaches, the inputs are replaced by $\lambda_r\to 1$, $\Lambda_r+\sum_r M_r\to U_r$, and $\Lambda_r\to U_r$, respectively. Note that TRILEX with perturbative input can be identified with a GW approach. Key to our method is the stabilization step through the Jacobian $\mathcal{J}$ (green arrow). When the $\mathcal{P}$ matrix is used instead of the conventional damping parameter $p$ [cf.\ Eq.~\eqref{eq:SOT}], an unstable fixed point (blue dot) is transformed to a stable one. The output is reached after obtaining (numerically) self-consistency of the vertices and depends on the respective method.}
    \label{fig:MBE-SOT-Scheme}
\end{figure*}

With an input of the fully irreducible vertex $\Lambda_r$, Eqs.~\eqref{eq:SBE-vertex}--\eqref{eq:G} constitute a set of self-consistent equations for the vertices $\Sigma, W_r, \lambda_r, M_r$ (see App.~\ref{sec:Jacobian-MBE} for the explicit iterative maps), which is diagrammatically visualized in Fig.~\ref{fig:MBE-SOT-Scheme}. In the self-consistency loop, the $U$-irreducible vertices $T_r$ are computed from $W_r$, $\lambda_r$, $M_r$, and the two-particle-irreducible rest function $\Lambda_r$, i.e., $T_r=T_r[W_r,\lambda_r,M_r,\Lambda_r]$ by inserting the BE decomposition of $F$ in Eq.~\eqref{eq:T_r_from_F_r} [see Eqs.~\eqref{eq:F_in_physical_channels} and Eqs.~\eqref{eq:T_r_from_MBE}]. We refer to this general set of relations as \emph{BE equations}.

We further distinguish the fully 2PI vertex $\Lambda_r$ from the fully $U$-irreducible vertex $\Lambda_r^U$. The latter can be obtained from $\tilde\Lambda_r = \Lambda_r-U_r$ by adding the $U$-irreducible MBE vertices $M_r$ [for details see Eq.~\eqref{eq:Lambda^U}].

\subsubsection{Different diagrammatic approaches}
\label{sec:GW_SBE_MBE}

\begin{comment}
\begin{table}[]
    \centering
    \begin{tabular}{c|ccc|ccc}
        \hline
        \hline
         & & weak & & & strong & \\
        quantity & 3TL & SBE & MBE & 3TL & SBE & MBE\\
        \hline
        $\Sigma$ & $\circlearrowleft$ & $\circlearrowleft$ & $\circlearrowleft$ & $\circlearrowleft$ & $\circlearrowleft$ & $\circlearrowleft$ \\
        $W_r$ & $\circlearrowleft$ & $\circlearrowleft$ & $\circlearrowleft$ & $\circlearrowleft$ & $\circlearrowleft$ & $\circlearrowleft$ \\
        $\lambda_r$ & $1$ & $\circlearrowleft$ & $\circlearrowleft$ & $\checkmark$ & $\circlearrowleft$ & $\circlearrowleft$\\
        $M_r$ & $0$ & $0$ & $\circlearrowleft$ & $\checkmark$ & $\checkmark$ & $\circlearrowleft$\\
        $\Lambda_r$ & $U_r$ & $U_r$ & $U_r$ & $\checkmark$ & $\checkmark$ & $\checkmark$ \\
        \hline
        \hline
    \end{tabular}
    \caption{Overview on different schemes. The circular arrow $\circlearrowleft$ means that the respective vertex is iterated in the self-consistent cycle and the checkmark $\checkmark$ means that the respective vertex is provided as exact input. In the weak approaches, the two-particle-irreducible vertex $\Lambda_r$ is given by the bare vertex $U_r$. Furthermore, if not iterated over, the Hedin vertices $\lambda_r$ are approximated as $1$ and the multi-boson vertices $M_r$ as $0$.}
    \label{tab:Overview}
\end{table}
\end{comment}

Within the general framework of the BE formalism, we consider three different classes of self-consistent schemes (TRILEX, SBE, MBE), which increase in algorithmic complexity depending on which quantities are iterated over (and which quantities are kept fixed). Figure~\ref{fig:MBE-SOT-Scheme} provides an overview of all the schemes, which are discussed below in more detail.

Before examining the three approaches, it is important to stress the following aspect: We will start our numerical investigation by setting the \emph{noniterated} quantities of TRILEX, SBE, and MBE to their exact values (exploiting the exact solutions of the HA and ZPM in Apps.~\ref{sec:HA-vertices}--\ref{sec:ZP_BE}). In this case, the correct outcome of the self-consistent procedure, for all the three classes of methods considered, should yield, in principle, the exact solution of the problem under investigation.
Evidently, analyzing under which condition this is really the case, and whether the convergence to the exact solution can be restored, is of pivotal importance for all self-consistent diagrammatic approaches designed for the strong-coupling/nonperturbative regimes, such as the diagrammatic extensions of DMFT \cite{Rohringer2018Diagrammatic}.

Otherwise, if the fixed quantities are approximated perturbatively (e.g., perturbatively in  $U$, as we will do in Sec.~\ref{sec:Weak-coupling_results}), the three classes of methods, from TRILEX to MBE, will directly correspond to different levels of perturbative approximations, from the most rough to the most precise ones.

The simplest approach we consider is known under the name of \emph{triply irreducible local expansion (TRILEX)}. Here, the input that is kept fixed are the three-leg Hedin vertices $\lambda_r$, Eq.~\eqref{eq:lambda_r}. This allows to iteratively calculate the self-energy $\Sigma$, Eqs.~\eqref{eq:Sigma_Hartree}--\eqref{eq:G}, and the bosonic propagators $W_r$, Eqs.~\eqref{eq:W_r}--\eqref{eq:P_r} (see yellow boxes in Fig.~\ref{fig:MBE-SOT-Scheme}).

Consistent with our discussion above, the value of $\lambda_r$ can be replaced by their exact values (as we will do in most of our calculations for the HA and ZPM) or, instead, approximated. This can be done either perturbatively, e.g. by setting $\lambda_r$ to $1$, or nonperturbatively, e.g., by exploiting the values obtained by DMFT.  
It is worth noticing here that TRILEX with perturbative input can be identified with an extended version of the widely used \emph{GW} approach, that takes into account multiple channels \cite{Vucicevic2017} \footnote{Conventional GW approaches only iterate over one or two of the three channels $c$, $m$, $s$ (usually $c$) for the bosonic propagator. We use the combination of the $c$ and $m$ channel, Eq.~\eqref{eq:Sigma_symmetrized}, for our numerical results.}. 
On the other hand, we must recall that the acronym TRILEX has been originally introduced in this latter (nonperturbative) context, to identify an agile diagrammatic extension of DMFT. Specifically, in conventional TRILEX for lattice systems, three-point vertices $\lambda_r$ are computed locally through DMFT and then given as input for a self-consistent loop over the bosonic propagators $W_r$, Eqs.~\eqref{eq:W_r}--\eqref{eq:P_r}, and the fermionic propagators $G$, Eqs.~\eqref{eq:Sigma_Hartree}--\eqref{eq:G}.

Evidently, the huge algorithmic advantage of the TRILEX-based schemes is that the four-point vertices $M_r$ and $\Lambda_r$ are \emph{not} needed as an input as well as along the iterations, since they are not included in Eqs.~\eqref{eq:W_r}--\eqref{eq:P_r} and \eqref{eq:Sigma_Hartree}--\eqref{eq:G}. In the version of TRILEX we implement here, the $c$ and $m$ channel feed into the self-energy through Eq.~\eqref{eq:Sigma_symmetrized}, which leaves the $s$ channel, i.e., $W_s$, decoupled from the self-consistent loop. 

As a next step of complexity, one can include the Hedin vertices in the iteration, keeping the MBE vertices $M_r$ fixed [therefore, not iterating the Bethe--Salpeter equations~\eqref{eq:M_r}]. This procedure, illustrated in the green boxes of the flow chart in Fig.~\ref{fig:MBE-SOT-Scheme},  corresponds to the \emph{single-boson exchange (SBE)} class of approaches.
The development of SBE-based schemes was primarily motivated from a numerical perspective. In fact, the slowly decaying frequency structures of the full two-particle vertex $F$ \cite{Rohringer2012Local,Wentzell2020High} can be directly encoded in the SBE vertices $\nabla_r^{\omega\nu\nu'}=\lambda_r^{\omega\nu}W_r^\omega\lambda_r^{\omega\nu'}$. Then, since $W_r^\omega$ and $\lambda_r^{\omega\nu}$ have a simpler frequency dependence, they can be stored on much larger grids than the MBE vertices $M_r^{\omega\nu\nu'}$, which explicitly depend on all of the three frequency variables. 
Similarly as above, different possible choices can be made for the input of this class of schemes. On a perturbative level, one could set $\Lambda_r=U_r$ and MBE vertices are completely neglected, i.e., $M_r=0$, which is known as \emph{SBE approximation} \cite{Gievers2022Multiloop,Gievers2025Subleading,AlEryani2026Functional}. On the other hand, one could take the sum of $U$-irreducible vertices $\Lambda_r^U = \tilde\Lambda_r+\sum_r M_r$ as exact or as a nonperturbative input [cf.\ Eqs.~\eqref{eq:Lambda^U}]. 
The latter approach can be identified with \textit{embedded MBE} from Ref.~\cite{Kiese2024Embedded}, where the $U$-irreducible vertex of a \mbox{(cluster-)}DMFT calculation is used as an input for the SBE self-consistency.

Finally, the solution of all BE equations \eqref{eq:SBE-vertex}--\eqref{eq:G} with some input for $G_0$, $U_r$, and $\Lambda_r$ is physically equivalent to that of the usual parquet equations \cite{Bickers2004Selfconsistent,Krien2022Plain}. 
This is because the same types of diagrams are included, they are only classified in a different way ($U$ reducibility instead of two-particle reducibility). We refer to it as the \emph{multi-boson exchange (MBE)} approach since all vertices ($W_r$, $\lambda_r$, $M_r$) are iterated over. In the perturbative limit, the fully 2PI vertex is set to its first-order approximation, i.e., $\Lambda_r\to U_r$, which is known as the \emph{parquet approximation}. 

In order to simulate the situation occurring in one of the most advanced diagrammatic extensions of DMFT (namely, the full/parquet-based dynamical vertex approximation \cite{Toschi2007,Valli2015Dynamical,Kauch2020Generic}), we perform computations where the input for $\Lambda_r$ is set to the exact $\Lambda_r$ of the system. For the HA, a purely local system, DMFT yields numerically the exact result. Thus, for the HA, MBE with the exact input of $\Lambda_r$ is formally equivalent to full parquet D$\Gamma$A, which uses the DMFT result as input for the parquet equations \cite{Toschi2007}. 

\subsection{Stabilized iteration}
\label{sec:Jacobian_formalism}

As illustrated in Fig.~\ref{fig:MBE-SOT-Scheme}, the various schemes iterate over the quantities $\Sigma$, $W_r$, $\lambda_r$, $M_r$ to find self-consistent solutions to the BE equations~\eqref{eq:SBE-vertex}--\eqref{eq:G}. In the following, we summarize Eqs.~\eqref{eq:SBE-vertex}--\eqref{eq:G} as $\Psi=f[\Psi]$, where the generic state vector $\Psi$ contains all iterated quantities, i.e., $\Psi=(\Sigma,W_r,\lambda_r,M_r)$ of the specific scheme considered \footnote{In TRILEX, $\Psi$ contains only $\Sigma$ and $W_r$, in SBE, it additionally includes $\lambda_r$, and, in MBE, it additionally includes $M_r$.}, and $f[\Psi]$ is the map given by the right-hand sides of Eqs.~\eqref{eq:Sigma_Hartree}--\eqref{eq:Sigma}, \eqref{eq:W_r}--\eqref{eq:Pi_r}, \eqref{eq:lambda_r}--\eqref{eq:T_r_from_F_r}, and \eqref{eq:M_r} [for details see Eqs.~\eqref{eq:f[Psi]}].

Conventionally, the BE equations~\eqref{eq:SBE-vertex}--\eqref{eq:G} in the different approaches are solved by the \emph{damped iteration},
\begin{align}
    \Psi_{n+1} = p\, f[\Psi_{n}] + (1-p) \Psi_{n}.
    \label{eq:Damped-iteration}
\end{align}
Hence, at the $(n+1)$-th iteration, $\Psi$ is updated as a mixture of the previous state $\Psi_{n}$ and the map $f[\Psi]$ given by the BE equations \eqref{eq:SBE-vertex}--\eqref{eq:G} 
evaluated at the previous state, i.e., $f[\Psi_{n}]$.
Hereby, $p\in (0,1]$ is a mixing (or damping) parameter, where $p=1$ refers to the case with no mixing and $p\to 0$ is the limit of no update.

Close to a self-consistent solution, i.e., to a fixed point of the map ($\Psi^* \!=\!  f[\Psi^*]$), the updates are essentially controlled by the $p$-dependent Jacobian,
\begin{align}
    \mathcal{J}^p[\Psi] = p\frac{\delta f}{\delta\Psi}+(1-p)\frac{\delta\Psi}{\delta\Psi}=p \, \mathcal{J}[\Psi] + (1-p)\,\mathds{1},
    \label{eq:J_p}
\end{align}
where $\mathcal{J}[\Psi]=\delta f/\delta\Psi$ is the $p$-independent Jacobian, directly defined by the differentiated BE equations [for details see Eqs.~\eqref{eq:dW_dW}--\eqref{eq:dSigma_dSigma}]. The eigenvalues of $\mathcal{J}^p$ determine the local stability of the fixed point $\Psi^*$. More precisely, a fixed point $\Psi^*$ is locally stable for the damped iteration, Eq.~\eqref{eq:Damped-iteration}, if all corresponding eigenvalues have an absolute value smaller than one:
\begin{align}
    |j^p_\alpha| = |p (j_\alpha-1) + 1| < 1,
    \label{eq:j_p}
\end{align}
where $j^p_\alpha$ denote the eigenvalues of $\mathcal{J}_p$ and $j_\alpha$ the eigenvalues of $\mathcal{J}$ \footnote{The index $\alpha$ iterates over all the eigenvalues $j^p_\alpha$ and $j_\alpha$. For a finite frequency box, we can set $\alpha = 1, 2, ..., N$ where $N$ is the dimension of the state $\Psi$.}.
In that case, $\Psi^*$ is an attractive fixed point, i.e., the damped iteration, Eq.~\eqref{eq:Damped-iteration}, converges towards $\Psi^*$, provided one starts close enough to $\Psi^*$.

As long as $\mathrm{Re}\,j_\alpha <1$ is fulfilled for all eigenvalues $j_\alpha$ of $\mathcal{J}$, there exists a damping parameter $p>0$ such that local stability of the damped iteration, Eq.~\eqref{eq:Damped-iteration}, is guaranteed, i.e., the condition, Eq.~\eqref{eq:j_p}, is satisfied for all $j_\alpha$. Evidently, the closer some eigenvalue $\mathrm{Re}\,j_\alpha$ approaches $1$, the smaller $p$ has to be chosen
(for details see App.~\ref{sec:Stability-conditions}).

On the other hand, if $\mathrm{Re}\, j_\alpha > 1$ for some eigenvalue $j_\alpha$, the fixed point $\Psi^*$ of the system is unstable and the damped iteration, Eq.~\eqref{eq:Damped-iteration}, is not applicable anymore. 
One possible way to proceed is to apply the algorithmic modification proposed by E{\ss}l et al.\ \cite{Essl2026Origin} for stabilizing unstable fixed points of iterative approaches. We refer to the corresponding modified algorithm as \emph{stabilized iteration}. Here, the mixing parameter $p$ is replaced by the \emph{stabilization matrix} $\mathcal{P}$:
\begin{align}
    \Psi_{n+1} &= \mathcal{P}\, f[\Psi_{n}] + (\mathds{1}-\mathcal{P})\Psi_{n}.
    \label{eq:SOT}
\end{align}
$\mathcal{P}$ is defined through the linear transform $\mathcal{U}$ that diagonalizes the Jacobian $\mathcal{J}$: 
\begin{subequations}
    \begin{align}
        \mathcal{P} &= \mathcal{U}\, \mathcal{D}\, \mathcal{U}^{-1},
        \label{eq:P-matrix}\displaybreak[2]\\
        \text{with~}\mathcal{D}_{\alpha\alpha^\prime} &= p\,\sgn(1-\mathrm{Re}\, j_\alpha)\,\delta_{\alpha\alpha'}
        \label{eq:D_aa}\\
        \text{and~} \mathcal{J} &= \mathcal{U} \left(\begin{matrix}
            j_1 & 0 & \cdots \\
            0 & j_2 & \\
            \vdots & & \ddots
        \end{matrix}\right)\mathcal{U}^{-1}.
    \end{align}
\end{subequations}
The signs in the diagonal matrix $\mathcal{D}$ must be chosen according to the stability of the respective eigendirection $\alpha$. If $\mathrm{Re}\,j_\alpha<1$, i.e., the eigendirection is stable, $+p$ is selected. On the other hand, if $\mathrm{Re}\,j_\alpha >1$, $-p$ is taken instead. Evidently, changing the sign for unstable eigendirections in Eq.~\eqref{eq:SOT} transforms a repulsive fixed point $\Psi^*$ into an attractive one. Then, the eigenvalues of the modified $p$-dependent Jacobian $\mathcal{J}^p$ corresponding to Eq.~\eqref{eq:SOT}, i.e., $\mathcal{D_{\alpha\alpha}}(j_\alpha-1)+1$, are thus simply stabilizable by properly adjusting the mixing parameter $p$ (for more details see App.~\ref{sec:Stability-conditions}) \footnote{At nonhyperbolic fixed points, where at least one eigenvalue of the Jacobian yields exactly $\mathrm{Re}\,j_\alpha = 1$ \cite{guckenheimer1983}, the stability of the fixed point depends on higher-order derivatives of $f[\Psi]$ and our method is not applicable.}. The mixing parameter $p$ in Eq.~\eqref{eq:D_aa} is chosen in such a way that local stability is guaranteed:
\begin{align}
    p &= c \min_\alpha \left\vert\mathrm{Re}\,\frac{2}{1-j_\alpha}\right\vert, 
    \label{eq:p_max}
\end{align}
with the additional parameter $c\in(0,1)$ set to $c=0.5$ in our numerical computations (cf.\ App.~\ref{sec:Stability-conditions}).

Besides the simplest choice of using a \emph{uniform} mixing, where $\mathcal{D}_{\alpha\alpha'}$ has the same magnitude everywhere, i.e., $p\,\mathrm{sgn}(1-\mathrm{Re}\,j_\alpha)\delta_{\alpha\alpha'}$ and $p$ is calculated using Eq.~\eqref{eq:p_max}, we also consider here, as alternative algorithmic path, a \emph{nonuniform} mixing $\mathcal{D}_{\alpha\alpha'}=p_\alpha\delta_{\alpha\alpha'}$, where the damping parameter is chosen individually for each eigendirection of $\mathcal{J}$ \cite{Essl2026Stabilizing}\footnote{Note that in contrast to Ref.~\cite{Essl2026Stabilizing}, here, the nonuniform mixing also allows values that are greater than one in magnitude.}:
\begin{align}
    p_\alpha = c \,\mathrm{Re}\frac{2}{1-j_\alpha}.
    \label{eq:nonuniform_p}
\end{align}
As one may expect, nonuniform mixing accelerates convergence in problematic regions, where some $\mathrm{Re}\,j_\alpha$ are very close to $1$. To optimize numerical convergence, we apply this procedure, Eq.~\eqref{eq:nonuniform_p}, in the case of TRILEX and SBE. In contrast, the uniform mixing, Eq.~\eqref{eq:p_max}, turns out to be more stable for MBE, which is highly affected by vertex divergences as in the usual parquet iteration \cite{Essl2026Stabilizing}. More details on the numerical implementation of $\mathcal{P}$ are provided in App.~\ref{sec:Schur}.

As for the determination of the Jacobian for the different classes of methods, we recall that the BE equations \eqref{eq:SBE-vertex}--\eqref{eq:G} provide an iterative map $f[\Psi]$ for all the quantities to be iterated (cf.\ Fig.~\ref{fig:MBE-SOT-Scheme}), i.e., we have $\Psi=(\Sigma,W_r)$ for TRILEX, $\Psi=(\Sigma,W_r,\lambda_r)$ for SBE, and $\Psi=(\Sigma,W_r,\lambda_r,M_r)$ for MBE. Evidently, the corresponding Jacobian $\mathcal{J}=\delta f/\delta\Psi$ contains all possible functional derivatives of the given quantities.
For instance, in the case of TRILEX, we have:
\begin{align}
        \mathcal{J}_{\mathrm{TRILEX}} &= \left(\begin{matrix}
        \frac{\delta f_\Sigma^\nu}{\delta\Sigma^{\tilde\nu}} &
        \frac{\delta f_\Sigma^\nu}{\delta W_{\tilde r}^{\tilde\omega}}\\
        \frac{\delta f_{W_r}^\omega}{\delta\Sigma^{\tilde\nu}} &
        \frac{\delta f_{W_r}^\omega}{\delta W_{\tilde r}^{\tilde\omega}} 
        \end{matrix}\right),
        \label{eq:J_GW}
   \end{align}
while for the case of SBE:
 \begin{align}
        \mathcal{J}_{\mathrm{SBE}} &= \left(\begin{matrix}
        \frac{\delta f_\Sigma^\nu}{\delta\Sigma^{\tilde\nu}} &
        \frac{\delta f_\Sigma^\nu}{\delta W_{\tilde r}^{\tilde\omega}} &
        \frac{\delta f_\Sigma^\nu}{\delta\lambda_{\tilde r}^{\tilde\omega\tilde\nu}}\\
        \frac{\delta f_{W_r}^\omega}{\delta\Sigma^{\tilde\nu}} &
        \frac{\delta f_{W_r}^\omega}{\delta W_{\tilde r}^{\tilde\omega}} &
        \frac{\delta f_{W_r}^\omega}{\delta\lambda_{\tilde r}^{\tilde\omega\tilde\nu}}\\
        \frac{\delta f_{\lambda_r}^{\omega\nu}}{\delta\Sigma^{\tilde\nu}} &
        \frac{\delta f_{\lambda_r}^{\omega\nu}}{\delta W_{\tilde r}^{\tilde\omega}} &
        \frac{\delta f_{\lambda_r}^{\omega\nu}}{\delta\lambda_{\tilde r}^{\tilde\omega\tilde\nu}}
        \end{matrix}\right),
            \label{eq:J_SBE}
       \end{align}
and, eventually, for the MBE approach:
\begin{align}
    \mathcal{J}_{\rm MBE}=\frac{\delta f}{\delta\Psi} = \left(\begin{matrix}
        \frac{\delta f_\Sigma^\nu}{\delta\Sigma^{\tilde\nu}} &
        \frac{\delta f_\Sigma^\nu}{\delta W_{\tilde r}^{\tilde\omega}} &
        \frac{\delta f_\Sigma^\nu}{\delta\lambda_{\tilde r}^{\tilde\omega\tilde\nu}} &
        \frac{\delta f_\Sigma^\nu}{\delta M_{\tilde r}^{\tilde\omega\tilde\nu\tilde\nu'}}\\
        \frac{\delta  f_{W_r}^\omega}{\delta\Sigma^{\tilde\nu}} &
        \frac{\delta  f_{W_r}^\omega}{\delta W_{\tilde r}^{\tilde\omega}} &
        \frac{\delta  f_{W_r}^\omega}{\delta\lambda_{\tilde r}^{\tilde\omega\tilde\nu}} &
        \frac{\delta  f_{W_r}^\omega}{\delta M_{\tilde r}^{\tilde\omega\tilde\nu\tilde\nu'}}\\
        \frac{\delta f_{\lambda_r}^{\omega\nu}}{\delta\Sigma^{\tilde\nu}} &
        \frac{\delta f_{\lambda_r}^{\omega\nu}}{\delta W_{\tilde r}^{\tilde\omega}} &
        \frac{\delta f_{\lambda_r}^{\omega\nu}}{\delta\lambda_{\tilde r}^{\tilde\omega\tilde\nu}} &
        \frac{\delta f_{\lambda_r}^{\omega\nu}}{\delta M_{\tilde r}^{\tilde\omega\tilde\nu\tilde\nu'}}\\
        \frac{\delta  f_{M_r}^{\omega\nu\nu'}}{\delta\Sigma^{\tilde\nu}} &
        \frac{\delta  f_{M_r}^{\omega\nu\nu'}}{\delta W_{\tilde r}^{\tilde\omega}} &
        \frac{\delta  f_{M_r}^{\omega\nu\nu'}}{\delta\lambda_{\tilde r}^{\tilde\omega\tilde\nu}} &
        \frac{\delta  f_{M_r}^{\omega\nu\nu'}}{\delta M_{\tilde r}^{\tilde\omega\tilde\nu\tilde\nu'}}
    \end{matrix}\right).
    \label{eq:Jacobian-MBE}
\end{align}
For numerical computations, we order the quantities in a flattened $N\times N$ matrix where $N$ is the number of all data points included in $\Psi$ (respecting different vertices, channels, and frequencies). 

Closed expressions for the Jacobian, Eqs.~\eqref{eq:J_GW}--\eqref{eq:Jacobian-MBE}, in terms of the respective iterated quantities, i.e., $\mathcal{J}=\mathcal{J}[\Psi]=\mathcal{J}[\Sigma,W_r,\lambda_r,M_r]$ are provided in App.~\ref{sec:Jacobian-MBE}. Moreover, in App.~\ref{sec:Numerics}, we give further details on how the size of $\mathcal{J}$ can be reduced in the BE formalism and we present a more robust and efficient way of constructing the stabilization matrix $\mathcal{P}$ beyond naive diagonalization of $\mathcal{J}$.
Other technical aspects of the stabilized-iteration method are discussed in Ref.~\cite{Essl2026Stabilizing}.

\section{Stability of the physical solution}
\label{sec:Stability}

In the following, we will systematically analyze the stability of the physical fixed point of the HA and ZPM in the self-consistent iteration schemes for the three classes of approaches considered, in order of increasing complexity (TRILEX, SBE, MBE). This study will be performed by directly exploiting the formalism discussed in the previous section. More precisely, we will  compute the eigenspectrum of the Jacobian for the respective scheme [cf.~Eqs.~(\ref{eq:J_GW})--(\ref{eq:Jacobian-MBE})] at the physical fixed point of the \emph{exact solution} of the HA and ZPM, across their whole phase diagram.

Before undertaking this task, however, it is convenient to examine the occurrence of divergences in the (corresponding) irreducible vertex functions, since the crossing of the instability threshold in the Jacobian  eigenspectrum often occurs in the presence of a simultaneous divergence of the vertex \cite{Essl2026Origin}. This was indeed the ``practical'' reason why the convergence of iterative diagrammatic approaches were generally considered as unfeasible after the crossing of the first vertex divergence line. 

\subsection{Divergences of the irreducible vertices}
\label{sec:Vertex_divergences}

For more than a decade it has been explored how entering the nonperturbative regime of the many-electron problem affects important sub-classes of diagrams contributing to the two-particle scattering amplitude $F$: the \emph{irreducible vertex} functions \cite{Schaefer2013Divergent,Janis2014Critical,Gunnarsson2016Parquet,Vucicevic2018Practical,Chalupa2018Divergences,Springer2020Interplay}. As mentioned in the previous sections, for the classes of approaches considered here, two kind of irreducibilities are of particular relevance: the \emph{two-particle irreducibility} (2PI) (when a vertex diagram can be split in two diagrams, by cutting two fermionic propagating lines) and the \emph{$U$ irreducibility} (when a vertex diagram can be split in two diagrams, by cutting a bare interaction $U$ line).
Evidently, the former class constitutes a crucial building block of the MBE/parquet formalism, while the latter are of special relevance for the SBE approaches.

According to the state-of-the-art understanding, nonperturbative scattering processes can drive the \emph{divergence} of 2PI vertices \cite{Schaefer2013Divergent,Janis2014Critical,Gunnarsson2016Parquet,Vucicevic2018Practical,Chalupa2018Divergences,Springer2020Interplay}, while no divergence of $U$-irreducible vertices \cite{Krien2019Single,Gievers2022Multiloop} was ever reported so far. As we will discuss below, however, the common wisdom on this latter aspect is \emph{not} correct.

We start our specific investigation, anyhow, from the more studied case of 2PI vertices. For those, it has been extensively demonstrated \cite{Chalupa2021Fingerprints,Adler2024Nonperturbative} that the physical origin for the divergence of their on-site part \footnote{More recent extensions of such studies have proven how the divergences observed in the nonlocal part of the 2PI vertices of cluster-DMFT \cite{Vucicevic2018Practical,Meixner2026Nonperturbative} and DCA \cite{Gunnarsson2016Parquet} solutions of the two-dimensional Hubbard model must be ascribed to strong nonperturbative short-range spin fluctuations.}
 directly lies in the gradual formation of local magnetic moments (and the unavoidably associated charge localization), which is driven by increasing values of the on-site electrostatic repulsion $U$. Hence, the proper inclusion of such a nonperturbative vertex behavior turns out to be absolutely crucial for the description of phenomena such as the Mott metal-insulator-transition in DMFT \cite{Pelz2023Highly} and the adjacent phase-separation instabilities \cite{Reitner2020Attractive,Reitner2024Protection,Meixner2026Nonperturbative}.

\begin{figure*}
    \centering
    \includegraphics[width=1.0\textwidth]{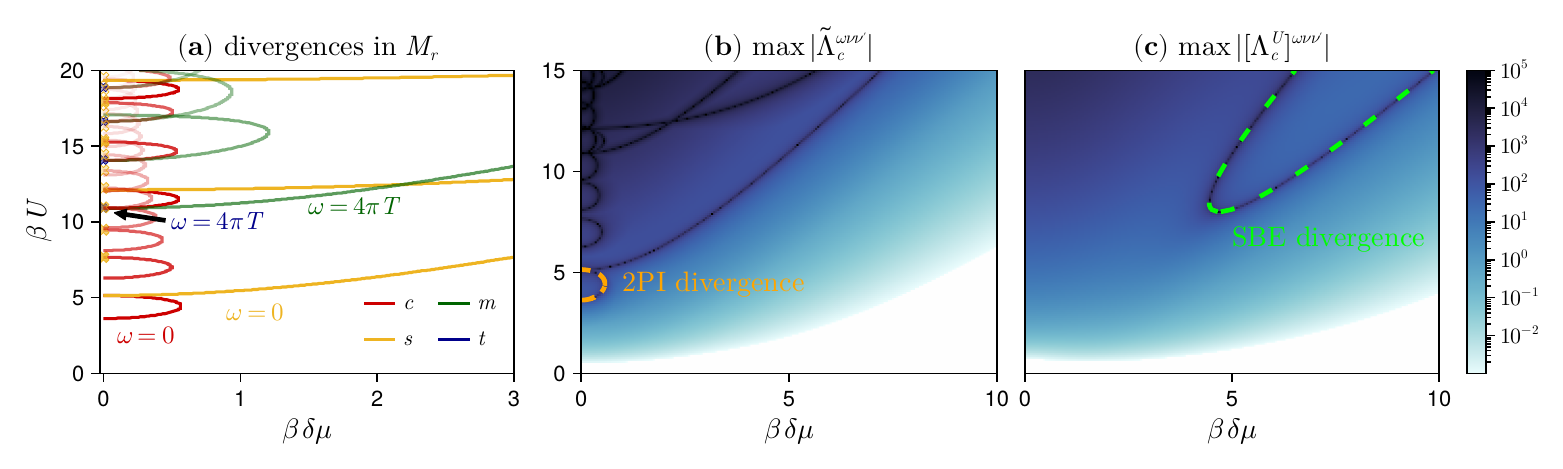}
    \caption{(a) Location of divergences of the MBE vertices $M_r$ in the phase space of the HA.
    The colors mark the corresponding channel $r=c,m,s,t$, while transparency of the colors indicate the bosonic frequency $\omega$ (the more transparent, the higher the value of $\omega$).
    Note that, for $\omega\neq0$, $M_r$ divergences in the singlet and triplet channel only appear at half-filling, i.e., $\delta\mu=0$ (marked by crosses). The colorful labels indicate the bosonic frequencies corresponding to the first vertex divergence occurring in the respective channel. (b) Maximal value of the two-particle-irreducible rest function $\max_{\omega,\nu,\nu'}|\tilde\Lambda_c^{\omega\nu\nu
    '}|$, where the orange line marks the first 2PI divergence in $M_c$, and (c) the $U$-irreducible rest function $\max_{\omega,\nu,\nu'}|[\tilde\Lambda_c^U]^{\omega\nu\nu
    '}|$, where the green line marks the SBE divergence of $M_c$ and $\lambda_c$ (see text).
    }
    \label{fig:Vertex_divergences}
\end{figure*}

Our specific results for the divergences of 2PI vertices, or short \emph{2PI divergences}, for the HA are reported in Fig.~\ref{fig:Vertex_divergences}(a)--(b).
In this respect, we note that, since the full vertex, which can be exactly decomposed as  $F =\Lambda + \sum_r (M_r+\nabla_r-U)$ [cf.\ Eq.~\eqref{eq:Diagrammatic_decomposition}] cannot diverge at any finite $T$ in the HA and since the SBE vertices $\nabla_r$ do not exhibit 2PI divergences, the divergences of the 2PI vertex $\Lambda$ must then also appear in the MBE vertices $M_r$. 

In Fig.~\ref{fig:Vertex_divergences}(a), we show the divergence lines of $M_r$ in the parameter space of the HA, entirely spanned by the interaction $U$ and the deviation from half-filling $\delta\mu$.
The colors mark the channels $r$ where the divergences occur ($c$ red, $m$ green, $s$ yellow, $t$ blue), while the transparency indicates the respective bosonic (transfer) frequency $\omega$ of the vertex, for which the respective divergence appears (the more transparent, the higher $\omega$).

At half-filling, i.e., $\delta\mu = 0$, we recover the known literature results \cite{Schaefer2016Nonperturbative,Thunstroem2018Analytical}, with the first vertex divergence of $M_r$ (and hence of the 2PI vertex $\Lambda_r$) appearing in the $c$ channel at $\beta U = 2\pi/\sqrt{3}\simeq 3.628$ for $\omega=0$ and, specifically for $\nu,\nu'=\pm\pi T$ \cite{Thunstroem2018Analytical} [cf.\ lowest red line in Fig.~\ref{fig:Vertex_divergences}(a)]. The second divergence at $\beta U \simeq 5.127$ occurs in the $c$ and $s$ channel at $\omega=0$ globally, i.e. for \emph{all} $\nu,\nu'$ \footnote{\emph{Local} vertex divergences occur in specific fermionic frequencies $\nu, \nu'$, whereas \emph{global} vertex divergences occur simultaneously in all fermionic frequencies \cite{Thunstroem2018Analytical}.} [cf.\ red and yellow lines in Fig.~\ref{fig:Vertex_divergences}(a)]. In the magnetic channel (green line) and triplet channel (blue cross), the first vertex divergence occurs at $\beta U=10.92$ only at $\omega=4\pi T$.

Figure~\ref{fig:Vertex_divergences}(a) also shows divergence lines out of half-filling ($\delta \mu \neq 0$) with results  coincident at $\omega =0$ with those of Ref.~\cite{Essl2024General}, whereas our study is extended to $\omega \neq 0$, too. We note that, in general, two vertex divergences connect at an \emph{exceptional point} \cite{Reitner2024Protection}, which appears at a specific value of $\delta\mu$, forming characteristic lobes in the phase space. Further, while the HA displays vertex divergences in \emph{all} channels, in the $s$ and $t$ channels, one also finds a specific kind of vertex divergences that only appear at $\omega\neq 0$ and at half-filling. Those divergences are marked by yellow and blue crosses \footnote{This behavior only occurs in the $s$ and $t$ channel for $\omega\neq 0$, since, there, the vertex functions are not $\kappa$-real matrices and therefore a real eigenvalue can become complex valued without the need of crossing another real eigenvalue. This lack of $\kappa$-real matrices also leads to the fact that the $s$ and $t$ channel do not correspond to physical response functions for $\omega\neq 0$.}.

The data reported in Fig.~\ref{fig:Vertex_divergences}(b) show the maximal absolute values of the 2PI vertex $\tilde\Lambda_c=\Lambda_c-U_c$ as intensity plot
\footnote{Specifically, we report here the maximal absolute value of $\tilde\Lambda_c^{\omega\nu\nu'}$ within our numerical data with finite frequency grids.}. This analysis readily confirms the anticipated fact that the 2PI vertex $\tilde\Lambda_c$ inherits vertex divergences from all channels of the MBE vertices $M_r$: The sharp dark blue regions of extremely large absolute values perfectly match the divergence lines of $M_r$ in Fig.~\ref{fig:Vertex_divergences}(a). For the sake of clarity, the lobe connecting the first and second vertex divergence is highlighted by an orange dashed line.

The simultaneous divergences of $M_r$ and $\Lambda_r$ reported in Fig.~\ref{fig:Vertex_divergences}(a)--(b), as expected, perfectly cancel each other, yielding a finite (nondivergent value) of the $U$-irreducible vertex $\Lambda^U=\Lambda-U+\sum_r M_r$ [cf.\ Eq.~\eqref{eq:Lambda^U}], as one can easily verify by comparing the corresponding parameter ranges with the intensity plot data of $\Lambda^U_c$ reported in Fig.~\ref{fig:Vertex_divergences}(c). Hence, in this parameter region, the SBE formalism, based on the $U$ reducibility, is \changed{protected} from the divergence of its diagrammatic building blocks, which is often mentioned as an additional reason to exploit the SBE class of approaches. Evidently, from a numerical point of view, it is preferable to compute and manipulate quantities, which are immune to divergences \cite{Lihm2026Finite-difference}.

However, the data reported in Fig.~\ref{fig:Vertex_divergences}(c) reveal that, by further increasing $U$, this hope is shattered: A \emph{different} kind of vertex divergence \changed{associated with the BE decomposition} is found to occur in the strong-coupling  ($\beta U\gtrsim 7$) and highly imbalanced regime ($\beta\delta\mu\gtrsim 4.5$) of the HA, which we dub \emph{SBE divergence} and illustrate in the figure by a green dashed line. More precisely, this complementary kind of divergence only occurs, at the level of the HA, in the $c$ channel, for $U>0$ and quite far away from half-filling $\delta\mu\neq 0$ at $\omega =0$ \footnote{Out of SU(2) symmetry, i.e., with an additional magnetic field $h$, the SBE divergence can also occur at $U<0$ due to the general Shiba mapping \cite{Essl2024General}.}. Hence, only $\lambda_c$ (and the related quantities $\nabla_c$, $M_c$, and $\Lambda_c^U$) is affected by the SBE divergence, whereas all the other vertices and channels are immune to it. Furthermore, such SBE divergences are exactly canceled in the sum $\nabla_c+M_c$ [as one can see by a direct comparison of the same parameter regimes in Figs.~\ref{fig:Vertex_divergences}(b) and (c)] and, thus, in the traditional parquet decomposition. These restrictions might explain why SBE divergences have remained unnoticed for a long time (see App.~\ref{sec:SBE_Vertex_divergece} for details).

While the investigation of the microscopical origin and the physical implications of the SBE divergences will certainly require a dedicated study (e.g., along the lines of Ref.~\cite{Adler2024Nonperturbative}), we strongly suspect that also the lattice Hubbard model may display similar singularities, e.g., in the proximity of charge fluctuation instabilities, i.e., where the charge susceptibility $\chi_c = W_c/U_c^2-[U_c]^{-1}$ fulfills $\chi_c= -[U_c]^{-1}$ (with $U_c = -U <0$) and $W_c=0$ (see App.~\ref{sec:SBE_Vertex_divergece}). \changed{We therefore suspect that SBE divergences may be related to the convergence problems reported for GW and TRILEX approaches near a zero of $W_r$ \cite{Ayral2012Spectral,Huang2014Extended,Ayral2015}. By contrast, the strong enhancement of the Hedin vertices $\lambda_r$ for $T\to 0$ discussed in Refs.~\cite{Harkov2021Parametrization,Yu2025Pairing} is of a different nature: it occurs at small fermionic frequencies in the local-moment regime, whereas the SBE divergence reported here occurs at finite $T$ and is tied to a zero of $W_r$.}

Coming back to the different diagrammatic methods presented in Sec.~\ref{sec:GW_SBE_MBE}, whose stability will be discussed in the next subsection, we conclude that MBE exhibits vertex divergences in all channels and frequencies for various values of $U$ and $\delta\mu$, while SBE approaches might be affected by the different (SBE-type) divergence, which, for the HA, only occurs in the $c$ channel at $\omega=0$ out of half-filling. Finally, per construction, TRILEX only iterates over $\Sigma$ and $W_r$ that do not exhibit any vertex divergences. Hence, it will only inherit the SBE divergences \emph{if} they are originally present in the corresponding input of the Hedin vertex $\lambda_c$.

\subsection{Identification of unstable regions}
\label{sec:Stability_regions}

\begin{figure*}
	\centering
    \includegraphics[width=\textwidth]{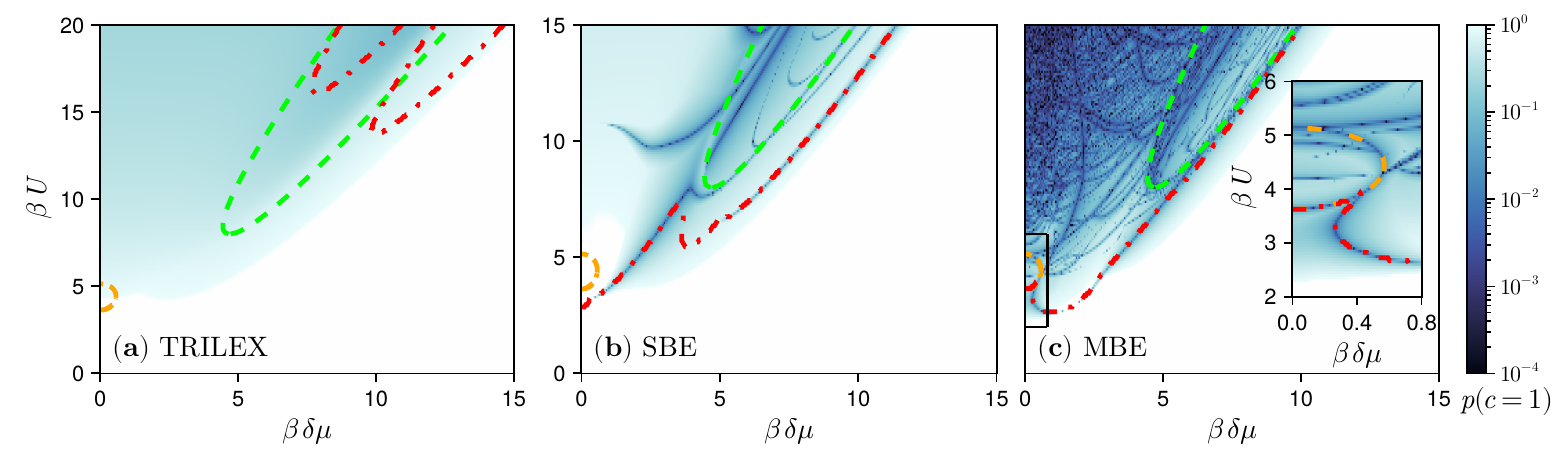}
	\caption{Threshold mixing parameter $p(c=1)$, Eq.~\eqref{eq:p_max}, which still stabilizes the physical fixed point of the respective diagrammatic schemes (a) TRILEX, (b) SBE, (c) MBE. The red dashed lines mark where the first crossing  $\mathrm{Re}\,j_\alpha = 1$ occurs, i.e., in the regions above/within the red lines, 
    the conventional damped iteration, Eq.~\eqref{eq:Damped-iteration}, does not converge to the physical solution. The orange and green dashed lines mark the first 2PI divergence and the SBE divergence.
    The inset in panel (c) magnifies the region around the first vertex divergence in MBE.
    }
	\label{fig:Stability_regions}
\end{figure*}

To provide a concise overview of our calculations and allow for an easy comparison of the fixed-point stability of the iterative solution in different diagrammatic schemes, we present in Fig.~\ref{fig:Stability_regions} an intensity plot of the \emph{threshold} damping parameter $p(c=1)$ of Eq.~\eqref{eq:p_max}. This is formally the maximal value of $p$ needed to locally stabilize the physical fixed point of the selected diagrammatic approach, where the stabilized iteration Eq.~\eqref{eq:SOT} instead of the conventional damped iteration Eq.~\eqref{eq:Damped-iteration} is needed beyond the red line.  
Specifically, the three panels of Fig.~\ref{fig:Stability_regions} show the corresponding results for $p(c=1)$ in the \mbox{$U$-$\delta \mu$} phase space of the HA for the diagrammatic schemes considered (TRILEX, SBE, MBE), where the Jacobian is evaluated at the respective physical fixed point. 

In the chosen color-intensity scale, the white regions are stable also without mixing, i.e., $p=1$, while a progressively bluish hue indicate increasingly lower values of $p$, and, hence, the need of a stronger mixing to achieve convergence.
Most importantly, in the regions enclosed \emph{above} the dashed red lines, the physical fixed point is unstable with respect to the conventional damped iteration Eq.~\eqref{eq:Damped-iteration}. In these regions, the stabilized iteration, Eq.~\eqref{eq:SOT}, should be used instead to stabilize the physical fixed point.

A closer inspection of the results indicates the following:
Figure~\ref{fig:Stability_regions}(a) clearly shows that TRILEX overall needs the least damping. The mixing parameter $p(c=1)$ [evaluated from Eqs.~\eqref{eq:p_max} and \eqref{eq:J_GW}] is of the order of $10^{-1}-10^0$ throughout the plotted region. Up to intermediate values of $\delta\mu$, there are also no unstable eigenvalues $j_\alpha$ of the Jacobian and, thus, the conventional damped iteration, Eq.~\eqref{eq:Damped-iteration}, can converge to the physical fixed point.

At the same time, our analysis does reveal the presence of two unstable regions for the fixed point of the damped iteration scheme above the dashed red lobes on the top right corner of Fig.~\ref{fig:Stability_regions}(a), i.e. at quite strong coupling $\beta U\gtrsim 13.0$ and high particle-hole imbalances  $\beta\delta\mu\gtrsim 8.0$. Hence, our results indicate that even the simplest class of approaches, based on TRILEX, can exhibit regimes, where the physical fixed point of the HA becomes unstable, inducing a misleading convergence to an unphysical self-consistent solution.

In various GW methods, convergence problems were found \cite{pokhilko2021,pokhilko2025,wang2026}, which are especially severe if vertex corrections are included. Some investigations included multiple self-consistent solutions \cite{Lani2012,Stan2015Unphysical,tandetzky2015,pokhilko2021}. It is left for future studies if those are connected to the unstable region in TRILEX that we observe.

Figure~\ref{fig:Stability_regions}(b) presents $p(c=1)$, computed from the Jacobian of SBE, Eq.~\eqref{eq:J_SBE}. 
The increased complexity of this method reflects in an enlarged region, where iteration instabilities are found, as they now include also the regime of small imbalance  (small values of $\delta\mu$), and even at half-filling. In fact, for $\delta\mu=0$, the SBE damped-iteration scheme becomes already unstable at $\beta U\simeq 2.8$ (marked by the red dashed line).
Consistent with the fact that the terms of the SBE decomposition are not directly subject to the 2PI divergences, the occurrence of this instability is decoupled from such divergences: The first instability at half-filling is encountered even before the first 2PI divergence (marked by the dashed orange lobe).
On the other hand, by comparing green and red dashed lines in Fig.~\ref{fig:Stability_regions}(b), also the SBE divergence in $\lambda_c$ lies in the unstable region.
Overall, all divergences are located beyond the line of first unstable eigenvalues, $\mathrm{Re}\,j_\alpha>1$ (red dashed line), and, thus are inaccessible by the conventional damped iteration, Eq.~\eqref{eq:Damped-iteration}. This is probably another reason why divergences in the Hedin vertices have not been discussed so far.

Interestingly, for the SBE, the region where the necessary mixing is largest ($p(c=1)$ is the lowest, dark blue zones) is located, for large $U$ values, rather near the line $U=2\,\delta\mu$ where the ground state of the HA changes from double to degenerate single occupation. Hence, it is plausible to assume that in the limit $T\to 0$, such a match can become more accurate even for smaller values of $U$. 

Eventually, Fig.~\ref{fig:Stability_regions}(c) provides the necessary mixing $p(c=1)$ for the MBE, Eq.~\eqref{eq:Jacobian-MBE}. 
Quite remarkably, although the decomposition of diagrams is performed in a different manner in the MBE scheme compared to the usual parquet decomposition and, therefore, also the iterative scheme is different, the shape of the instability region of MBE turns out to be identical to that of the conventional self-consistent parquet  formalism \cite{Essl2026Stabilizing}.
In both cases, the stability properties are evidently connected to the appearance of 2PI divergences. 
In particular, at half-filling, i.e., $\delta\mu=0$, the first crossing of an eigenvalue $\mathrm{Re}\,j_\alpha>1$ (red dashed line) is \emph{exactly} on top of the first vertex divergence at $\beta U \simeq 3.628$ (orange dashed arc). This is displayed more clearly in the inset of Fig.~\ref{fig:Stability_regions}(c). 
The coincidence in parameter space of 2PI divergences and misleading convergence is the reason why it was believed for a long time that those imply each other. However, as discussed in Ref.~\cite{Essl2026Stabilizing}, misleading convergences can also occur independently from vertex divergences, making the 2PI divergences a sufficient but not necessary condition for an instability.  The inset panel reveals that for $\beta\delta\mu\gtrsim 0.3$ and $\beta U\gtrsim 3.0$, the unstable region with $\mathrm{Re}\,j_\alpha>1$ already exists before the first vertex divergence. In general, the unstable region of MBE exhibits many lines of small $p(c=1)$, i.e., huge mixing is needed. This can be traced back to 2PI divergences that come along with many large eigenvalues $j_\alpha$ leading to small values of $1/(1-j_\alpha)$. Specifically, the strong-coupling region near half-filling $\beta U\gtrsim 6.0$ and $\beta\delta\mu\lesssim 2.0$ is pervaded with numerous lines of small values of $p(c=1)$.

In App.~\ref{sec:app_stability_regions}, we provide some additional results for the stability of diagrammatic methods in the phase space.

\begin{figure*}
    \centering
    \includegraphics[width=\linewidth]{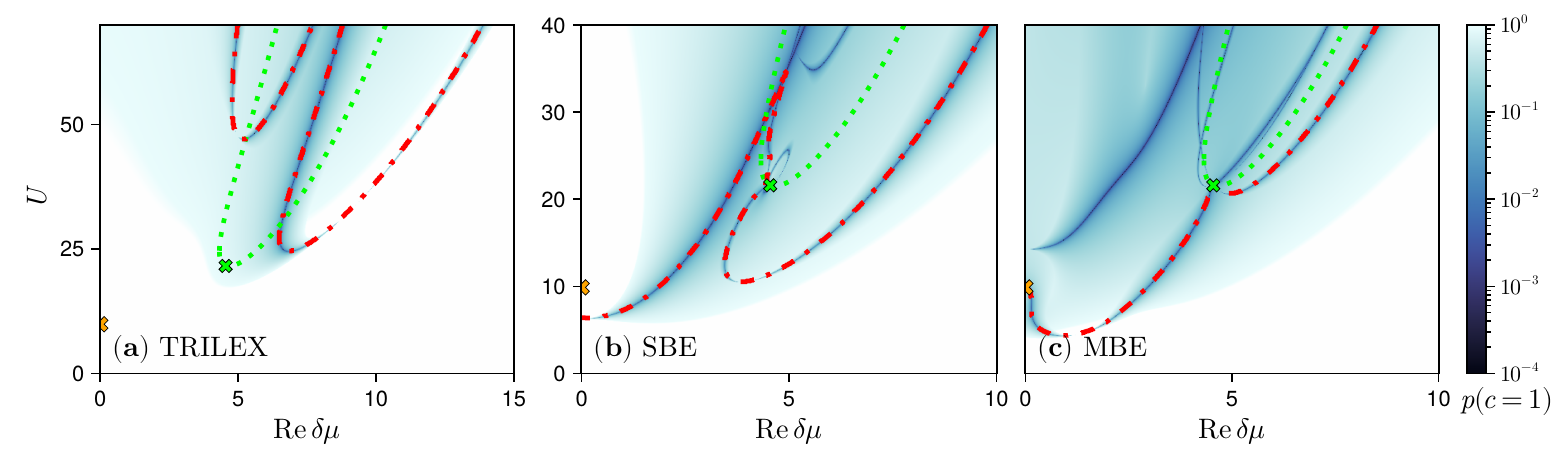}
    \caption{Same as Fig.~\ref{fig:Stability_regions} but for the ZPM. The orange cross marks the 2PI divergence in the charge channel at $\mathrm{Re}\,\delta\mu=0$, $U=\pi^2$. The green cross marks the SBE divergence in the charge channel, i.e., $W_c=0$, while the dotted green line marks the points where $\mathrm{Re}\,W_c=0$ but $\mathrm{Im}\,W_c\neq0$ (see App.~\ref{sec:SBE_div_ZP}).}
    \label{fig:Stability_ZP}
\end{figure*}

While the stability results reported above are specific to the HA, we suspect that several of the qualitative features of the instability regions are more general,  so that the conclusion we draw might be a useful compass for the correct algorithmic implementation of various self-consistent schemes also for more realistic models. As a supporting argument for that, in Fig.~\ref{fig:Stability_ZP}, we analyze instabilities, in an analogous fashion as above, for the ZPM, a different exactly solvable problem. Despite the different expression of its action, Eq.~\eqref{eq:ZPM}, the similarities to Fig.~\ref{fig:Stability_regions} are undeniable. TRILEX [Fig.~\ref{fig:Stability_regions}(a)] reveals two regions with unstable eigenvalues and, in SBE [Fig.~\ref{fig:Stability_regions}(b)], the 2PI divergence (orange cross) is located in the unstable region. In MBE [Fig.~\ref{fig:Stability_regions}(c)], the crossing of the first unstable eigenvalue exactly starts at the 2PI divergence (orange cross) at half-filling. For more details on the similarities between HA and ZPM, see App.~\ref{sec:ZP_additional}.

\section{Application of stabilized iteration}
\label{sec:Results}

In this section, we demonstrate how the stabilization procedure, discussed in Sec.~\ref{sec:Jacobian_formalism} works successfully for all the diagrammatic schemes considered (cf.\ Fig.~\ref{fig:MBE-SOT-Scheme}) independently of whether one starts from an exact/nonperturbative input or from a perturbative one. The first set of results is evidently relevant for intermediate-to-strong-coupling approximations (such as the diagrammatic extension of DMFT \cite{Rohringer2018Diagrammatic}), while the second set is significant for self-consistent diagrammatic calculations in the weak-to-intermediate regime.

Based on these results, we will highlight important practical aspects, relevant for the future method development in the many-electron theory.

\subsection{Results with nonperturbative input}
\label{sec:Stabilized_iteration_strong}

We start by  demonstrating how the stabilized iteration with $\mathcal{P}$, Eq.~\eqref{eq:SOT}, for TRILEX, SBE, and MBE, with an exact input for respective noniterated quantities, does actually lead to convergence to the physical fixed point, where the conventional damped iteration with $p$, Eq.~\eqref{eq:Damped-iteration}, fails.

For this, we iteratively solve the BE equations~\eqref{eq:SBE-vertex}--\eqref{eq:G} for TRILEX, SBE, and MBE using \emph{both} the conventional (damped iteration) scheme, Eq.~\eqref{eq:Damped-iteration}, \emph{and} the stabilized iteration, Eq.~\eqref{eq:SOT}, for selected paths in the phase space with increasing interaction $U$ and fixed $\delta\mu$. Before the first crossing $\mathrm{Re}\,j_\alpha>1$, those two methods are equivalent and converge to the same result. Beyond that, convergence to the physical fixed point is only ensured by the stabilized iteration. To make evident how successful the schemes are, we compare all obtained fixed-point results against the exact solution $\Psi_\mathrm{phys}$. Corresponding plots (cf.\ Figs.~\ref{fig:GW_SOT_dmu=12.0}--\ref{fig:W_r_comparison_weak} and Fig.~\ref{fig:MBE_SOT_dmu=1.0} in the appendix) follow the format defined in Fig.~\ref{fig:GW_SOT_dmu=12.0}(d): Vertex functions are shown at specific Matsubara frequencies as a function of the interaction $U$. The exact solution (cf.\ App.~\ref{sec:HA-vertices}) is illustrated by solid lines while converged solutions from the conventional damped iteration are depicted by circles and those from the stabilized iteration by pluses. Different colors refer to the channels ($c$ red, $m$ green, $s$ yellow, $t$ blue) and parts of the self-energy ($\Sigma_\mathrm{H}$ light blue, $\mathrm{Im}\,\tilde\Sigma$ pink, $\mathrm{Re}\,\tilde\Sigma$ purple). Further, we present the real parts of the eigenvalues $j_\alpha$ in dependence on $U$ (colors indicate if $j_\alpha\in\mathbb{C}$ or $j_\alpha\in\mathbb{R}$), the (vertical red dashed) lines of instabilities $\mathrm{Re}\,j_\alpha=1$ and (vertical yellow dashed) lines where the mixing parameter becomes $p(c=1)<1$. Values between $-1$ and $1$ (shaded region) are scaled linearly while values beyond are scaled logarithmically.

Beginning with the simplest scheme, we note how the convergence of TRILEX at half-filling, ($\delta\mu = 0$), is not affected by unstable eigenvalues $\mathrm{Re}\, j_\alpha$ [cf.\ Fig.~\ref{fig:Stability_regions}(a)]. Thus, the damped-iteration scheme with $p$ is sufficient to converge to the physical solution and the stabilized-iteration scheme with $\mathcal{P}$ is not needed (not explicitly shown here).

\begin{figure}
	\centering
    \includegraphics[width=0.9\columnwidth]{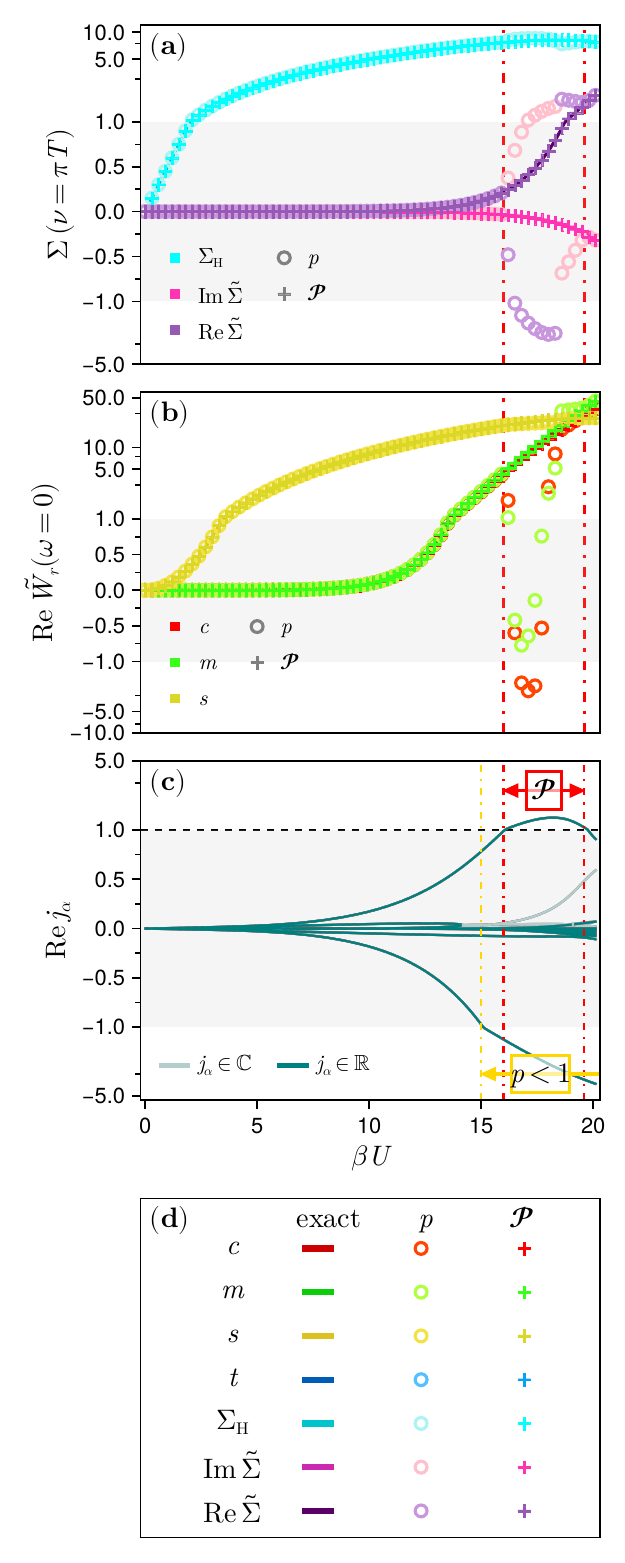}
	\caption{Converged quantities at specific Matsubara frequencies in dependence on $U$ from TRILEX at $\beta\delta\mu=12.0$. (a) Self-energies $\Sigma$ at $\nu=\pi T$, (b) bosonic propagators $\tilde W_r$ at $\omega=0$. Colors indicate channels ($c$ red, $m$ green, $s$ yellow, $t$ blue) and parts of the self-energy (light blue $\Sigma_\mathrm{H}$, pink $\mathrm{Im}\,\tilde\Sigma$, purple $\mathrm{Re}\,\tilde\Sigma$). Lines and markers with different shades indicate the convergence methods used (lines for exact values, circles for damped iteration $p$, pluses for stabilized iteration $\mathcal{P}$) [see panel (d)]. (c) Real parts of the eigenvalues $j_\alpha$ (teal) (dark for purely real $j_\alpha$ and light for complex-valued $j_\alpha$). Dashed vertical lines and arrows mark regions with finite mixing (yellow) and instabilities $\mathrm{Re}\,j_\alpha > 1$ (red).}
	\label{fig:GW_SOT_dmu=12.0}
\end{figure}

Out of half-filling, instead,
for large imbalances $\delta\mu$ and strong interactions $U$, TRILEX can become unstable, too [cf.\ Fig.~\ref{fig:Stability_regions}(a)]. 
To illustrate a specific case, we show converged results at $\beta\delta\mu=12.0$ in Fig.~\ref{fig:GW_SOT_dmu=12.0} for (a) $\Sigma$ and (b) $\tilde W_r=W_r - U_r$. As anticipated from the stability phase diagram [cf.\ Fig.~\ref{fig:Stability_regions}(a)], we observe an instability for $\beta U\in[16.0,19.6]$ (red lines in Fig.~\ref{fig:GW_SOT_dmu=12.0}). In that region, the results from damped iteration [circles in Figs.~\ref{fig:GW_SOT_dmu=12.0}(a)--(b)] do not match with the physical fixed point. Remarkably (and unlike MBE and SBE, see below), the TRILEX scheme becomes stable again above $\beta U\gtrsim 19.6$. 
Nonetheless, consistent with our statements in Sec.~\ref{sec:Jacobian_formalism}, we are able to cure successfully all observed convergence issues of TRILEX, by exploiting our stabilized-iteration method with $\mathcal{P}$, Eq.~\eqref{eq:SOT}, as we can directly appreciate by looking at the corresponding data [pluses in Figs.~\ref{fig:GW_SOT_dmu=12.0}(a)--(b)]. The eigenvalues plotted $\mathrm{Re}\,j_\alpha$ in Fig.~\ref{fig:GW_SOT_dmu=12.0}(c) nicely underline that the unstable region is characterized by the crossing of an eigenvalue $j_\alpha$.

\begin{figure}
	\centering
    \includegraphics[width=0.9\columnwidth]{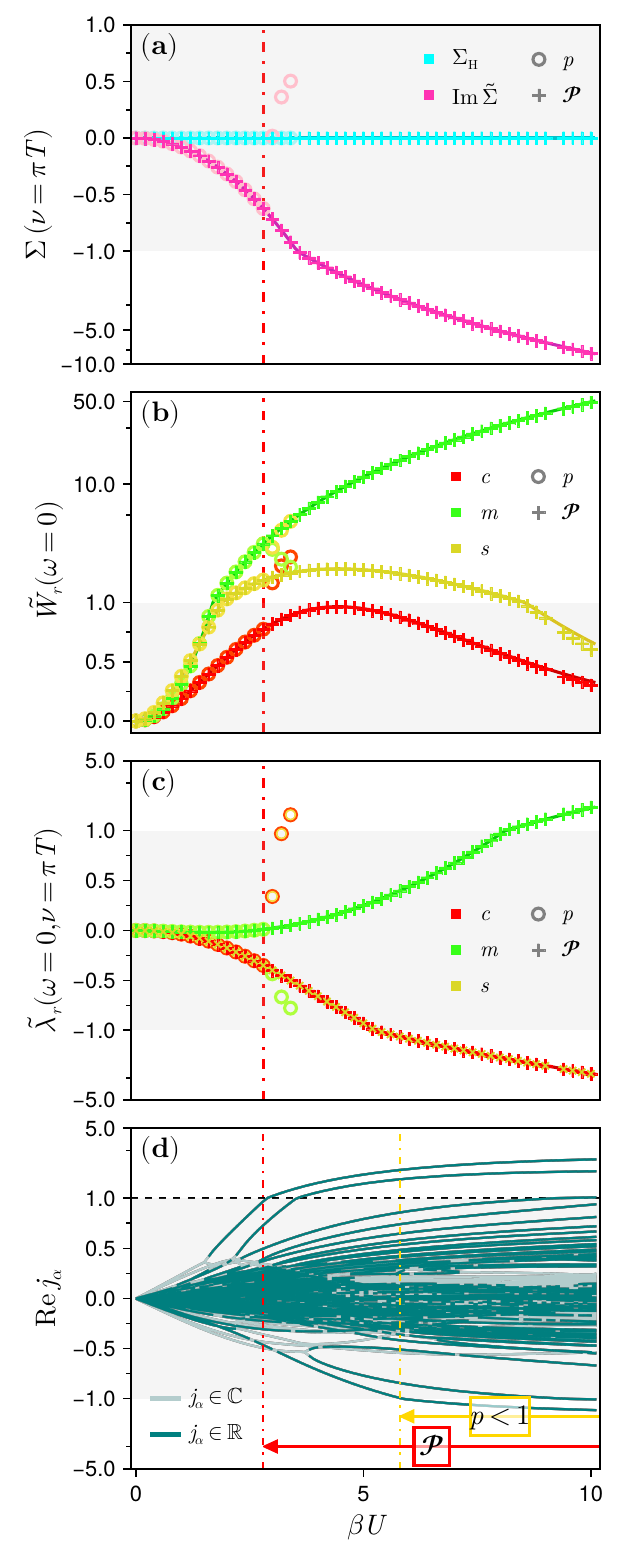}
	\caption{Converged quantities at specific Matsubara frequencies in dependence on $U$ from SBE at half-filling. Details about colors and markers are provided in Fig.~\ref{fig:GW_SOT_dmu=12.0}(d). (a) Self-energies $\Sigma$ at $\nu=\pi T$, (b) bosonic propagators $\tilde W_r$ at $\omega=0$, (c) Hedin vertices $\tilde\lambda_r$ at $\omega=0$ and $\nu=\pi T$, (d) real parts of the eigenvalues $j_\alpha$. 
    }
	\label{fig:SBE_SOT_dmu=0.0}
\end{figure}

By moving a step deeper in the diagrammatics, we show in Fig.~\ref{fig:SBE_SOT_dmu=0.0} our converged results for the SBE approach at half-filling, obtained from both the damped iteration, Eq.~\eqref{eq:Damped-iteration}, (circles) and the stabilized iteration, Eq.~\eqref{eq:SOT}, (pluses). At specific frequencies, i.e., $\omega=0$ and $\nu=\pi T$, the quantities (a) $\Sigma$, (b) $\tilde W_r$ (b), and (c) $\tilde\lambda_r=\lambda_r-1$ are plotted against the interaction $U$ and compared to the exact results (lines). In panel (d), we show the respective eigenvalues $\mathrm{Re}\,j_\alpha$. We can immediately note two crossings of $\mathrm{Re}\, j_\alpha = 1$, the first of those located at $\beta U\simeq 2.8$, i.e., quite before the first vertex divergence takes place at $\beta U\simeq 3.6$ [cf.\ panel (d)]. Indeed, at $\beta U\simeq 2.8$, the conventional damped iteration with $p$ (circles) begins to converge to an unphysical solution. This misleading convergence is obtained up to $\beta U\lesssim 3.6$, the location of the first vertex divergence, where surprisingly the second eigenvalue becomes also unstable. Beyond that, damped iteration does not converge anymore 
\footnote{This behavior of the misleading convergence is curious because SBE is agnostic to the 2PI divergence at half-filling, but it appears that the unphysical solution is indirectly affected by it.}.

Hence, for $\beta U\gtrsim 2.8$, the stabilized iteration, Eq.~\eqref{eq:SOT}, is needed.  The corresponding results, with $\mathcal{P}$, Eq.~\eqref{eq:SOT}, do converge successfully \footnote{We note as a curiosity, that up to $\beta U\lesssim 6.0$ mixing is not needed, i.e., $p(c=1)=1$. So there is an intermediate regime that requires stabilization of the Jacobian without the necessity of finite mixing.} to the physical solution (crosses).  We only observe that the region around $\beta U \simeq 9.3$ is hard to converge since, there, some eigenvalues vary only flatly around $\mathrm{Re}\, j_\alpha\simeq 1$. We improved the convergence in that region by using the nonuniform mixing, Eq.~\eqref{eq:nonuniform_p}.

Out of half-filling, as discussed before, the SBE diagrammatic can be plagued by SBE divergences [cf.\ Fig.~\ref{fig:Vertex_divergences}(c)], which also affect the stability condition in Fig.~\ref{fig:Stability_regions}(b).
To explicitly illustrate the effect of the newly found SBE divergences on the iteration, in analogy to Fig.~\ref{fig:SBE_SOT_dmu=0.0}, we show converged results for SBE at $\beta\delta\mu = 6.0$ in Fig.~\ref{fig:SBE_SOT_dmu=6.0}. Here, $\lambda_c$ exhibits a divergence at $\beta U\simeq 9.14$ and $\beta U\simeq 13.68$, which is visible in Fig.~\ref{fig:SBE_SOT_dmu=6.0}(c) (red lines). The converged results from stabilized iteration (pluses) match the exact ones in between and beyond both SBE divergences while the results from damped iteration (circles) deviate from the exact results behind the first instability. The second crossing of an eigenvalue [cf.\ Fig.~\ref{fig:SBE_SOT_dmu=6.0}(d)], coincides with the first SBE divergence at $\beta U\simeq 9.14$ and marks the point beyond which convergence of the damped iteration is not possible anymore.

\begin{figure}
	\centering
    \includegraphics[width=0.9\columnwidth]{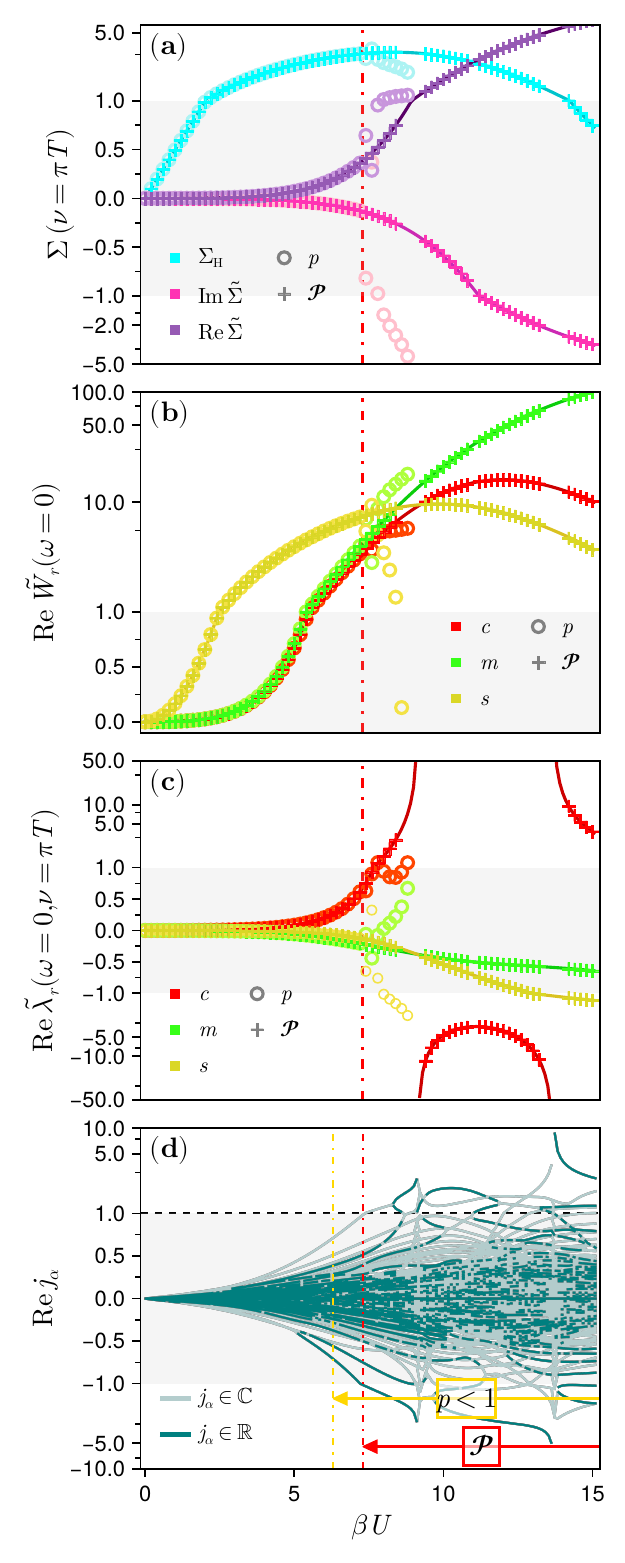}
	\caption{Converged quantities at specific Matsubara frequencies in dependence on $U$ from SBE at $\beta\delta\mu=6.0$. Details about colors and markers are provided in Fig.~\ref{fig:GW_SOT_dmu=12.0}(d). (a) Self-energies $\Sigma$ at $\nu=\pi T$, (b) bosonic propagators $\tilde W_r$ at $\omega=0$, (c) Hedin vertices $\tilde\lambda_r$ at $\omega=0$ and $\nu=\pi T$, (d) real parts of the eigenvalues $\mathrm{Re}\,j_\alpha$. 
    }
	\label{fig:SBE_SOT_dmu=6.0}
\end{figure}

\begin{figure*}
	\centering
    \includegraphics[width=\textwidth]{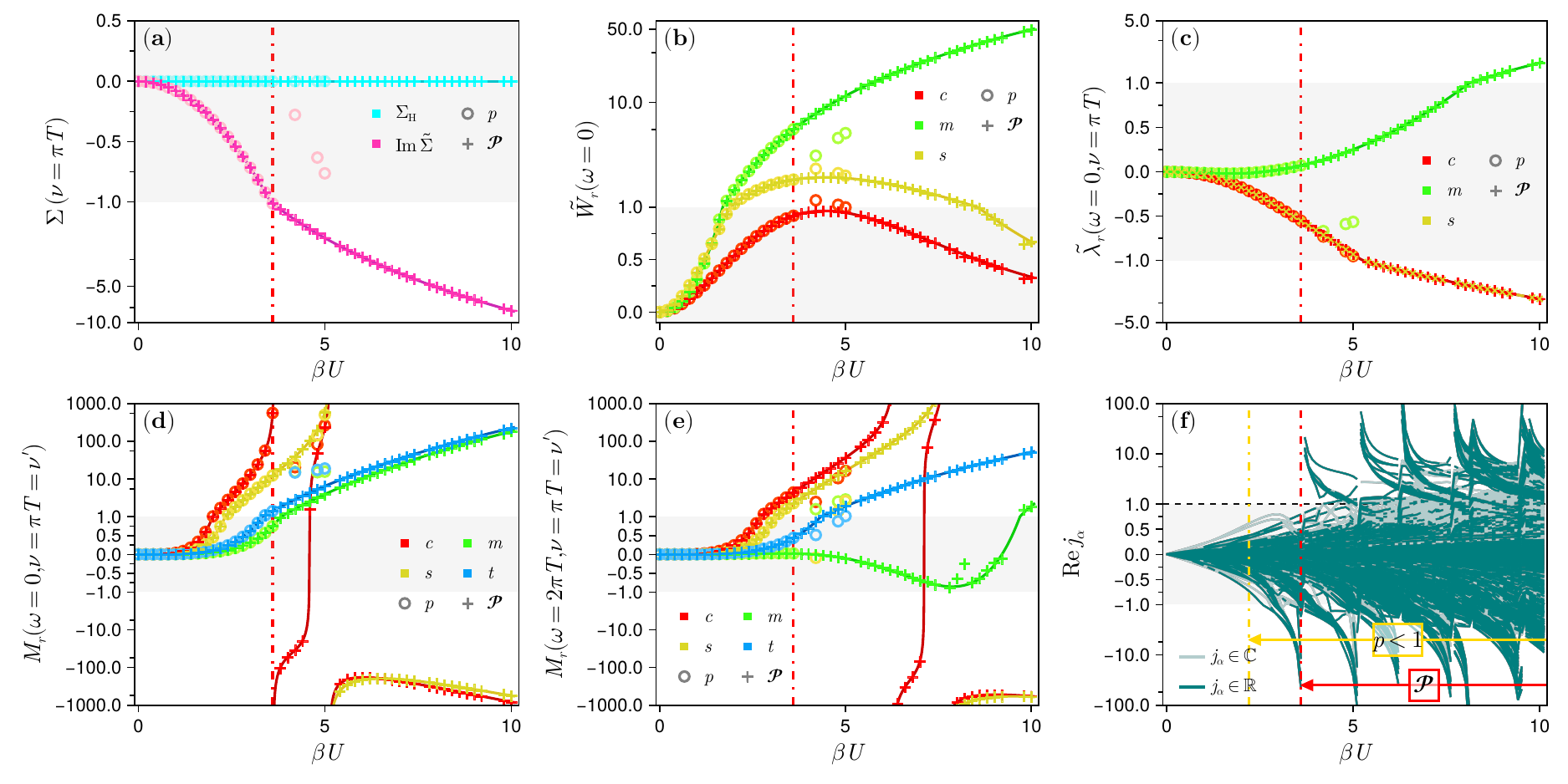}
	\caption{Converged quantities at specific Matsubara frequencies in dependence on $U$ from MBE at half-filling. Details about colors and markers are provided in Fig.~\ref{fig:GW_SOT_dmu=12.0}(d). (a) Self-energies $\Sigma$ at $\nu=\pi T$, (b) bosonic propagators $\tilde W_r$ at $\omega=0$, (c) Hedin vertices $\tilde\lambda_r$ at $\omega=0$ and $\nu=\pi T$, (d) MBE vertices $M_r$ at $\omega=0$ and $\nu=\pi T=\nu'$, (e) MBE vertices $M_r$ at $\omega=2\pi T$ and $\nu=\pi T=\nu'$, (f) real parts of the eigenvalues $j_\alpha$.
    }
	\label{fig:MBE_SOT_dmu=0.0}
\end{figure*}

Finally, we consider the more elaborate MBE scheme, which includes the (numerically expensive) iterative calculation of the MBE vertices $M_r$. Further, its Jacobian $\mathcal{J}$, Eq.~\eqref{eq:Jacobian-MBE}, is heavily affected by 2PI divergences [cf.\ Figs.~\ref{fig:Vertex_divergences}(a)--(b) and \ref{fig:Stability_regions}(c)].

Our MBE results at half-filling are presented in Fig.~\ref{fig:MBE_SOT_dmu=0.0}. Here, once again, we compare the exact values (lines) with results from the damped iteration with $p$ (circles) and the stabilized iteration with $\mathcal{P}$ (pluses) for (a) $\Sigma$, (b) $\tilde W_r$, (c) $\tilde \lambda_r$, and (d)--(e) $M_r$.
Panel (f) shows the corresponding real parts of the eigenvalues $\mathrm{Re}\,j_\alpha$.

Up to the first 2PI vertex divergence at $\beta U\simeq 3.6$, both the conventional damped iteration with $p$ (circles) and the stabilized iteration with $\mathcal{P}$ (pluses) 
converge to the exact physical solution (lines), while a deviation of the fixed point in the conventional damped-iteration scheme is observed for larger $U$ values.

This result is not surprising because, as already mentioned, the eigenvalues of the MBE scheme directly inherit the features of the 2PI divergences [cf.\ Fig.~\ref{fig:MBE_SOT_dmu=0.0}(e)]. 
Importantly, the first instability $\mathrm{Re}\, j_\alpha>1$ (marked by a red dashed line) occurs exactly at the first vertex divergence at $\beta U\simeq 3.6$ [see also again the inset of Fig.~\ref{fig:Stability_regions}(c)]. 
This explains why for a long time misleading convergence was interpreted to be directly connected to the occurrence of vertex divergences. However, our investigation shows that vertex divergences are only a sufficient but not a necessary condition for an unstable physical fixed point. An example for a parameter path at $\delta\mu=1$, where the first instability occurs before the vertex divergence, is given in the appendix in Fig.~\ref{fig:MBE_SOT_dmu=1.0}.

Taking a closer look to the data, between the first and second vertex divergence at $\beta U\in (3.6,5.2)$, the conventional scheme with $p$ converges to an unphysical solution. Beyond that, there are no further converged solutions with $p$. We suppose that for $\beta U\gtrsim 5.2$ the unphysical fixed point of the damped iteration, Eq.~\eqref{eq:Damped-iteration}, is located too far away from the starting point, such that convergence to this unphysical fixed point fails. 

On the other hand, with stabilized iteration, Eq.~\eqref{eq:SOT}, including $\mathcal{P}$ (pluses), we manage to obtain converged solutions up to particularly high, nonperturbative values of the electronic interaction, i.e., even \emph{beyond} the \emph{sixth} vertex divergence line at $\beta U\simeq 9$, with the only exception of the $U$ values in close proximity to the 2PI divergences \footnote{We attribute this to the fact that the damping parameter $p$ needed to locally stabilize the fixed point goes to zero when approaching a vertex divergence. Furthermore, it empirically appears that the basin of attraction to the (stabilized) physical fixed-point seems to become very small in the vicinity of vertex divergences.}. 

\subsection{Perturbative approximations}
\label{sec:Weak-coupling_results}

As is well known, nonperturbative computations are not always feasible. Hence, especially in the framework of realistic/ab-initio calculations, diagrammatic approaches are also performed with a perturbative input. For this reason, it is useful to investigate the general stability conditions, as well as the performance of the proposed stabilization procedure also in the realm of \emph{perturbative approximations}.

In particular, in this subsection, we show converged results of self-consistent TRILEX, SBE, and MBE with (lowest-order) perturbative inputs, i.e., $\lambda_r\to 1$, $\Lambda_r+\sum_r M_r\to U_r$, and  $\Lambda_r\to U_r$ (cf.\ Fig.~\ref{fig:MBE-SOT-Scheme}).

From the practical point of view, the key difference to the analogous study above with exact input (cf.\ Sec.~\ref{sec:Stabilized_iteration_strong}) is that the Jacobian cannot be calculated exactly and needs to be composed from the solution $\Psi$ at another $U$ value. 
We start the self-consistent loop from a converged solution at previous $U$ values $\Psi(U-\delta U)$ or predicted states $\Psi_\mathrm{extra}(U)$, which are extrapolated from previous solutions. Depending on the success of consecutive runs, we also adapt the step size $\delta U$ (for further details see App.~\ref{sec:Weak-coupling_details} and Fig.~\ref{fig:Perturbative_algorithm} therein).

Our detailed analysis, reported below, shows the occurrence of misleading convergence in \emph{all} diagrammatic approaches considered.
This might appear surprising as \emph{none} of the (SBE/2PI) vertex divergences discussed in Sec.~\ref{sec:Vertex_divergences} can be generated in such a perturbative framework.  In reality, this provides further evidence that, though features of misleading convergence might be triggered by such divergences, they generally occur totally independently from those.

\begin{figure*}
	\centering
    \includegraphics[width=\textwidth]{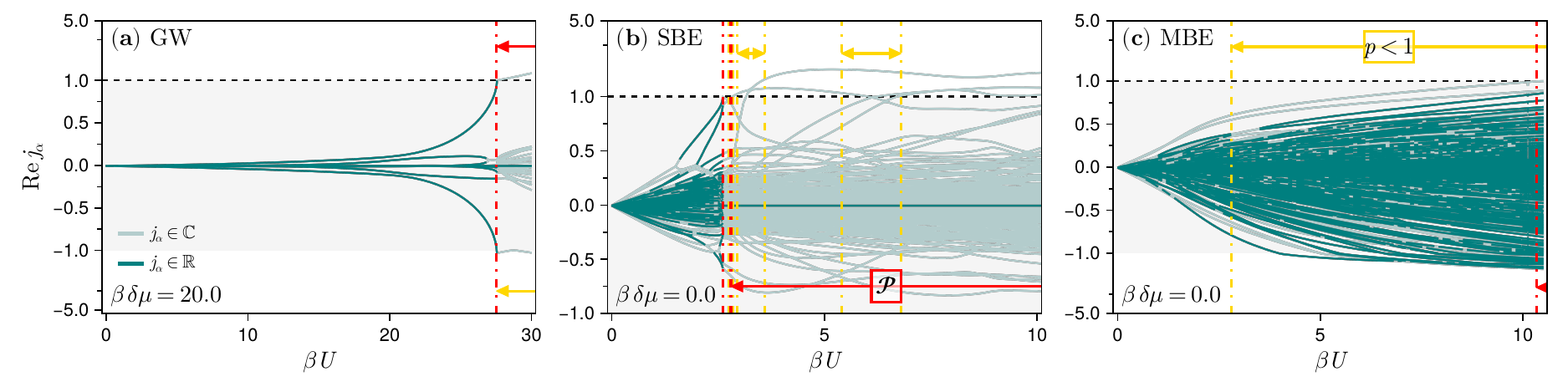}
	\caption{Real parts of the eigenvalues $j_\alpha$ of the Jacobian, Eqs.~\eqref{eq:J_GW}--\eqref{eq:Jacobian-MBE}, from the different perturbative diagrammatic schemes (a) GW, (b) SBE, (c) MBE. The lines are generated in dependence of $U$ at different values of $\delta\mu$. Dark teal lines represent purely real $j_\alpha$ and light teal ones complex-valued $j_\alpha$. Dashed vertical lines and arrows mark regions with finite mixing (yellow) and instabilities $\mathrm{Re}\,j_\alpha > 1$ (red). 
    }
	\label{fig:evals_dmu=0.0_weak}
\end{figure*}

\begin{figure*}
	\centering
    \includegraphics[width=\textwidth]{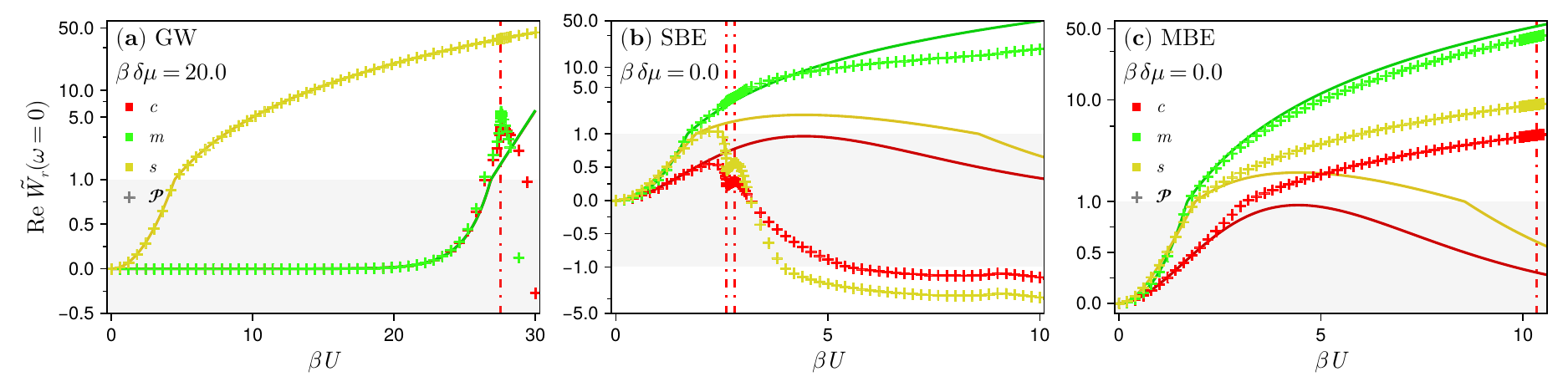}
	\caption{Converged bosonic propagators $W_r$ at $\omega=0$ in dependence on $U$ for the different diagrammatic schemes (a) GW, (b) SBE, (c) MBE with perturbative input (cf.\ Fig.~\ref{fig:MBE-SOT-Scheme}). The data are shown in dependence on $U$ at different values of $\delta\mu$. Details about colors and markers are provided in Fig.~\ref{fig:GW_SOT_dmu=12.0}(d).}
	\label{fig:W_r_comparison_weak}
\end{figure*}

Specifically, we show in Figs.~\ref{fig:evals_dmu=0.0_weak} and \ref{fig:W_r_comparison_weak} the eigenvalues as well as the bosonic propagators $\tilde W_r=W_r-U_r$ from self-consistent solutions with perturbative inputs (cf.\ Fig.~\ref{fig:MBE-SOT-Scheme}) for different dopings $\beta\delta\mu$ that exhibit instabilities in the Jacobian (cf.\ dashed red lines).

We can immediately note that for the perturbative TRILEX [cf.\ Figs.~\ref{fig:evals_dmu=0.0_weak}(a) and \ref{fig:W_r_comparison_weak}(a)], which can be identified, to a general extent, as a GW approach in multiple channels, 
we find instabilities only at extremely high couplings and imbalances, i.e., $\beta U\simeq 27.5$ and $\beta\delta\mu=20.0$. This is similar to the approach with exact input (cf.\ Figs.~\ref{fig:Stability_regions}(a) and \ref{fig:GW_SOT_dmu=12.0}). The fact that the dominant eigenvalue approaches unity $\mathrm{Re}\,j_\alpha\to 1$ very steeply (apparently with a diverging derivative), $\mathrm{d}j_\alpha/\mathrm{d}U \to \infty$ makes it hard to converge beyond the line of instability. Nonetheless, this is eventually possible by allowing a \emph{finite imaginary part} in $W_c$ and $W_m$ (cf.\ App.~\ref{sec:Weak-coupling_details}), which displays a clearly \emph{unphysical feature} in contrast to the exact solution. 
At the same time, it is quite evident from our data that the converged results for $W_r$  deviate clearly from the exact result \emph{already before} the instability is reached (cf.\ Fig.~\ref{fig:W_r_comparison_weak}(a) and App.~\ref{sec:Weak_coupling_appendix}).
Empirically, the instability of the fixed point appears to occur only in the (large $U$) parameter region, where the  perturbative self-consistent TRILEX/GW has already become, {\sl per se}, an \emph{unreliable} approximation.

For the perturbative SBE scheme [cf.\ Figs.~\ref{fig:evals_dmu=0.0_weak}(b) and \ref{fig:W_r_comparison_weak}(b)], a first instability at half-filling already occurs at $\beta U=2.61$, which surprisingly lies near the $U$ value where an instability occurs in nonperturbative SBE [cf.\ Fig.~\ref{fig:SBE_SOT_dmu=0.0}]. Similarly to the instability in GW, we observe that the respective dominant eigenvalue approaches $\mathrm{Re}\,j_\alpha\to 1$ with an apparently infinite slope. 
However, the respective bosonic propagators $W_r$ already deviate significantly from the exact result in this regime [cf.\ Fig.~\ref{fig:W_r_comparison_weak}(b)]. Furthermore, beyond the first instability $\beta U>2.61$, where the stabilization procedure is successfully applied, we again only find solutions with finite imaginary part of the bosonic propagators. 
Thus, stabilized iteration for the SBE approach with perturbative input does formally allow for convergence beyond the instability of its fixed point, but the obtained results lie extremely far from the physical solution, with little or no practical use (see also Fig.~\ref{fig:GW_SBE_SOT_weak}). Indeed, such a strong deviation of perturbative SBE from the exact result was also observed in other models \cite{Gievers2025Subleading}.
In Fig.~\ref{fig:W_r_comparison_weak_Im} in App.~\ref{sec:Weak_coupling_appendix}, we show the imaginary parts of $\tilde W_r$, which prominently highlights the unphysicality of the converged results after the instability.
Further, we note that we find a similar behavior also in the ZPM: Since the bosonic propagators $W_r$ of the ZPM are complex out of particle-hole symmetry (see App.~\ref{sec:ZP_BE}), we only find the behavior of a suddenly appearing imaginary part only at particle-hole symmetry.

Finally, the perturbative MBE scheme [cf.\ Figs.~\ref{fig:evals_dmu=0.0_weak}(c) and \ref{fig:W_r_comparison_weak}(c)] exhibits its first instability at half-filling only at $\beta U\simeq 10.3$. While MBE with nonperturbative/exact input is highly affected by 2PI divergences (cf.\ Figs.~\ref{fig:Stability_regions}(c) and \ref{fig:MBE_SOT_dmu=0.0}), the eigenvalues of the perturbative approach are much smaller than in the corresponding nonperturbative implementation, ensuring the stability of the conventional damped iteration algorithm for relatively larger values of $U$. In fact, similarly to the other diagrammatic schemes, the instability only appears in  parameter regions, where the approximation is not justified anymore and the perturbative solution deviates strongly from the physical one (see also Fig.~\ref{fig:MBE_SOT_dmu=0.0_weak}).

\subsection{Consequences for practical calculations}
\label{sec:Outlook}

Throughout this work, we have restricted our analysis to the HA and the ZPM, which allow for controlled benchmarks of the stabilization method illustrated in Sec.~\ref{sec:Jacobian_formalism}. It is important to discuss then which relevant piece of information can be extracted from our results to be used as a guidance for future developments and the applications of cutting-edge diagrammatic approaches to the many-electron problem.

The perturbative implementation of all diagrammatic approaches considered displays, systematically, instabilities of the corresponding fixed point (Figs.~\ref{fig:evals_dmu=0.0_weak}--\ref{fig:W_r_comparison_weak}). However, these instabilities occur only at sufficiently strong coupling, i.e., in regions where the converged solution is already quite far from the exact one (cf.\ also Figs.~\ref{fig:GW_SBE_SOT_weak}--\ref{fig:W_r_comparison_weak_Im}). Hence, even though the perturbative fixed point of the conventional iteration procedure can become unstable for all considered cases, such a problem is merely academic,
since the instability appears only when the corresponding perturbative approximation has already become quite unreliable.

The instabilities of the nonperturbative schemes, instead, have severe consequences for practical applications, e.g., for all the diagrammatic approaches, where the exact input is replaced by a numerically obtained \mbox{(cluster-)}DMFT result \cite{Rohringer2018Diagrammatic}.
Here, we start  by noting the following: When comparing the stability regions of the diagrammatic schemes for both models (cf.\ Figs.~\ref{fig:Stability_regions} and \ref{fig:Stability_ZP}), we find that for all three schemes (from TRILEX to MBE), the instability regions agree qualitatively well between the HA and the ZPM: The location of the instabilities relative to the vertex divergences, the overall shape of the instability regions, and the type of instability that occurs are all quite consistent across the two models. In App.~\ref{sec:ZP_additional}, we support these claims by showing the respective eigenvalues of the ZPM (cf.\ Figs.~\ref{fig:ZP_evals_dmu=0.0} and \ref{fig:ZP_evals_dmu!=0.0}).
Such an essential agreement in the stability features of these two intrinsically different models leads us to conjecture that correspondingly similar situations will occur, \emph{mutatis mutandis}, when applying these diagrammatic approaches to more realistic cases (such as, e.g., the lattice Hubbard model) in their strong-coupling regimes [e.g., by replacing the exact input currently used in the diagrammatic schemes with a \mbox{(cluster-)}DMFT input].

Specifically, for MBE this corresponds to a parquet-D$\Gamma$A calculation \cite{Toschi2007, Valli2015Dynamical,Kauch2020Generic,Eckhardt2020Truncated, Bippus2026a} and for SBE to the embedded multi-boson exchange (eMBEX) scheme introduced in Ref.~\cite{Kiese2024Embedded}. Our findings further show that the TRILEX method \cite{Ayral2015,Ayral2016} is also affected by the instabilities of the self-consistency. This raises the question whether also GW+EDMFT \cite{Kotliar2002,Biermann2003,Ayral2013,yeh2026}, which can be seen as a further approximation of the fermion-boson vertex in TRILEX \cite{Vucicevic2017}, might suffer from the same problem in specific parameter regimes. Those could be related to convergence problems and the existence of multiple solutions reported in different GW-based schemes \cite{Lani2012,Stan2015Unphysical,tandetzky2015,pokhilko2021,pokhilko2025,wang2026}.

Hence, our proposed stabilization scheme, or a similar Jacobian-based iteration, may turn out to be the crucial missing ingredient in the current implementations of several cutting-edge self-consistent diagrammatic approaches. Indeed, its inclusion could allow their reliable convergence in the physically most intriguing, and numerically most challenging, strong-coupling regimes.

\section{Conclusion}
\label{sec:Conclusion}

Our work provides  a thorough investigation of the fixed-point stability for different classes of self-consistent diagrammatic many-electron approaches, based on the boson-exchange (BE) formalism \cite{Krien2019Single,Krien2019Parquetlike,Krien2020Boson,Krien2020Tiling,Krien2022Plain,Krien2022Explaining,Bonetti2022Single,Gievers2022Multiloop,Fraboulet2022Single,Adler2024Nonperturbative,Kiese2024Embedded,meixner2025,Patricolo2025Single,Gievers2025Subleading,Meixner2026Nonperturbative,meixner2026}.
We underline here that the approaches derived in this framework, from the simpler TRILEX \cite{Ayral2015,Vucicevic2017,Stepanov2019Consistent} (which includes the GW \cite{Hedin1965new,Aryasetiawan1998GW,Vucicevic2017,Golze2019GW} as an approximation), to the SBE \cite{Krien2019Single,Bonetti2022Single,Kiese2024Embedded} and the more complex MBE \cite{Krien2020Tiling,Krien2022Plain,Fraboulet2022Single,Gievers2022Multiloop,Fraboulet2026Multiloop,AlEryani2026Functional}, are at the forefront of method development in the many-electron theory \cite{Rohringer2018Diagrammatic,Qin2022Hubbard}, as they generally allow for a better numerical performance and a clearer physical interpretation of the obtained results.

As unveiled by our systematic applications of these approaches to testbed, exactly solvable models (such as the HA and the ZPM), the conventional damped-iteration algorithm, which is broadly adopted to achieve the self-consistent solutions for all these BE-based approaches, displays significant problems of \emph{``misleading convergence''}: In quite relevant (intermediate-to-strong-coupling) regions of the phase space, the iterative algorithm ceases to converge to the physical fixed point, since an unphysical solution becomes stable while the physical fixed point becomes unstable (or the convergence turns out to be impossible at all).

This confirms the suspicion, recently advanced in Ref.~\cite{Essl2026Origin} and verified in the specific case of the parquet-based methods \cite{Essl2026Stabilizing}, that the misleading convergence can plague, as a general  ``algorithmic disease'', several self-consistent many-electron schemes even if they are not derived within the framework of the Luttinger--Ward functional (LWF). This means, even in the cases, when the diagrammatic approaches are not directly affected by the intrinsic multivaluedness problems \cite{Kozik2015Nonexistence} of the LW formalism.

Our analysis yields a robust proof for this expectation: We do not only show that, for all the general BE classes of approaches we considered,  misleading convergence problems are systematically encountered, independently of whether one starts from a nonperturbative/exact input or from a weak-coupling/perturbative one, but we also clarify the relation between the observed convergence issues with the divergences of two-particle-irreducible vertex functions \cite{Schaefer2013Divergent}, which are a typical (nonperturbative) manifestation of the LW multivaluedness \cite{Gunnarsson2017Breakdown}. Further, we also investigate the role of a complementary class of vertex divergences, \changed{not explicitly} discussed \changed{before}, namely the divergences of the $U$-irreducible vertex functions.  Specifically, we find that these divergences are a \emph{sufficient}, but not necessary \emph{condition} for the instability of the physical fixed point.

Such a clear-cut outcome of our study offers very good news to the field of many-electron simulations. In fact, while there is currently no known solution to the misleading convergence issues directly arising from the LW multivaluedness, i.e., from the breakdown of self-consistent perturbation theory, \cite{Kozik2015Nonexistence,Stan2015Unphysical,Rossi2015Skeleton,Schaefer2016Nonperturbative,Gunnarsson2017Breakdown,Tarantino2017Selfconsistent,Essl2026Origin},  when convergence problems are merely originated by the stability condition of the fixed point of the conventional (damped) iteration algorithm, as in the case of the BE diagrammatic approaches, there exists a clear remedy: A modification of the iteration scheme \cite{Essl2026Origin}, defined on the basis of the  corresponding Jacobian, capable to stabilize the physical solution, where it is unstable.
Indeed, our calculations on the HA and ZPM demonstrate that, through the implementation of this modified iteration procedure, the physical fixed point of all the BE-diagrammatic schemes considered can be stabilized, {\sl de facto}, in the \emph{whole} phase-space, from weak to strong coupling.

This achievement provides a particularly useful compass for the state-of-the-art development of performant (BE-based) diagrammatic algorithms, such as, e.g., the diagrammatic extensions \cite{Rohringer2018Diagrammatic} of DMFT \cite{Georges1998Dynamical}  and/or cluster-DMFT \cite{Maier2005Quantum}. In fact, this important piece of information can enable several diagrammatic approaches to eventually access the hardest regions of the parameter space, where electronic correlations are known to drive most fascinating and exotic phenomena.

The recent rapid advances in data compression tools, such as quantics tensor trains \cite{Shinaoka2023Multiscale,Ritter2024Quantics,Rohshap2025Two,Grosso2026Adaptive,rohshap2026,Rohshap2026a}, provide a quite promising route to handle the huge computational effort needed for evaluating the Jacobian and the stabilization matrix in future  applications to realistic lattice-model systems.

\begin{acknowledgments}
The authors thank \changed{Aiman Al-Eryani,} Samuel Badr, Jan von Delft, Evgeny Kozik, Fabian Kugler, Jae-Mo Lihm, and Markus Wallerberger for insightful discussions. This work was funded in part by the Austrian Science Fund (FWF) projects through Grant DOI 10.55776/P36332, 10.55776/V1018, 10.55776/PIN4372024,  10.55776/I5487, as well as  10.55776/KIN2563725. For open access purposes, the authors have applied a CC BY public copyright license to any author-accepted manuscript version arising from this submission. Calculations have been partly performed using Austrian Scientific
Computing (ASC). 
The authors used Claude (Anthropic) to assist with proofreading and copyediting of the manuscript text.
\end{acknowledgments}

\changed{
\section*{Data and code availability}
A dataset containing all numerical data and scripts for calculating and plotting these data will be made publicly available in the TU Wien Research Data repository. %\cite{data}.
}

\onecolumngrid
\appendix

\section{Frequency and channel conventions}
\label{sec:Frequencies}

As a four-point object, the full vertex $F$ originally depends on four frequencies and spin indices. Generically, we can arrange its constituents as
\begin{align}
    F = \underbrace{\Lambda}_{\text{2PI}} + \underbrace{\sum_r (M_r + \nabla_r - U)}_{\text{2PR}} = \underbrace{\Lambda^U}_{U\text{~irr.}} + \underbrace{\sum_r\nabla_r - 2U}_{U\text{~red.}},
    \label{eq:Diagrammatic_decomposition}
\end{align}
where the sums run over the three diagrammatic channels $r=\ph,\pp,\phx$ (particle-hole $\ph$, particle-particle $\pp$, and particle-hole crossed $\phx$) of two-particle reducibility and the dependence on frequencies and spin arguments is not made explicit. The separation into fully two-particle-irreducible (2PI) $\Lambda$ and two-particle-reducible (2PR) vertices $M_r$, $\nabla_r$ yields the parquet decomposition, and the separation into $U$-irreducible vertices $\Lambda^U = \Lambda-U+\sum_r M_r$ and $U$-reducible vertices $\nabla_r$, Eq.~\eqref{eq:SBE-vertex}, yields the SBE decomposition \cite{Gievers2022Multiloop}. 

Energy conservation leads to a dependence on only three frequencies, which we order in three different ways according to the diagrammatic channels $r=\ph,\pp\,\phx$. For this, we use the same conventions as in Ref.~\cite{Gievers2025PhD}:
\begin{align}
    F_\ph^{\omega\nu\nu'} =F^{(\nu'+\omega)\nu\nu'(\nu+\omega)},
    \quad
    F_\pp^{\omega\nu\nu'} =F^{(-\nu)(\nu+\omega)(-\nu')(\nu'+\omega)},
    \quad
    F_\phx^{\omega\nu\nu'} =F^{\nu(\nu'+\omega)\nu'(\nu+\omega)}.
    \label{eq:Frequencies}
\end{align}
For the spin indices [cf.\ Eq.~\eqref{eq:F}], we use the short-hand notation:
\begin{align}
    F^\updn = F^{\updn\updn},\quad F^\updnx = F^{\updn\dnup}, \quad F^{\upup} = F^{\upup\upup}.
\end{align}
Using SU(2)-spin symmetry, i.e., $F^\updn=F^\dnup$ and $F^\upup=F^\updn+F^\updnx$, we express all vertices in the four physical channels,
\begin{align}
    F_{c,m}^{\omega\nu\nu'} = [F^\upup]_\ph^{\omega\nu\nu'}\pm [F^\updn]_\ph^{\omega\nu\nu'},%\\
    \quad 
    F_{s,t}^{\omega\nu\nu'} = [F^\updn]_\ph^{\omega\nu\nu'}\mp [F^\updnx]_\ph^{\omega\nu\nu'},
    \label{eq:Channels}
\end{align}
which diagonalize the Bethe--Salpeter equations~\eqref{eq:M_r}.
Using the channel-specific notation, Eqs.~\eqref{eq:Frequencies}--\eqref{eq:Channels}, $F_r^{\omega\nu\nu'}$, Eq.~\eqref{eq:Diagrammatic_decomposition}, can be written as:
\begin{subequations}
    \label{eq:F_in_physical_channels}
	\begin{align}
		\nonumber F_c(\bm{\nu}_\ph)
		\nonumber &= \Lambda_c(\bm{\nu}_\ph) + M_c(\bm{\nu}_\ph) -\tfrac{1}{2} M_c(\bm{\nu}_\phx(\bm{\nu}_\ph)) -\tfrac{3}{2}M_m(\bm{\nu}_\phx(\bm{\nu}_\ph)) + \tfrac{1}{2}M_s(\bm{\nu}_\pp(\bm{\nu}_\ph)) + \tfrac{3}{2}M_t(\bm{\nu}_\pp(\bm{\nu}_\ph))\\ &\phantom{=}+\nabla_c(\bm{\nu}_\ph)
        -\tfrac{1}{2}\nabla_c(\bm{\nu}_\phx(\bm{\nu}_\ph)) - \tfrac{3}{2}\nabla_m(\bm{\nu}_\phx(\bm{\nu}_\ph)) + \tfrac{1}{2}\nabla_s(\bm{\nu}_\pp(\bm{\nu}_\ph)) -3U_c,\displaybreak[2]\\
		\nonumber F_m(\bm{\nu}_\ph)
		\nonumber&= \Lambda_m(\bm{\nu}_\ph) + M_m(\bm{\nu}_\ph) - \tfrac{1}{2}M_c(\bm{\nu}_\phx(\bm{\nu}_\ph)) + \tfrac{1}{2}M_m(\bm{\nu}_\phx(\bm{\nu}_\ph)) -\tfrac{1}{2}M_s(\bm{\nu}_\pp(\bm{\nu}_\ph)) + \tfrac{1}{2}M_t(\bm{\nu}_\pp(\bm{\nu}_\ph))\\
		&\phantom{=}+\nabla_m(\bm{\nu}_\ph)- \tfrac{1}{2}\nabla_c(\bm{\nu}_\phx(\bm{\nu}_\ph)) + \tfrac{1}{2}\nabla_m(\bm{\nu}_\phx(\bm{\nu}_\ph)) -\tfrac{1}{2}\nabla_s(\bm{\nu}_\pp(\bm{\nu}_\ph)) -3U_m,\displaybreak[2]\\
		\nonumber F_s(\bm{\nu}_\pp)
		\nonumber &= \Lambda_s(\bm{\nu}_\pp) + M_s(\bm{\nu}_\pp) + \tfrac{1}{2}M_c(\bm{\nu}_\phx(\bm{\nu}_\pp)) - \tfrac{3}{2}M_m(\bm{\nu}_\phx(\bm{\nu}_\pp)) + \tfrac{1}{2}M_c(\bm{\nu}_\ph(\bm{\nu}_\pp)) - \tfrac{3}{2}M_m(\bm{\nu}_\ph(\bm{\nu}_\pp))\\
		&\phantom{=} +\nabla_s(\bm{\nu}_\pp)+\tfrac{1}{2}\nabla_c(\bm{\nu}_\phx(\bm{\nu}_\pp)) - \tfrac{3}{2}\nabla_m(\bm{\nu}_\phx(\bm{\nu}_\pp)) + \tfrac{1}{2}\nabla_c(\bm{\nu}_\ph(\bm{\nu}_\pp)) - \tfrac{3}{2}\nabla_m(\bm{\nu}_\ph(\bm{\nu}_\pp)) -3U_s,\displaybreak[2]\\
		\nonumber F_t(\bm{\nu}_\pp)
		\nonumber&= \Lambda_t(\bm{\nu}_\pp) + M_t(\bm{\nu}_\pp) - \tfrac{1}{2}M_c(\bm{\nu}_\phx(\bm{\nu}_\pp)) - \tfrac{1}{2}M_m(\bm{\nu}_\phx(\bm{\nu}_\pp)) + \tfrac{1}{2}M_c(\bm{\nu}_\ph(\bm{\nu}_\pp)) + \tfrac{1}{2}M_m(\bm{\nu}_\ph(\bm{\nu}_\pp))\\
		&\phantom{=}-\tfrac{1}{2}\nabla_c(\bm{\nu}_\phx(\bm{\nu}_\pp)) - \tfrac{1}{2}\nabla_m(\bm{\nu}_\phx(\bm{\nu}_\pp)) + \tfrac{1}{2}\nabla_c(\bm{\nu}_\ph(\bm{\nu}_\pp)) + \tfrac{1}{2}\nabla_m(\bm{\nu}_\ph(\bm{\nu}_\pp)),
	\end{align}
\end{subequations}
where the $U$-irreducible vertices are $\nabla_r^{\omega\nu\nu'}=\lambda^{\omega\nu}_rW^\omega_r\lambda^{\omega\nu'}_r$ and $\bm{\nu}_r(\bm{\nu}_{r'})$ refers to the transformation from the natural frequencies of channel $r'$ to those of channel $r$. The transformations $\bm{\nu}_r(\bm{\nu}_{r'})$ between channels are explicitly given by
\begin{subequations}
    \label{eq:Channel_transformations}
    \begin{alignat}{4}
        &\bm{\nu}_\ph &&= \left(\begin{matrix}
            \omega \\ \nu \\ \nu'
        \end{matrix}\right)_{\!\ph} &&= \left(\begin{matrix}
            0 & 1 & -1 \\
            1 & 0 & \phantom{-}1 \\
            0 & 0 & \phantom{-}1 
        \end{matrix}\right)\!\left(\begin{matrix}
            \omega \\ \nu \\ \nu'
        \end{matrix}\right)_{\!\phx} &&= \left(\begin{matrix}
            0 & -1 & \phantom{-}1 \\
            1 & \phantom{-}1 & \phantom{-}0 \\
            0 & \phantom{-}0 & -1
        \end{matrix}\right)\!\left(\begin{matrix}
            \omega \\ \nu \\ \nu'
        \end{matrix}\right)_{\!\pp},\displaybreak[2]\\
        &\bm{\nu}_\pp &&= \left(\begin{matrix}
            \omega \\ \nu \\ \nu'
        \end{matrix}\right)_{\!\pp} &&= \left(\begin{matrix}
            1 & \phantom{-}1 & \phantom{-}1 \\
            0 & -1 & \phantom{-}0 \\
            0 & \phantom{-}0 & -1 
        \end{matrix}\right)\!\left(\begin{matrix}
            \omega \\ \nu \\ \nu'
        \end{matrix}\right)_{\!\phx} &&= \left(\begin{matrix}
            \phantom{-}1 & 1 & \phantom{-}1 \\
            -1 & 0 & -1 \\
            \phantom{-}0 & 0 & -1
        \end{matrix}\right)\!\left(\begin{matrix}
            \omega \\ \nu \\ \nu'
        \end{matrix}\right)_{\!\ph},\displaybreak[2]\\
        &\bm{\nu}_\phx &&=\left(\begin{matrix}
            \omega \\ \nu \\ \nu'
        \end{matrix}\right)_{\!\phx} &&= \left(\begin{matrix}
            1 & \phantom{-}1 & \phantom{-}1 \\
            0 & -1 & \phantom{-}0 \\
            0 & \phantom{-}0 & -1 
        \end{matrix}\right)\!\left(\begin{matrix}
            \omega \\ \nu \\ \nu'
        \end{matrix}\right)_{\!\pp} &&= \left(\begin{matrix}
            0 & 1 & -1 \\
            1 & 0 & \phantom{-}1 \\
            0 & 0 & \phantom{-}1
        \end{matrix}\right)\!\left(\begin{matrix}
            \omega \\ \nu \\ \nu'
        \end{matrix}\right)_{\!\ph} .
    \end{alignat}
\end{subequations}

Consequently, in the channel-specific form, the fully $U$-irreducible vertex $\Lambda^U=\Lambda-U+\sum_rM_r$ in Eq.~\eqref{eq:Diagrammatic_decomposition} obtains the following reformulations:
\begin{subequations}
    \label{eq:Lambda^U}
	\begin{align}
		\Lambda_c^U(\bm{\nu}_\ph) &= \tilde\Lambda_c(\bm{\nu}_\ph)+ M_c(\bm{\nu}_\ph) -\tfrac{1}{2} M_c(\bm{\nu}_\phx(\bm{\nu}_\ph)) -\tfrac{3}{2}M_m(\bm{\nu}_\phx(\bm{\nu}_\ph)) + \tfrac{1}{2}M_s(\bm{\nu}_\pp(\bm{\nu}_\ph)) + \tfrac{3}{2}M_t(\bm{\nu}_\pp(\bm{\nu}_\ph)),\displaybreak[2]\\
        \Lambda_m^U(\bm{\nu}_\ph) &= \tilde\Lambda_m(\bm{\nu}_\ph)  + M_m(\bm{\nu}_\ph) - \tfrac{1}{2}M_c(\bm{\nu}_\phx(\bm{\nu}_\ph)) + \tfrac{1}{2}M_m(\bm{\nu}_\phx(\bm{\nu}_\ph)) -\tfrac{1}{2}M_s(\bm{\nu}_\pp(\bm{\nu}_\ph)) + \tfrac{1}{2}M_t(\bm{\nu}_\pp(\bm{\nu}_\ph)),\displaybreak[2]\\
		\Lambda_s^U(\bm{\nu}_\pp) &= \tilde\Lambda_s(\bm{\nu}_\pp)  + M_s(\bm{\nu}_\pp) + \tfrac{1}{2}M_c(\bm{\nu}_\phx(\bm{\nu}_\pp)) - \tfrac{3}{2}M_m(\bm{\nu}_\phx(\bm{\nu}_\pp)) + \tfrac{1}{2}M_c(\bm{\nu}_\ph(\bm{\nu}_\pp)) - \tfrac{3}{2}M_m(\bm{\nu}_\ph(\bm{\nu}_\pp)),\displaybreak[2]\\
		\Lambda^U_t(\bm{\nu}_\pp) &= \tilde\Lambda_t(\bm{\nu}_\pp)+ M_t(\bm{\nu}_\pp) - \tfrac{1}{2}M_c(\bm{\nu}_\phx(\bm{\nu}_\pp)) - \tfrac{1}{2}M_m(\bm{\nu}_\phx(\bm{\nu}_\pp)) + \tfrac{1}{2}M_c(\bm{\nu}_\ph(\bm{\nu}_\pp)) + \tfrac{1}{2}M_m(\bm{\nu}_\ph(\bm{\nu}_\pp)).
	\end{align}
\end{subequations}
Here, $\tilde \Lambda_r=\Lambda_r-U_r$ are the fully 2PI vertices excluding the bare contributions $U_r$, Eq.~\eqref{eq:U_r}.
Finally, inserting Eqs.~\eqref{eq:F_in_physical_channels} and \eqref{eq:Lambda^U} into Eqs.~\eqref{eq:T_r_from_F_r}, the $U$-irreducible vertices $T_r^{\omega\nu\nu'}$ can be brought in terms of $\Lambda_r^U$, $M_r$, and $\nabla_r$:
\begin{subequations}
    \label{eq:T_r_from_MBE}
	\begin{align}
        T_c(\bm{\nu}_\ph)
        &= \Lambda^U_c(\bm{\nu}_\ph) 
        -\tfrac{1}{2}\nabla_c(\bm{\nu}_\phx(\bm{\nu}_\ph)) - \tfrac{3}{2}\nabla_m(\bm{\nu}_\phx(\bm{\nu}_\ph)) + \tfrac{1}{2}\nabla_s(\bm{\nu}_\pp(\bm{\nu}_\ph)) -2U_c,\displaybreak[2]\\
        T_m(\bm{\nu}_\ph) 
        &= \Lambda^U_m(\bm{\nu}_\ph)
        - \tfrac{1}{2}\nabla_c(\bm{\nu}_\phx(\bm{\nu}_\ph)) + \tfrac{1}{2}\nabla_m(\bm{\nu}_\phx(\bm{\nu}_\ph)) -\tfrac{1}{2}\nabla_s(\bm{\nu}_\pp(\bm{\nu}_\ph)) -2U_m,\displaybreak[2]\\
        T_s(\bm{\nu}_\pp) 
        &= \Lambda^U_s(\bm{\nu}_\pp) 
        +\tfrac{1}{2}\nabla_c(\bm{\nu}_\phx(\bm{\nu}_\pp)) - \tfrac{3}{2}\nabla_m(\bm{\nu}_\phx(\bm{\nu}_\pp)) + \tfrac{1}{2}\nabla_c(\bm{\nu}_\ph(\bm{\nu}_\pp)) - \tfrac{3}{2}\nabla_m(\bm{\nu}_\ph(\bm{\nu}_\pp)) -2U_s,\displaybreak[2]\\
        T_t(\bm{\nu}_\pp) 
        &=\Lambda^U_t(\bm{\nu}_\pp) 
        -\tfrac{1}{2}\nabla_c(\bm{\nu}_\phx(\bm{\nu}_\pp)) - \tfrac{1}{2}\nabla_m(\bm{\nu}_\phx(\bm{\nu}_\pp)) + \tfrac{1}{2}\nabla_c(\bm{\nu}_\ph(\bm{\nu}_\pp)) + \tfrac{1}{2}\nabla_m(\bm{\nu}_\ph(\bm{\nu}_\pp)).
	\end{align}
\end{subequations}
This is the explicit form used in the numerical implementation of the self-consistent cycle.

\section{Solution of the Hubbard atom (HA)}
\label{sec:HA-vertices}

Following Ref.~\cite{Gievers2025PhD}, the SBE vertices of the HA can be exactly derived from the spectral representation \cite{Kugler2021Multipoint,Halbinger2023Spectral}.

The partition function $Z$ and density per spin $n$ are given as
\begin{align}
    Z = 1+2\,\ee^{\beta\mu}+\ee^{\beta(2\mu-U)},\quad
    n = \frac{1}{Z}\left(\ee^{\beta\mu}+\ee^{\beta(2\mu-U)}\right).
    \label{eq:HA_Z}
\end{align}
For the Green's function $G$ and self-energy $\Sigma$, we have
\begin{align}
    \label{eq:HA_G}
        G^\nu = \frac{1}{\ii\nu+\mu-\Sigma^\nu},\quad
        \Sigma^\nu = U n + \frac{U^2(1-n)n}{\ii\nu+\mu+U(n-1)}.
\end{align}
Conveniently, $U/2$ is absorbed in the Hartree term $\Sigma_\mathrm{H}$ such that the original noninteracting Green's function $G_0^\nu =1/(\ii\nu+\mu)$ is defined as
\begin{align}
    G_0^\nu &= \frac{1}{\ii\nu+\mu-U/2},
\end{align}
and the Hartree term is $\Sigma_\mathrm{H}=\lim_{\nu\to\infty}(\Sigma_\nu-U/2)=U(n-1/2)$, which exactly vanishes at half-filling.

The bosonic propagators $\tilde W_r = W_r - U_r$ are given by
\begin{subequations}
    \label{eq:HA_W_r}
    \begin{align}
        \tilde W_c^\omega &= \frac{U^2}{Z^2}\beta\delta_\omega\ee^{\beta\mu}\left(1+\ee^{\beta(2\mu-U)}+2\ee^{\beta(\mu-U)}\right)= -U^2\beta\delta_\omega\left[(1-n)(1-2n)-1/Z\right],\displaybreak[2]\\
        \tilde W_m^\omega &= \frac{U^2}{Z}\beta\delta_\omega\ee^{\beta\mu}=U^2\beta\delta_\omega\left[1-n-1/Z\right],\\
        \tilde W_s^\omega &= \frac{2U^2}{Z}\frac{\ee^{\beta(2\mu-U)}-1}{\ii\omega-U+2\mu}=-\frac{2U^2(1-2n)}{\ii\omega-U+2\mu}.
    \end{align}
\end{subequations}
Note that, at half-filling, $\lim_{\mu\to U/2}\tilde W_s^\omega = 2U^2\beta\delta_\omega/ Z$.

We define the factors $x(\nu)=\ii\nu+\mu$ and $y(\nu) = \ii\nu + \mu - U$ \cite{Pairault2000Strong}. With this, the Hedin vertices $\tilde \lambda_r = \lambda_r - 1$ yield
\begin{subequations}
	\label{eq:HA_lambda}
	\begin{align}
        \tilde\lambda_c^{\omega,\nu}&=U\frac{\tilde W_c^\omega}{W_c^\omega}\frac{1-2n}{y(\nu)+Un}
        -U^2\frac{U+\tilde W_c^\omega}{W_c^\omega}\frac{(1-n)n}{(y(\nu)+Un)(y(\omega+\nu)+Un)},\displaybreak[2]\\
        \tilde\lambda_m^{\omega,\nu}&=U\frac{\tilde W_m^\omega}{W_m^\omega}\frac{1-2n}{y(\nu)+Un}
        -U^2\frac{-U+\tilde W_m^\omega}{W_m^\omega}\frac{(1-n)n}{(y(\nu)+Un)(y(\omega+\nu)+Un)},\displaybreak[2]\\
        \tilde\lambda_s^{\omega,\nu}&=U\frac{\tilde W_s^\omega}{W_s^\omega}\left[\frac{1-n}{y(-\nu)+Un}+\frac{1-n}{y(\omega+\nu)+Un}\right]
        +U^2\frac{U+\tfrac{1}{2}\tilde W_s^\omega}{\tfrac{1}{2}W_s^\omega}\frac{(1-n)^2}{(y(-\nu)+Un)(y(\omega+\nu)+Un)}.
	\end{align}
\end{subequations}
At half-filling $\mu = U/2$, we have $y(\nu)+Un=\ii\nu$ such that the expressions above are simplified.

From the connected four-point correlators,
\begin{subequations}
	\label{eq:HA_G_con^(4)}
	\begin{align}
		\nonumber&[G^{(4)}_\mathrm{con}]^{\updn}(\nu_1,\nu_2,\nu_4,\nu_3)
		= \left\langle c_\uparrow^{\nu_1}c_\downarrow^{\nu_2}\bar c_\downarrow^{\nu_3}\bar c_\uparrow^{\nu_4}\right\rangle - \beta\delta_{\nu_1\nu_4}G^{\nu_1}G^{\nu_2}\\
		\nonumber&=
		\frac{-[\tilde W_\phx^\updnx]^{\nu_2-\nu_3}}{x(\nu_1)x(\nu_2)y(\nu_3)y(\nu_4)}+\frac{[\tilde W_\pp^\updn]^{\nu_1+\nu_2}+U(1-2n)}{y(\nu_1)y(\nu_2)y(\nu_3)y(\nu_4)}
        +\frac{1}{y(\nu_1)x(\nu_3)x(\nu_4)y(\nu_4)}\left[[\tilde W_\phx^\updn]^{\nu_3-\nu_1}+\frac{U^3n}{x(\nu_1)y(\nu_3)}\right]\\
		\nonumber&\phantom{=}-\frac{U}{Z}\left[\frac{1}{x(\nu_1)x(\nu_2)x(\nu_3)y(\nu_4)}+\frac{1}{y(\nu_1)y(\nu_2)y(\nu_3)x(\nu_4)}\right]
        -\frac{U}{Z}\left[\frac{Z(1-n)-1}{x(\nu_1)y(\nu_2)y(\nu_3)y(\nu_4)}+\frac{Z(1-n)-1}{y(\nu_1)x(\nu_2)x(\nu_3)x(\nu_4)}\right]\\
		&\phantom{=}-\frac{U^3}{Zx(\nu_1)y(\nu_1)x(\nu_4)y(\nu_4)}\left[\frac{1}{x(\nu_2)y(\nu_2)}+\frac{Z-1}{x(\nu_3)y(\nu_3)}\right],\displaybreak[2]\\
		&[G^{(4)}_\mathrm{con}]^{\hat{\updn}}(\nu_1,\nu_2,\nu_4,\nu_3)
		= \left\langle c_\uparrow^{\nu_1}c_\downarrow^{\nu_2}\bar c_\uparrow^{\nu_3}\bar c_\downarrow^{\nu_4}\right\rangle + \beta\delta_{\nu_1\nu_3}G^{\nu_1}G^{\nu_2}
        = -[G^{(4)}_\mathrm{con}]^{\updn}(\nu_1,\nu_2,\nu_3,\nu_4),\displaybreak[2]\\
		&[G^{(4)}_\mathrm{con}]^{\upup}(\nu_1,\nu_2,\nu_4,\nu_3)
		= \left\langle c_\uparrow^{\nu_1}c_\uparrow^{\nu_2}\bar c_\uparrow^{\nu_3}\bar c_\uparrow^{\nu_4}\right\rangle + \beta(\delta_{\nu_1\nu_3}-\delta_{\nu_1\nu_4})G^{\nu_1}G^{\nu_2}
        =\frac{[\tilde W_\phx^\upup]^{\nu_1-\nu_3}-[\tilde W_\phx^\upup]^{\nu_2-\nu_3}}{x(\nu_1)y(\nu_1)x(\nu_2)y(\nu_2)},
	\end{align}
\end{subequations}
we gain direct access to the full vertices $F_r$ and $U$-irreducible vertices $T_r$ using Eqs.~\eqref{eq:F}, \eqref{eq:SBE-vertex}, and \eqref{eq:T_r_from_F_r}.

To fulfill time-reversal and complex-conjugation symmetry numerically exactly \cite{Gievers2025PhD}, we actually compute the Bethe--Salpeter equations~\eqref{eq:M_r} for $M_r$ in a symmetrized form:
\begin{align}
    M^{\omega\nu\nu'}_r &= \tfrac{1}{\beta}\sum_{\nu''}T_r^{\omega\nu\nu''}\Pi_r^{\omega\nu''}T_r^{\omega\nu''\nu'}%\\
    -\tfrac{1}{2\beta}\sum_{\nu''}\left[M^{\omega\nu\nu''}\Pi_r^{\omega\nu''}T^{\omega\nu''\nu'}_r
    +T_r^{\omega\nu\nu''}\Pi_r^{\omega\nu''}M_r^{\omega\nu''\nu'}\right].
    \label{eq:M_r_symmetrized}
\end{align}
Since $T_r$ and $\Pi_r$ are known exactly, we obtain the numerical exact input for $M_r$ by solving \eqref{eq:M_r_symmetrized} for $M_r$:
\begin{align}
    M_r^{\omega\nu\nu'} &=\tfrac{1}{2\beta^2}\sum_{\nu_1,\nu_2}\left\{T_r^{\omega\nu\nu_1}\Pi_r^{\omega\nu_1}T_r^{\omega\nu_1\nu_2}[(\mathds{1}-\Pi_r T_r)^{-1}]^{\omega\nu_2\nu'}
    + [(\mathds{1}-\Pi_r T_r)^{-1}]^{\omega\nu\nu_1} T_r^{\omega\nu_1\nu_2}\Pi_r^{\omega\nu_2}T_r^{\omega\nu_2\nu'}\right\}.
    \label{eq:BSE_inverted}
\end{align}
Here, the matrix $[\mathds{1}-\Pi_rT_r]^{\omega}$ is inverted in the space of its two fermionic frequencies $\nu$ and $\nu'$ for each bosonic frequency $\omega$. The finite grid size of $[\mathds{1}-\Pi_rT_r]^{\omega}$ generates a small numerical error on $M_r^{\omega\nu\nu'}$ (cf.\ App.~\ref{sec:Frequency-boxes}).

\section{Solution of the zero-point model (ZPM)}
\label{sec:ZP_BE}

Since all quantities in the ZPM are just complex scalars, we can calculate the bosonic propagators $W_r$, Hedin vertices $\lambda_r$, and MBE vertices $M_r$ analytically.
This is done by using the relation between two-particle correlation functions and SBE vertices \cite{Gievers2022Multiloop,Gievers2025PhD} together with the analytical expressions for the two-particle quantities given in Ref.~\cite{Essl2026Stabilizing}.

The bosonic propagators $W_r$ are given by
    \begin{align}
        W_c = -U - \frac{U^2(U+\delta\mu^2)}{(\delta\mu^2-U)^2}, \quad
		W_m = U+\frac{U^2}{U-\delta\mu^2}, \quad
		W_s =-2W_m, \quad W_t=0 %U-\frac{2U^2}{U-\delta\mu^2}
        .
        \label{eq:ZP_W}
    \end{align}
Note that, out of half-filling, all $W_r$ are complex due to the fact that we only consider a single Matsubara frequency. This unphysical behavior could be overcome by considering two decoupled ZPMs with complex conjugated chemical potentials that represent a negative and a positive Matsubara frequency.

The Hedin vertices $\lambda_r$ are given by
 \begin{align}
        \lambda_c = \frac{(U+\delta\mu^2)(U - \delta\mu^2)^2}{2 U^2 \delta\mu^2 - U \delta\mu^4 + \delta\mu^6},\quad
		\lambda_m = \frac{(U - \delta\mu^2)^2}{\delta\mu^4-2 U \delta\mu^2}=
		\lambda_s,\quad \lambda_t=1.
        \label{eq:ZP_hedin}
    \end{align}

Finally, the MBE vertices $M_r$ are given by
%\begin{subequations}
    %\label{eq:ZP_M}
    \begin{align}
        M_c &= -\frac{U^4(3\delta\mu^2-U)^2}{\delta\mu^4(\delta\mu^2+U)(\delta\mu^4-U\delta\mu^2+2U^2)}, \quad M_m = -\frac{U^4}{\delta\mu^4(\delta\mu^2-2U)}, \quad
        M_s = -2M_m, \quad M_t = 0.        
        %\frac{U^2 \left(U-\delta\mu ^2\right) \left(2 \delta\mu ^4+3 U^2-3 \delta\mu ^2 U\right)^2}{\delta\mu ^4 \left(\delta\mu ^4+3 U^2-2 \delta\mu ^2 U\right) \left(\delta\mu ^4+2 U^2-\delta\mu ^2 U\right)},\displaybreak[2]\\
		%M_m&= -\frac{U^4 \left(U^2-\delta\mu ^4\right)^2}{\delta\mu ^4 \left(\delta\mu ^4+2 U^2-\delta\mu ^2 U\right) \left(\delta\mu ^6+U^3+3 \delta\mu ^2 U^2-\delta\mu ^4 U\right)},\displaybreak[2]\\
		%M_s&= \frac{U^2 \left(-3 \delta\mu ^6+5 U^3-9 \delta\mu ^2 U^2+7 \delta\mu ^4 U\right)^2}{\delta\mu ^4 \left(\delta\mu ^4+2 U^2-\delta\mu ^2 U\right) \left(2 \delta\mu ^6+5 U^3+\delta\mu ^4 U\right)},\displaybreak[2]\\
        %M_t&= \frac{U^2 \left(U-\delta\mu ^2\right)^6}{\delta\mu ^4 \left(\delta\mu ^4+2 U^2-\delta\mu ^2 U\right) \left(2 \delta\mu ^6+U^3+2 \delta\mu ^2 U^2-\delta\mu ^4 U\right)}.
        \label{eq:ZP_M}
    \end{align}
%\end{subequations}

For completeness, we also reproduce the self-energy $\Sigma$ and the fully 2PI vertices $\Lambda_r$ from Ref.~\cite{Essl2026Stabilizing} (note the different sign convention for $\Lambda_r$):
\begin{subequations}
    \begin{align}
        \label{eq:ZP_Sigma}
        \Sigma&= U\frac{\delta\mu}{\delta\mu^2-U} -\frac{U^2}{\delta\mu\,(\delta\mu^2-U)}=\frac{U}{\delta\mu},\displaybreak[2]\\
        \Lambda_c &=-U + \frac{2U^4}{\delta\mu^4(\delta\mu^2+U)}, \quad
        \Lambda_m 
        =-\Lambda_c, \quad
        \Lambda_s 
        =2\Lambda_c, \quad \Lambda_t = 0.
        \label{eq:ZP_Lambda}
    \end{align}
\end{subequations}
The self-energy $\Sigma$ is a sum of $\Sigma_\mathrm{H}$, Eq.~\eqref{eq:Sigma_Hartree}, and $\tilde\Sigma$, Eq.~\eqref{eq:Sigma}.

\section{Divergence in the Hedin vertex}
\label{sec:SBE_Vertex_divergece}

From Eqs.~\eqref{eq:W_r} and \eqref{eq:P_r}, we conclude that the Hedin vertex $\lambda_r$ exactly diverges where the bosonic propagator $W_r$ vanishes (apart from the trivial solution where $U=0$ and thus $W_r=0=\tilde\lambda_r$). This is very similar to the fact that 2PI divergences in $\Gamma_r=F_r-M_r-(\nabla_r-U_r)$ are related to a vanishing eigenvalue of the generalized susceptibility $\chi_r^{\omega\nu\nu'}$ \cite{Schaefer2016Nonperturbative}.
In the following, we will investigate the location of the SBE divergences for the HA and the ZPM by exploring the regions where $W_r$ vanishes. \changed{Note that the divergence of $\lambda_r$ is compensated by the zero of $W_r$ in the product $\lambda_r W_r$, which remains finite. Consequently, the SBE divergences do not appear in the high-frequency asymptotic classes of $F$ \cite{Wentzell2020High,Gievers2022Multiloop,Gievers2025PhD}, whereas $\nabla_r=\lambda_rW_r\lambda_r$ and $M_r$ do diverge.}

\subsection{SBE divergence for the HA}

Let us investigate the roots of $W_r$. First of all, for finite $\omega$, the bosonic propagators $W_c$, $W_m$, and $W_s$ do not vanish apart from the trivial case $U=0$ (see Eq.~\ref{eq:HA_W_r}). Therefore, it is sufficient to investigate the case $\omega=0$.

\begin{comment}
In the charge channel, we have at half-filling, i.e., $\delta\mu = 0$:
\begin{align}
    \lim_{\delta\mu\to 0}W^{\omega=0}_c =U \left[-1+U\beta \frac{\ee^{\beta U/2}2(1+\ee^{-\beta U/2})}{(2+2\ee^{\beta U/2})^2}\right] =U\left[-1+\frac{U\beta}{2(1+\ee^{\beta U/2})}\right].
\end{align}
From $\ee^{\beta U/2}>\beta U/2$, it follows:
\begin{align}
    \ee^{\beta U/2} + 1 > \beta U / 2
    \Rightarrow2(\ee^{\beta U/2} + 1) > \beta U
    \Rightarrow 1 > \frac{\beta U}{2(\ee^{\beta U/2}+1)}.
\end{align}
So at half-filling, there is only the trivial root at $U=0$. Out of half-filling, the situation 
\end{comment}

In the charge channel, we have:
\begin{align}
    W^{\omega=0}_c = U\left[-1+U\beta\frac{\ee^{\beta(U/2+\delta\mu)}(1+\ee^{2\beta\delta\mu}+2\ee^{\beta(\delta\mu-U/2)})}{(1+2\ee^{\beta(U/2+\delta\mu)}+\ee^{2\beta\delta\mu})^2}\right].
\end{align}
We can directly follow that, for $U<0$, the whole term in the bracket becomes negative so nontrivial roots can only appear for $U>0$. To find a closed form for the roots, let us introduce the substitutions $x=\ee^{\beta U/2}$ and $y=\ee^{\beta\delta\mu}$. Then the equation inside the bracket becomes a quartic equation in $y$,
\begin{align}
    -1+2\ln x\frac{xy(1+y^2+2y/x)}{(1+2xy+y^2)^2}=0 \Rightarrow y^4 + 2 x (2-\ln x)(y^3+y)+2y^2(1+2x^2-2\ln x)+1 = 0,
\end{align}
whose four roots
\begin{align}
    \nonumber & y(x) = \tfrac{1}{2}\left(x(-2+\ln x) -\sigma_2\sqrt{\ln x}\sqrt{4-4x^2+x^2\ln x}\right.\\
    &\phantom{=}\left.+\sigma_1\sqrt{4(-1+\ln x)+\sigma_2(4x\sqrt{\ln x}\sqrt{4-4x^2+x^2\ln x}-2x(\ln x)^{3/2}\sqrt{4-4x^2+x^2\ln x})+2x^2(2+(-4+\ln x)\ln x)}\right),
\end{align}
with $\sigma_1,\sigma_2=\pm 1$ can be easily transformed to a relation between $\beta\delta\mu$ and $\beta U$, namely $\beta\delta\mu = \ln[y(\ee^{\beta U/2})]$. From this, we follow one curved line of zeros for $\delta\mu >0$ and $\delta\mu <0$ each, which envelopes the line of density change $\delta\mu = \pm U/2$. In particular, there is no nontrivial root at half-filling, i.e., $\delta\mu = 0$.

In the magnetic channel, we have:
\begin{align}
    W^{\omega=0}_m = U\left[ 1+U\beta\frac{\ee^{\beta(U/2+\delta\mu)}}{1+2\ee^{\beta(U/2+\delta\mu)}+\ee^{2\beta\delta\mu}}\right].
    \label{eq:W_m^0}
\end{align}
Let us show that the term inside the brackets is strictly positive. The global minimum of $\ee^{\beta U/2}(2+\beta U)$ is $-2\ee^{-2}$ and the global maximum of $-2\cosh\beta\delta\mu=-\ee^{\beta\delta\mu}-\ee^{-\beta\delta\mu}$ is $-2$. From $-2\ee^{-2}>-2$, we conclude:
\begin{align}
    \ee^{\beta U/2}(2+\beta U) > -\ee^{\beta\delta\mu}-\ee^{-\beta\delta\mu} &> 0\\
    \Rightarrow 1+ U\beta\frac{\ee^{\beta(U/2+\delta\mu)}}{1+2\ee^{\beta(U/2+\delta\mu)}+\ee^{2\beta\delta\mu}} > 0,
\end{align}
where in the last step we divided by $Z>0$, Eq.~\eqref{eq:HA_Z}. From there, we conclude that $W_m^{\omega=0}$, Eq.~\eqref{eq:W_m^0}, only has the trivial root at $U=0$.

In the singlet channel, we have:
\begin{align}
    W^{\omega=0}_s = 2U\left[-1+\frac{U}{2\delta\mu}\frac{\ee^{2\beta\delta\mu}-1}{1+2\ee^{\beta(U/2+\delta\mu)}+\ee^{2\beta\delta\mu}}\right].
     \label{eq:W_s^0}
\end{align}
We want to prove that the term inside the bracket is negative. First, we put the term to a common denominator:
\begin{align}
    -1+\frac{U}{2\delta\mu}\frac{\ee^{2\beta\delta\mu}-1}{1+2\ee^{\beta(U/2+\delta\mu)}+\ee^{2\beta\delta\mu}} 
    &= \frac{-\cosh(\beta\delta\mu)-\ee^{\beta U/2}+\frac{U\beta}{2}\frac{\sinh(\beta\delta\mu)}{\beta\delta\mu}}{\cosh(\beta\delta\mu)+\ee^{\beta U/2}}.
\end{align}
Since the denominator is strictly positive, we exclusively investigate the numerator. In the following, we make use of
the tangential-line inequality $f(x)\geq f(x_0)+f'(x_0)(x-x_0)$, which is valid for convex functions $f(x)$ and for all $x_0\in\mathbb{R}$. Concretely, we use it for the exponential function: $\ee^{x}\geq \ee^{x_0}+\ee^{x_0}(x-x_0)$. 
We can estimate the numerator as:
\begin{align}
    -\cosh(\beta\delta\mu)-\ee^{\beta U/2}+\frac{U\beta}{2}\frac{\sinh(\beta\delta\mu)}{\beta\delta\mu} &\leq -\cosh(\beta\delta\mu)-\ee^{x_0}-\ee^{x_0}\left(\frac{\beta U}{2}-x_0\right)+\frac{U\beta}{2}\frac{\sinh(\beta\delta\mu)}{\beta\delta\mu}.
\end{align}
Without loss of generality, we choose $x_0=\ln\left(\frac{\sinh(\beta\delta\mu)}{\beta\delta\mu}\right)$ so
\begin{align}
    -\cosh(\beta\delta\mu)-\ee^{\beta U/2}+\frac{U\beta}{2}\frac{\sinh(\beta\delta\mu)}{\beta\delta\mu} \leq -\cosh(\beta\delta\mu)-\frac{\sinh(\beta\delta\mu)}{\beta\delta\mu}\left(1-\ln\frac{\sinh(\beta\delta\mu)}{\beta\delta\mu}\right)\leq -2.
\end{align}
The function $g(x)=-\cosh x -\frac{\sinh x}{x}\left(1-\ln\frac{\sinh x}{x}\right)$ has the global maximum $g(x=0)=-2$ and has negative curvature everywhere. 
Thus, $W_s^{\omega=0}$, Eq.~\eqref{eq:W_s^0}, only has the trivial root at $U=0$.

\subsection{SBE divergence for the ZPM}
\label{sec:SBE_div_ZP}

By solving Eq.~\eqref{eq:ZP_W} for $W_r=0$, we find isolated points in the phase space of $\mathrm{Re}\,\delta\mu$ and $U$ for which $W_r$ vanishes. 
For the charge channel, we find four points, symmetric in $\mathrm{Re}\,\delta\mu$: 
\begin{align}
    \mathrm{Re}\,\delta\mu = \frac{\sigma_1+2\sqrt{2}\sigma_2}{\sqrt{7}}\mathrm{Im}\,\delta\mu,\quad
    U = \frac{4(1+2\sqrt{2}\sigma_1\sigma_2)}{7}(\mathrm{Im}\,\delta\mu)^2,
    \quad \text{with~} \sigma_1, \sigma_2=\pm 1.
\end{align}
The two combinations $\sigma_1\sigma_2=1$ are at positive $U$ and the two combinations $\sigma_1\sigma_2=-1$ at negative $U$.

For the magnetic and singlet channel, we find only one point each, both located at particle-hole symmetry and negative $U$: $(\textrm{Re}\,\delta\mu,U)=(0,-\tfrac12(\mathrm{Im}\,\delta\mu)^2)$.

As already discussed in Sec.~\ref{sec:ZP_BE}, the bosonic propagators are complex valued. Therefore, in Fig.~\ref{fig:Stability_ZP}, we also show the lines in the phase space where $\mathrm{Re}\,W_r=0$ but $\mathrm{Im}\,W_r\neq0$, which originate from the singularities.

\section{Jacobian in the MBE formalism}
\label{sec:Jacobian-MBE}

The iterative map $f[\Psi]$ follows from the BE equations~\eqref{eq:SBE-vertex}--\eqref{eq:G} and contains the terms:
\begin{subequations}
    \label{eq:f[Psi]}
    \begin{align}
        f_{\Sigma}^\nu[\Sigma,W_r,\lambda_r] &= U\tfrac{1}{\beta}\sum_{\nu''}G^{\nu''}[\Sigma]+\tfrac{1}{2\beta}\sum_{\nu''}\left[ W_c^{\nu''-\nu}\lambda_c^{\nu''-\nu,\nu}+W_m^{\nu''-\nu}\lambda_m^{\nu''-\nu,\nu}\right]G^{\nu''}[\Sigma],\label{eq:f_Sigma}\displaybreak[2]\\
        f_{W_r}^\omega[\Sigma,\lambda_r] &= \frac{U_r}{1-U_r\tfrac{1}{\beta}\sum_{\nu''}\lambda_r^{\omega\nu''}\Pi_r^{\omega\nu''}[\Sigma]},\label{eq:f_W}\displaybreak[2]\\
        f_{\lambda_r}^{\omega\nu}[\Sigma,W_r,\lambda_r]&=1 + \tfrac{1}{\beta}\sum_{\nu''}\Pi_r^{\omega\nu''}[\Sigma]\,T_r^{\omega\nu''\nu}[W_r,\lambda_r,M_r],\label{eq:f_lambda}\displaybreak[2]\\
        \nonumber f_{M_r}^{\omega\nu\nu'}[\Sigma,W_r,\lambda_r,M_r]
        &=\tfrac{1}{\beta}\sum_{\nu''}T_r^{\omega\nu\nu''}[W_r,\lambda_r,M_r]\,\Pi_r^{\omega\nu''}[\Sigma]\,T_r^{\omega\nu''\nu'}[W_r,\lambda_r,M_r]\\
        &\phantom{=} -\tfrac{1}{2\beta}\sum_{\nu''}\left[M^{\omega\nu\nu''}\,\Pi_r^{\omega\nu''}[\Sigma]\,T^{\omega\nu''\nu'}_r[W_r,\lambda_r,M_r]
        +T_r^{\omega\nu\nu''}[W_r,\lambda_r,M_r]\,\Pi_r^{\omega\nu''}[\Sigma]\,M_r^{\omega\nu''\nu'}\right].\label{eq:f_M}
    \end{align}
\end{subequations}
The state $\Psi$ consists of the iterated quantities $\Sigma$, $W_r$, $\lambda_r$, and $M_r$.
The constituents of $\Psi$ are viewed as independent variables for the iteration while the remaining quantities $G$, Eq.~\eqref{eq:G}, $\Pi_r$, Eq.~\eqref{eq:Pi_r}, $P_r$, Eq.~\eqref{eq:P_r}, and $T_r$, \eqref{eq:T_r_from_MBE}, denoted as \emph{auxiliary quantities} in Fig.~\ref{fig:MBE-SOT-Scheme}, are given in terms of these quantities.
In the following, we list the entries of the Jacobian $\mathcal{J}$, Eqs.~\eqref{eq:J_GW}--\eqref{eq:Jacobian-MBE}, which follow directly from taking derivatives of the iterative map, Eq.~\eqref{eq:f[Psi]}.

We start with the derivatives of the bosonic propagator, Eq.~\eqref{eq:f_W}:
\begin{align}
    \label{eq:dW_dW}
    \frac{\delta f_{W_r}^\omega}{\delta W^{\tilde\omega}_{\tilde r}} = 0, \quad
    \frac{\delta f_{W_r}^\omega}{\delta \lambda^{\tilde\omega\tilde\nu}_{\tilde r}} =
    \delta_{r\tilde r}\delta_{\omega\tilde\omega}\tfrac{1}{\beta}(W_r^\omega)^2\Pi^{\omega\tilde{\nu}}_r, \quad
    \frac{\delta f_{W_r}^\omega}{\delta M^{\tilde\omega\tilde\nu\tilde\nu'}_{\tilde r}} = 0.
\end{align}

The derivatives of the Hedin vertices, Eq.~\eqref{eq:f_lambda},
\begin{align}
    \frac{\delta f_{\lambda_r}^{\omega\nu}}{\delta X^{\tilde\omega\tilde\nu\tilde\nu'}_{\tilde r}} &= \tfrac{1}{\beta}\sum_{\nu''}\Pi_r^{\omega\nu''}\frac{\delta T^{\omega\nu''\nu}_r}{\delta X^{\tilde\omega\tilde\nu\tilde\nu'}_{\tilde r}},
    \label{eq:dM_dM}
\end{align}
and MBE vertices, Eq.~\eqref{eq:f_M},
\begin{subequations}
    \begin{align}
        \frac{\delta f_{M_r}^{\omega\nu\nu'}}{\delta X^{\tilde\omega\tilde\nu\tilde\nu'}_{\tilde r}}&=\tfrac{1}{\beta}\sum_{\nu''}\left[\frac{\delta T^{\omega\nu\nu''}_r}{\delta X^{\tilde\omega\tilde\nu\tilde\nu'}_{\tilde r}}\Pi_r^{\omega\nu''}(T_r^{\omega\nu''\nu'}-\tfrac{1}{2}M_r^{\omega\nu''\nu'}) + (T_r^{\omega\nu\nu''}-\tfrac{1}{2}M_r^{\omega\nu''\nu'})\Pi_r^{\omega\nu''}\frac{\delta T^{\omega\nu''\nu'}_r}{\delta X^{\tilde\omega\tilde\nu\tilde\nu'}_{\tilde r}}\right] \text{ for } X\neq M,\displaybreak[2]\\
        \nonumber\frac{\delta f_{M_r}^{\omega\nu\nu'}}{\delta M^{\tilde\omega\tilde\nu\tilde\nu'}_{\tilde r}}&=\tfrac{1}{\beta}\sum_{\nu''}\left[\frac{\delta T^{\omega\nu\nu''}_r}{\delta M^{\tilde\omega\tilde\nu\tilde\nu'}_{\tilde r}}\Pi_r^{\omega\nu''}(T_r^{\omega\nu''\nu'}-\tfrac{1}{2}M_r^{\omega\nu''\nu'})+(T_r^{\omega\nu\nu''}-\tfrac{1}{2}M_r^{\omega\nu''\nu'})\Pi_r^{\omega\nu''}\frac{\delta T^{\omega\nu''\nu'}_r}{\delta M^{\tilde\omega\tilde\nu\tilde\nu'}_{\tilde r}}\right]\\
        &\phantom{=} -\tfrac{1}{2\beta}\delta_{r\tilde r}\delta_{\omega\tilde\omega}\left[\delta_{\nu\tilde\nu}\Pi^{\omega\tilde\nu'}T_r^{\omega\nu\tilde\nu}+T_r^{\omega\nu\tilde\nu}\Pi^{\omega\tilde\nu}\delta_{\nu'\tilde\nu'}\right],
    \end{align}
\end{subequations}
involve derivatives of the $U$-irreducible vertices $T_r^{\omega\nu\nu'}$, Eq.~\eqref{eq:T_r_from_MBE}, where $X^{\tilde\omega\tilde\nu\tilde\nu'}_{\tilde r}$ is a placeholder for $W^{\tilde\omega}_{\tilde r}$, $\lambda^{\tilde\omega\tilde\nu}_{\tilde r}$, $M^{\tilde\omega\tilde\nu\tilde\nu'}_{\tilde r}$. $T_r^{\omega\nu\nu'}$, on the other hand, includes SBE vertices $\nabla_r^{\omega\nu\nu'}$, Eq.~\eqref{eq:SBE-vertex}, in different channel combinations. To write the following expressions in a more compact form, we apply the frequency vectors $\bm{\nu}=(\omega,\nu,\nu')$ and $\tilde{\bm{\nu}}=(\tilde\omega,\tilde\nu,\tilde\nu')$ and their transformations $\bm{\nu}_r(\bm{\nu})$, Eq.~\eqref{eq:Channel_transformations}. The derivatives of the SBE vertices read:
\begin{subequations}
\label{eq:dnabla_dX}
\begin{align}
    \nabla_{r_2}^{\bm{\nu}_{r_1}(\bm{\nu})}
    &\equiv\nabla_{r_2}^{\omega_{r_1}(\bm{\nu}),\nu_{r_1}(\bm{\nu}),\nu'_{r_1}(\bm{\nu})} =\lambda_{r_2}^{\omega_{r_1}(\bm{\nu}),\nu_{r_1}(\bm{\nu})}W_{r_2}^{\omega_{r_1}(\bm{\nu})}\lambda_{r_2}^{\omega_{r_1}(\bm{\nu}),\nu'_{r_1}(\bm{\nu})}\\
    \Rightarrow\frac{\delta\nabla_{r_2}^{\bm{\nu}_{r_1}(\bm{\nu})}}{\delta W_{\tilde r}^{\tilde\omega}} &= \lambda_{r_2}^{\omega_{r_1}(\bm{\nu}),\nu_{r_1}(\bm{\nu})}\delta_{r_2\tilde r}\delta_{\omega_{r_1}(\bm{\nu}),\tilde\omega}\lambda_{r_2}^{\omega_{r_1}(\bm{\nu}),\nu'_{r_1}(\bm{\nu})},\displaybreak[2]\\
    \frac{\delta\nabla_{r_2}^{\bm{\nu}_{r_1}(\bm{\nu})}}{\delta \lambda_{\tilde r}^{\tilde\omega\tilde\nu}} &= \delta_{r_2\tilde r}\delta_{\omega_{r_1}(\bm{\nu}),\tilde\omega}\delta_{\nu_{r_1}(\bm{\nu}),\tilde\nu}W_{r_2}^{\omega_{r_1}(\bm{\nu})}\lambda_{r_2}^{\omega_{r_1}(\bm{\nu}),\nu'_{r_1}(\bm{\nu})} + \lambda_{r_2}^{\omega_{r_1}(\bm{\nu}),\nu_{r_1}(\bm{\nu})}W_{r_2}^{\omega_{r_1}(\bm{\nu})}\delta_{r_2\tilde r}\delta_{\omega_{r_1}(\bm{\nu}),\tilde\omega}\delta_{\nu'_{r_1}(\bm{\nu}),\tilde\nu}.
\end{align}
\end{subequations}
From this, we obtain the derivatives of the U-irreducible vertices $T_r^{\omega\nu\nu'}$, ordered in matrix form depending on the channels $r= (c,m,s,t)$:
\begin{subequations}
    \label{eq:dT_dX}
	\begin{align}
		\frac{\delta T_i^{\omega\nu\nu'}}{\delta M_j^{\tilde\omega\tilde\nu\tilde\nu'}}&=\left(\begin{matrix}
			\delta_{\bm{\nu}\tilde{\bm{\nu}}}-\tfrac{1}{2}\delta_{\bm{\nu}_\phx(\bm{\nu}),\tilde{\bm{\nu}}} & -\tfrac{3}{2}\delta_{\bm{\nu}_\phx(\bm{\nu}),\tilde{\bm{\nu}}}&
			\tfrac{1}{2}\delta_{\bm{\nu}_\pp(\bm{\nu}),\tilde{\bm{\nu}}}&
			\tfrac{3}{2}\delta_{\bm{\nu}_\pp(\bm{\nu}),\tilde{\bm{\nu}}}\\
			-\tfrac{1}{2}\delta_{\bm{\nu}_\phx(\bm{\nu}),\tilde{\bm{\nu}}}&
			\delta_{\bm{\nu}\tilde{\bm{\nu}}}+\tfrac{1}{2}\delta_{\bm{\nu}_\phx(\nu),\tilde{\bm{\nu}}} &
			-\tfrac{1}{2}\delta_{\bm{\nu}_\pp(\bm{\nu}),\tilde{\bm{\nu}}}&
			\tfrac{1}{2}\delta_{\bm{\nu}_\pp(\bm{\nu}),\tilde{\bm{\nu}}}\\
			\tfrac{1}{2}\delta_{\bm{\nu}_\phx(\bm{\nu}),\tilde{\bm{\nu}}}+\tfrac{1}{2}\delta_{\bm{\nu}_\ph(\bm{\nu}),\tilde{\bm{\nu}}}&
			-\tfrac{3}{2}\delta_{\bm{\nu}_\phx(\bm{\nu}),\tilde{\bm{\nu}}}-\tfrac{3}{2}\delta_{\bm{\nu}_\ph(\bm{\nu}),\tilde{\bm{\nu}}}&
			\delta_{\bm{\nu}\tilde{\bm{\nu}}}&
			0\\
			-\tfrac{1}{2}\delta_{\bm{\nu}_\phx(\bm{\nu}),\tilde{\bm{\nu}}}+\tfrac{1}{2}\delta_{\bm{\nu}_\ph(\bm{\nu}),\tilde{\bm{\nu}}}&
			-\tfrac{1}{2}\delta_{\bm{\nu}_\phx(\bm{\nu}),\tilde{\bm{\nu}}}+\tfrac{1}{2}\delta_{\bm{\nu}_\ph(\bm{\nu}),\tilde{\bm{\nu}}}&
			0&
			\delta_{\bm{\nu}\tilde{\bm{\nu}}}
		\end{matrix}\right),\displaybreak[2]\\
		\nonumber
        \frac{\delta T_i^{\omega\nu\nu'}}{\delta \lambda_j^{\tilde\omega\tilde\nu}}&=
		\left(\begin{matrix}
			-\tfrac{1}{2}\delta_{\bm{\nu}_\phx(\bm{\nu}),\tilde{\bm{\nu}}}W_c\lambda_c & -\tfrac{3}{2}\delta_{\bm{\nu}_\phx(\bm{\nu}),\tilde{\bm{\nu}}}W_m\lambda_m&
			\tfrac{1}{2}\delta_{\bm{\nu}_\pp(\bm{\nu}),\tilde{\bm{\nu}}}W_s\lambda_s&
			0\\
			-\tfrac{1}{2}\delta_{\bm{\nu}_\phx(\bm{\nu}),\tilde{\bm{\nu}}}W_c\lambda_c&
			\tfrac{1}{2}\delta_{\bm{\nu}_\phx(\nu),\tilde{\bm{\nu}}}W_m\lambda_m &
			-\tfrac{1}{2}\delta_{\bm{\nu}_\pp(\bm{\nu}),\tilde{\bm{\nu}}}W_s\lambda_s&
			0\\
			\tfrac{1}{2}\delta_{\bm{\nu}_\phx(\bm{\nu}),\tilde{\bm{\nu}}}W_c\lambda_c+\tfrac{1}{2}\delta_{\bm{\nu}_\ph(\bm{\nu}),\tilde{\bm{\nu}}}W_c\lambda_c&
			-\tfrac{3}{2}\delta_{\bm{\nu}_\phx(\bm{\nu}),\tilde{\bm{\nu}}}W_m\lambda_m-\tfrac{3}{2}\delta_{\bm{\nu}_\ph(\bm{\nu}),\tilde{\bm{\nu}}}W_m\lambda_m&
			0&
			0\\
			-\tfrac{1}{2}W_c\lambda_c\delta_{\bm{\nu}_\phx(\bm{\nu}),\tilde{\bm{\nu}}}+\tfrac{1}{2}W_c\lambda_c\delta_{\bm{\nu}_\ph(\bm{\nu}),\tilde{\bm{\nu}}}&
			-\tfrac{1}{2}W_m\lambda_m\delta_{\bm{\nu}_\phx(\bm{\nu}),\tilde{\bm{\nu}}}+\tfrac{1}{2}W_m\lambda_m\delta_{\bm{\nu}_\ph(\bm{\nu}),\tilde{\bm{\nu}}}&
			0&
			0
		\end{matrix}\right)
        \\
            &\phantom{=}+
		\left(\begin{matrix}
			-\tfrac{1}{2}\lambda_cW_c\delta_{\bm{\nu}_\phx(\bm{\nu}),\tilde{\bm{\nu}}} & -\tfrac{3}{2}\lambda_mW_m\delta_{\bm{\nu}_\phx(\bm{\nu}),\tilde{\bm{\nu}}}&
	        \tfrac{1}{2}\lambda_sW_s\delta_{\bm{\nu}_\pp(\bm{\nu}),\tilde{\bm{\nu}}}&
			0\\
			-\tfrac{1}{2}\lambda_cW_c\delta_{\bm{\nu}_\phx(\bm{\nu}),\tilde{\bm{\nu}}}&
			\tfrac{1}{2}\lambda_mW_m\delta_{\bm{\nu}_\phx(\nu),\tilde{\bm{\nu}}} &
			-\tfrac{1}{2}\lambda_sW_s\delta_{\bm{\nu}_\pp(\bm{\nu}),\tilde{\bm{\nu}}}&
			0\\
			\tfrac{1}{2}\lambda_cW_c\delta_{\bm{\nu}_\phx(\bm{\nu}),\tilde{\bm{\nu}}}+\tfrac{1}{2}\lambda_cW_c\delta_{\bm{\nu}_\ph(\bm{\nu}),\tilde{\bm{\nu}}}&
			-\tfrac{3}{2}\lambda_mW_m\delta_{\bm{\nu}_\phx(\bm{\nu}),\tilde{\bm{\nu}}}-\tfrac{3}{2}\lambda_mW_m\delta_{\bm{\nu}_\ph(\bm{\nu}),\tilde{\bm{\nu}}}&
			0&
			0\\
			-\tfrac{1}{2}\lambda_cW_c\delta_{\bm{\nu}_\phx(\bm{\nu}),\tilde{\bm{\nu}}}+\tfrac{1}{2}\lambda_cW_c\delta_{\bm{\nu}_\ph(\bm{\nu}),\tilde{\bm{\nu}}}&
			-\tfrac{1}{2}\lambda_mW_m\delta_{\bm{\nu}_\phx(\bm{\nu}),\tilde{\bm{\nu}}}+\tfrac{1}{2}\lambda_mW_m\delta_{\bm{\nu}_\ph(\bm{\nu}),\tilde{\bm{\nu}}}&
			0&
			0
		\end{matrix}\right),\displaybreak[2]\\
		\frac{\delta T_i^{\omega\nu\nu'}}{\delta W_j^{\tilde\omega}}&=
		\left(\begin{matrix}
			-\tfrac{1}{2}\lambda_c\delta_{\omega_\phx(\bm{\nu}),\tilde{\omega}}\lambda_c & -\tfrac{3}{2}\lambda_m\delta_{\omega_\phx(\bm{\nu}),\tilde{\omega}}\lambda_m
            & \tfrac{1}{2}\lambda_s\delta_{\omega_\pp(\bm{\nu}),\tilde{\omega}}\lambda_s
            &
			0\\
			-\tfrac{1}{2}\lambda_c\delta_{\omega_\phx(\bm{\nu}),\tilde{\omega}}\lambda_c&
			\tfrac{1}{2}\lambda_m\delta_{\omega_\phx(\bm{\nu}),\tilde{\omega}}\lambda_m
            &
			-\tfrac{1}{2}\lambda_s\delta_{\omega_\pp(\bm{\nu}),\tilde{\omega}}\lambda_s &
			0\\
			\tfrac{1}{2}\lambda_c\delta_{\omega_\phx(\bm{\nu}),\tilde{\omega}}\lambda_c+\tfrac{1}{2}\lambda_c\delta_{\omega_\ph(\bm{\nu}),\tilde{\omega}}\lambda_c
            &
			-\tfrac{3}{2}\lambda_m\delta_{\omega_\phx(\bm{\nu}),\tilde{\omega}}\lambda_m-\tfrac{3}{2}\lambda_m\delta_{\omega_\ph(\bm{\nu}),\tilde{\omega}}\lambda_m
            &
			0
            &
			0\\
			-\tfrac{1}{2}\lambda_c\delta_{\omega_\phx(\bm{\nu}),\tilde{\omega}}\lambda_c+\tfrac{1}{2}\lambda_c\delta_{\omega_\ph(\bm{\nu}),\tilde{\omega}}\lambda_c
            &
			-\tfrac{1}{2}\lambda_m\delta_{\omega_\phx(\bm{\nu}),\tilde{\omega}}\lambda_m+\tfrac{1}{2}\lambda_m\delta_{\omega_\ph(\bm{\nu}),\tilde{\omega}}\lambda_m
            &
			0
            &
			0
		\end{matrix}\right).
	\end{align}
\end{subequations}
Here, the frequency arguments of $W_r$ and $\lambda_r$ are not explicitly shown. They result from Eqs.~\eqref{eq:T_r_from_MBE} and \eqref{eq:dnabla_dX}. The compressed frequency dependencies of Kronecker delta symbols are analogous to those in Eqs.~\eqref{eq:dnabla_dX}.

The derivatives with respect to the self-energy $\Sigma_\nu$ result from the derivatives of the bubbles Eq.~\eqref{eq:Pi_r} with respect to $G$,
\begin{align}
    \frac{\delta\Pi^{\omega\nu}_{r=c,m}}{\delta G^{\tilde\nu}} = -\left[\delta_{\nu\tilde\nu}G^{\omega+\nu}+\delta_{(\omega+\nu)\tilde\nu}G^{\nu}\right],%\\
    \quad
    \frac{\delta\Pi^{\omega\nu}_{r=s,t}}{\delta G^{\tilde\nu}} = \tfrac{1}{2}\left[\delta_{(-\nu)\tilde\nu}G^{\omega+\nu}+\delta_{(\omega+\nu)\tilde\nu}G^{-\nu}\right],
\end{align}
and the derivative of the Dyson equation \eqref{eq:G},
\begin{align}
    \frac{\delta G^\nu}{\delta\Sigma^{\tilde\nu}}&=\delta_{\nu\tilde\nu}[G^{\nu}]^2.
\end{align}
For the derivative of the iterative maps with respect to the self-energy, we thus obtain
\begin{subequations}
    \begin{align}
        \frac{\delta f_{W_{r=c,m}}^\omega}{\delta \Sigma^{\tilde\nu}}&=-\tfrac{1}{\beta}\left[\lambda_r^{\omega\tilde\nu}G^{\omega+\tilde\nu}+\lambda_r^{\omega(\tilde\nu-\omega)}G^{\tilde\nu-\omega}\right](G^{\tilde\nu})^2(W^\omega_r)^2,\displaybreak[2]\\
        \frac{\delta f_{W_{r=s,t}}^\omega}{\delta \Sigma^{\tilde\nu}}&=\tfrac{1}{2}\tfrac{1}{\beta}\left[\lambda_r^{\omega(-\tilde\nu)}G^{\omega-\tilde\nu}+\lambda_r^{\omega(\tilde\nu-\omega)}G^{\omega-\tilde\nu}\right](G^{\tilde\nu})^2(W^\omega_r)^2,\displaybreak[2]\\
        \frac{\delta f_{\lambda_{r=c,m}}^{\omega\nu'}}{\delta\Sigma^{\tilde\nu}} &= -\tfrac{1}{\beta}\left[T^{\omega\tilde\nu\nu'}_r G^{\omega+\tilde\nu}+T_r^{\omega(\tilde\nu-\omega)\nu'}G^{\tilde\nu-\omega}\right](G^{\tilde\nu})^2,\displaybreak[2]\\
		\frac{\delta f_{\lambda_{r=s,t}}^{\omega\nu'}}
        {\delta\Sigma^{\tilde\nu}} &= \tfrac{1}{2\beta}\left[T_r^{\omega(-\tilde\nu)\nu'} G^{\omega-\tilde\nu}+T_r^{\omega(\tilde\nu-\omega)\nu'}G^{\omega-\tilde\nu}\right](G^{\tilde\nu})^2,\displaybreak[2]\\
        \nonumber\frac{\delta f_{M_{r=c,m}}^{\omega\nu\nu'}}{\delta\Sigma^{\tilde\nu}} &= -\tfrac{1}{2\beta}\left[(T_r^{\omega\nu\tilde\nu}-M_r^{\omega\nu\tilde\nu})G^{\omega+\tilde\nu}T_r^{\omega\tilde\nu\nu'}+(T_r^{\omega\nu(\tilde\nu-\omega)}-M_r^{\omega\nu(\tilde\nu-\omega)})G^{\tilde\nu-\omega}T_r^{\omega(\tilde\nu-\omega)\nu'}\right.\\
        &\phantom{=}\quad\quad \left. +T_r^{\omega\nu\tilde\nu}G^{\omega+\tilde\nu}(T_r^{\omega\tilde\nu\nu'}-M_r^{\omega\tilde\nu\nu'})+ T_r^{\omega\nu(\tilde\nu-\omega)}G^{\tilde\nu-\omega}(T_r^{\omega(\tilde\nu-\omega)\nu'}-M_r^{\omega(\tilde\nu-\omega)\nu'})\right](G^{\tilde\nu})^2,\displaybreak[2]\\
		\nonumber\frac{\delta f_{M_{r=s,t}}^{\omega\nu\nu'}}{\delta\Sigma^{\tilde\nu}} &= \tfrac{1}{4\beta}\left[(T_r^{\omega\nu(-\tilde\nu)}-M_r^{\omega\nu(-\tilde\nu)})G^{\omega-\tilde\nu}T_r^{\omega(-\tilde\nu)\nu'} +(T_r^{\omega\nu(\tilde\nu-\omega)}-M_r^{\omega\nu(\tilde\nu-\omega)})G^{\omega-\tilde\nu}T_r^{\omega(\tilde\nu-\omega)\nu'}\right.\\
        &\phantom{=}\quad\quad\left. +T_r^{\omega\nu(-\tilde\nu)}G^{\omega-\tilde\nu}(T_r^{\omega(-\tilde\nu)\nu'}-M_r^{\omega(-\tilde\nu)\nu'})+T_r^{\omega\nu(\tilde\nu-\omega)}G^{\omega-\tilde\nu}(T_r^{\omega(\tilde\nu-\omega)\nu'}-M_r^{\omega(\tilde\nu-\omega)\nu'})\right](G^{\tilde\nu})^2.
    \end{align}
\end{subequations}

For the self-energy, we give four different versions of Schwinger--Dyson equations~\eqref{eq:Sigma} in the main text that generate different iterative maps $f^\nu_{\tilde\Sigma}$ depending on the channels used. For Eqs.~\eqref{eq:Sigma_c}--\eqref{eq:Sigma_m}, derivatives with respect to $W_{r=c,m}$ and $\lambda_{r=c,m}$ are finite while, for Eq.~\eqref{eq:Sigma_s}, the derivative with respect to $W_s$ and $\lambda_s$ is finite:
\begin{subequations}
	\begin{alignat}{4}
		&\frac{\delta f_{\tilde\Sigma}^\nu}{\delta W_{r=c,m}^{\tilde\omega}} &&= \tfrac{1}{\beta}\lambda_{r}^{\tilde\omega,\nu}G^{\tilde\omega+\nu},\quad
        &&\frac{\delta f_{\tilde\Sigma}^\nu}{\delta\lambda_{r=c,m}^{\tilde\omega\tilde\nu}} &&= \tfrac{1}{\beta} W_{r}^{\tilde\omega}\delta_{\nu\tilde\nu}G^{\tilde\omega+\nu},\displaybreak[2]\\
		&\frac{\delta f_{\tilde\Sigma}^\nu}{\delta W_s^{\tilde\omega}} &&= -\tfrac{1}{2\beta}\lambda_s^{\tilde\omega,-\nu}G^{\tilde\omega-\nu},\quad
        &&\frac{\delta f_{\tilde\Sigma}^\nu}{\delta\lambda_s^{\tilde\omega\tilde\nu}} &&= -\tfrac{1}{2\beta}W_s^{\tilde\omega}\delta_{\nu,-\tilde\nu}G^{\tilde\omega-\nu}.
	\end{alignat}
\end{subequations}
Derivatives with respect to the MBE vertices vanish, i.e., $\delta\tilde\Sigma_\nu/\delta M^{\tilde\omega\tilde\nu\tilde\nu'}_{\tilde r}=0$. Also the derivative with respect to the self-energy depends on the channel chosen in the Schwinger--Dyson equation~\eqref{eq:Sigma}:
\begin{align}
	\frac{\delta f_{\tilde\Sigma}^\nu}{\delta\tilde\Sigma^{\tilde\nu}}&=\tfrac{1}{\beta}W_{r=c,m}^{\tilde\nu-\nu}\lambda_{r=c,m}^{\tilde\nu-\nu,\nu}G^2_{\tilde\nu}-\tfrac{1}{\beta}U_{r=c,m}(G^{\tilde\nu})^2=-\tfrac{1}{2\beta}W_s^{\tilde\nu+\nu}\lambda_s^{\tilde\nu+\nu,-\nu}G^2_{\tilde\nu}+\tfrac{1}{2\beta}U_s(G^{\tilde\nu})^2.
    \label{eq:dSigma_dSigma}
\end{align}
The derivatives of the symmetrized iterative map, Eq.~\eqref{eq:f_Sigma}, can be formed analogously.

The Jacobians for all considered diagrammatic schemes are $\kappa$-real matrices. It follows that the eigenvalues of the Jacobians are either real or come in complex conjugated pairs \cite{Hill1992}. 
We can prove this property by using the fact that all quantities that are part of the Jacobian are complex conjugated under a change of all Matsubara frequencies to their negatives \cite{Gievers2025PhD}, i.e., $(\omega,\nu,\nu^\prime)\rightarrow(-\omega,-\nu,-\nu^\prime)$, as done for the parquet equations in Ref.~\cite{Essl2026Stabilizing}.

Equations~\eqref{eq:dW_dW}--\eqref{eq:dSigma_dSigma} enable a direct evaluation of the Jacobian in terms of the already computed quantities of the algorithm, i.e., $\Psi$, which is in contrast to other methods, e.g., continuous-time quantum Monte Carlo, where it is more elaborate to determine the Jacobian.
In practice, the Jacobian can be determined from both exact physical values $\Psi_\mathrm{phys}$ and numerically computed data at self-consistency $\Psi_\mathrm{num}$. To minimize numerical costs, we evaluate $\mathcal{J}$ from the exact values once before we run the self-consistent scheme, Eq.~\eqref{eq:SOT}. This way, we can determine sufficient damping parameters $p$, Eqs.~\eqref{eq:p_max}--\eqref{eq:nonuniform_p}, before we actually run the self-consistent scheme and guarantee that the physical fixed point is always attractive and reachable. In a more realistic setting, $\mathcal{J}$ and thus the sufficient $p$ and $\mathcal{P}$ have to be computed from a numerically obtained solution at slightly different physical parameters (cf.\ procedure in App.~\ref{sec:Weak-coupling_details}).

\section{Stability conditions}
\label{sec:Stability-conditions}

\begin{figure*}
	\centering
    \includegraphics[width=\textwidth]{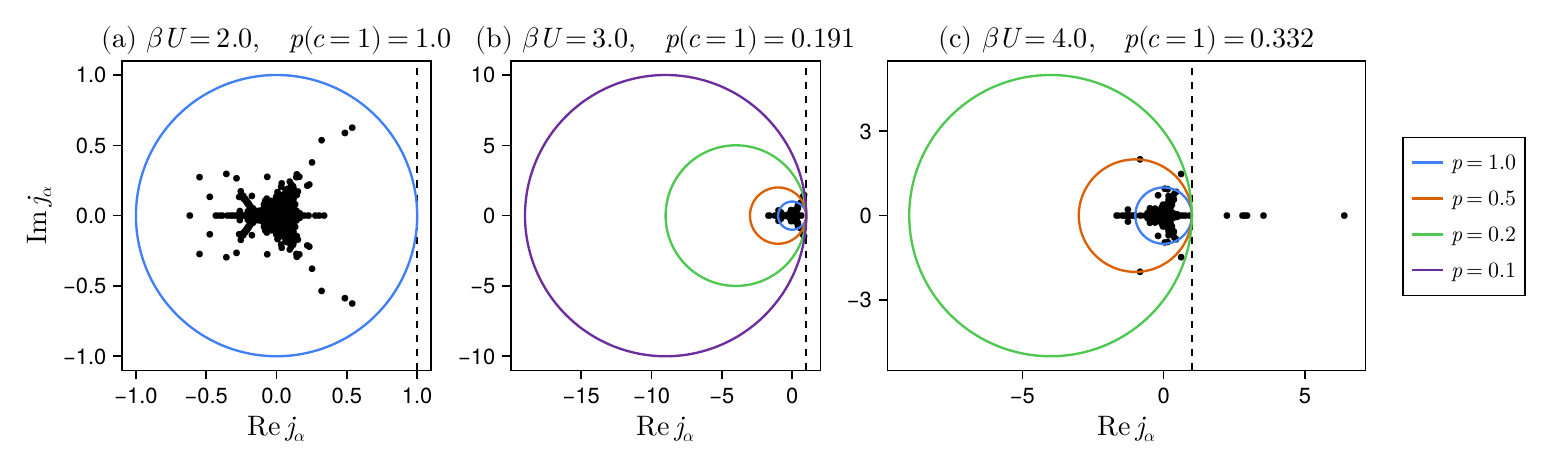}
	\caption{Eigenvalues $j_\alpha$ of the Jacobian, Eq.~\eqref{eq:Jacobian-MBE}, in the complex plane for MBE at half-filling. (a) At $\beta U=2.0$, all eigenvalues are within the circle given by Eq.~\eqref{eq:Circle-condition}. Here, mixing is not needed to stabilize the convergence. (b) At $\beta U=3.0$, mixing is needed to stabilize the convergence. The smaller $p$, the more eigenvalues are brought into the stable region. (c) Critical situation where some eigenvalues are beyond the vertical gray dashed line, i.e., $\mathrm{Re}\,j_\alpha>1$, and, thus, unstable. Here, stabilized iteration is needed to converge to the physical fixed point.}
	\label{fig:circles_evals_dmu=0.0}
\end{figure*}

In this appendix, we provide more details on the determination of the stability regions and the necessary damping parameter $p$ to guarantee local stability similar to Ref.~\cite{Essl2026Stabilizing}. Equation~\eqref{eq:j_p} defines the region where the eigenvalues $j_\alpha$ of $\mathcal{J}$ are stable:
\begin{align}
    |j^p_\alpha|^2= |p(j_\alpha-1)+1|^2<1
    \Rightarrow (\mathrm{Re}\, j_\alpha - 1 + 1/p)^2 + (\mathrm{Im}\, j_\alpha)^2<1/p^2.
    \label{eq:Circle-condition}
\end{align}
For a conventional damping parameter $p>0$ this yields a circle of radius $1/p$ around the center $z_0 = 1-1/p$, which is illustrated in Fig.~\ref{fig:circles_evals_dmu=0.0} \cite{Essl2026Stabilizing}. Here, the eigenvalues $j_\alpha$ of $\mathcal{J}_\mathrm{MBE}$, Eq.~\eqref{eq:Jacobian-MBE}, are shown for different values of $\beta U=2.0, 3.0, 4.0$ at half-filling. Indeed, for $\beta U=2.0$, all eigenvalues are within a circle of radius $p=1.0$, which means that no mixing is needed [cf.\ Fig.~\ref{fig:circles_evals_dmu=0.0}(a)]. Unstable eigenvalues with $\mathrm{Re}\, j_\alpha < 1$ can be stabilized by reducing the mixing parameter $p$, i.e., by increasing the circle defining the stable region [cf.\ Fig.~\ref{fig:circles_evals_dmu=0.0}(b)]. For eigenvalues with $\mathrm{Re}\, j_\alpha > 1$, on the contrary, $\Psi^*$ is a repulsive fixed point and the self-consistent scheme cannot be stabilized by damping. No matter how small the mixing parameter $p\in(0,1]$ is taken, the region where $\mathrm{Re}\, j_\alpha > 1$ cannot be included by the circle condition, Eq.~\eqref{eq:Circle-condition} [cf.\ Fig.~\ref{fig:circles_evals_dmu=0.0}(c)].

Multiplying Eq.~\eqref{eq:Circle-condition} with $p^2$ and using the binomial formula yields a quadratic inequality for $p$:
\begin{align}
    p^2(\mathrm{Re}\,j_\alpha-1)^2+2 p (\mathrm{Re}\,j_\alpha-1)+1+p^2(\mathrm{Im}\,j_\alpha)^2<1
    \Rightarrow p^2|j_\alpha-1|^2+2p(\mathrm{Re}\,j_\alpha-1)<0.
    \label{eq:p_intermediate}
\end{align}
The extremal damping parameters $p_1 = 0$ and $p_2=2(1-\mathrm{Re}\,j_\alpha)/|j_\alpha-1|^2 = 2\,\mathrm{Re}[1/(1-j_\alpha)]$ set a parabola with a minimum at $p_0 = \mathrm{Re}[1/(1-j_\alpha)]$. For nonuniform mixing, Eq.~\eqref{eq:nonuniform_p}, we just set individual $p_\alpha$ in each eigendirection $\alpha$ that are located between $p_1$ and $p_2$. This is the origin of the additional parameter $c\in(0,1)$. Setting $c=0.5$ in our numerical computations corresponds to $p_0$, the damping parameter that fulfills the inequality~\eqref{eq:p_intermediate} the most.

If there exists an unstable eigenvalue, i.e., $\mathrm{Re}\,j_\alpha> 1$, Eq.~\eqref{eq:p_intermediate} is not applicable since a negative $p$ would be required. In the following, we derive the expression that is valid if damped iteration, Eq.~\eqref{eq:Damped-iteration}, fails. In the stabilized method, Eq.~\eqref{eq:SOT}, the critical eigenvalue $\mathrm{Re}\,j_\alpha>1$ is stabilized by reflecting $j_\alpha$ with respect to the line at $\mathrm{Re}\,j_\alpha=1$ so the reflected value lies inside the circular stability region [cf.\ Fig.~\ref{fig:Critical_regions}(c)]. In other words, we solve the stability condition, Eq.~\eqref{eq:Circle-condition}, with an additional sign, i.e., $(j_\alpha-1)\to\sgn(1-\mathrm{Re}\,j_\alpha)(j_\alpha-1)$ [cf.\ Eq.~\eqref{eq:D_aa}]. Since $|\sgn(1-\mathrm{Re}\,j_\alpha)|^2=1$, repeating the derivation of Eq.~\eqref{eq:p_intermediate} with this substitution simply replaces $\mathrm{Re}\,j_\alpha-1$ by its absolute value, giving
\begin{align}
    |p\,\sgn(1-\mathrm{Re}\,j_\alpha)(j_\alpha-1)+1|^2<1%\\
    \Rightarrow p^2|j_\alpha-1|^2-2p|\mathrm{Re}\,j_\alpha-1|<0.
    \label{eq:Stabilized_eigenvalue_condition}
\end{align}
which, for $\mathrm{Re}\,j_\alpha\neq 1$, is equivalent to
\begin{align}
    p < \frac{2|\mathrm{Re}\,j_\alpha-1|}{|j_\alpha-1|^2} = \left\vert\mathrm{Re}\frac{2}{1-j_\alpha}\right\vert.
\end{align}
Taking the minimum over all eigendirections $\alpha$ and multiplying by $c\in(0,1)$ yields the uniform damping parameter presented in Eq.~\eqref{eq:p_max}. An upper limit of $1$ prevents overdamping.

\begin{comment}
Finally, we show how the shape of $\mathcal{P}$ is obtained from the stability condition, Eq.~\eqref{eq:Stabilized_eigenvalue_condition}. From the Jacobian associated with the stabilized method, Eq.~\eqref{eq:SOT}, we follow:
\begin{align}
    \mathcal{J}^\mathcal{P}[\Psi] &= \mathcal{P}\frac{\delta f}{\delta \Psi} + (\mathds{1}-\mathcal{P})\frac{\delta \Psi}{\delta \Psi}
    = \mathcal{U}\left[\mathcal{D}(\mathcal{U}^{-1}\mathcal{J}\mathcal{U}-\mathds{1})+\mathds{1}\right]\mathcal{U}^{-1}.
    \label{eq:Diagonalized_stabilized_Jacobian}
\end{align}
The corresponding eigenvalues $\mathcal{D}_{\alpha\alpha}(j_\alpha-1)+1$ can be read off from the diagonal matrix $\mathcal{D}(\mathcal{U}^{-1}\mathcal{J}\mathcal{U}-\mathds{1})+\mathds{1}$. By comparing to Eq.~\eqref{eq:Stabilized_eigenvalue_condition}, the elements of $\mathcal{D}$ are given as $\mathcal{D}_{\alpha\alpha}=p\,\sgn(1-\mathrm{Re}\,j_\alpha)$.
\end{comment}

\section{Numerical details}
\label{sec:Numerics}

\subsection{Frequency boxes}
\label{sec:Frequency-boxes}

The self-consistent computations are performed with an extension of the code used in Refs.~\cite{Gievers2025Subleading,Gievers2025PhD}, which is based on the Julia package \texttt{MatsubaraFunctions.jl} \cite{Kiese2024MatsubaraFunctions}. The iterative vertices are saved on finite grids of Matsubara frequencies: $\Sigma^\nu$ with $N^\nu_\Sigma$ fermionic frequencies, $W_r^\omega$ and $P_r^\omega$ with $N^\omega_W$ bosonic frequencies, $\lambda_r^{\omega\nu}$ on a $N^\omega_\lambda\times N^\nu_\lambda$ frequency box with $N^\omega_\lambda$ bosonic frequencies and $N^\nu_\lambda$ fermionic frequencies, and $M_r^{\omega\nu\nu'}$ on a $N^\omega_M\times N^\nu_M\times N^\nu_M$ frequency box with $N^\omega_M$ bosonic frequencies and $N^\nu_M$ fermionic frequencies. The total number $N$ of entries in the state $\Psi$ yields
\begin{align}
    N = 1 + N^\nu_\Sigma + 4 N^\omega_W + 4 N^\omega_\lambda N^\nu_\lambda + 4 N^\omega_M (N^\nu_M)^2.
    \label{eq:N}
\end{align}
The summand $1$ originates from the Hartree term $\Sigma_\mathrm{H}$, Eq.~\eqref{eq:Sigma_Hartree}, which is treated separately, and the factor $4$ comes from the diagrammatic channels ($c$, $m$, $s$, $t$). For the input quantities, we use $N_G^\nu$ fermionic frequencies for $G_0^\nu$ and a box of size $N^\omega_M\times N^\nu_M\times N^\nu_M$ for $\tilde\Lambda_r$ and $\Lambda_r^U$. For the auxiliary quantities, we take $N_G^\nu$ fermionic frequencies for $G^\nu$ and a box size of $N^\omega_\Pi\times N^\nu_\Pi$ for $\Pi_r^{\omega\nu}$ and $N^\omega_T\times N^\nu_T\times N^{\nu''}_T$ or $N^\omega\times N^{\nu''}_T\times N^{\nu'}_T$ for $T_r^{\omega\nu\nu'}$ (depending on whether the first or second fermionic Matsubara frequency of $T_r$ is summed over in the respective BE equation). Hereby, we can afford larger frequency grids for quantities which depend on less frequencies. The precise numbers of frequencies we use for the shown results are listed in Tab.~\ref{tab:Box-sizes}. For frequency points outside the frequency boxes we use extrapolation methods as outlined in App.~\ref{sec:Asymptotics}.

\begin{table}[]
    \centering
    \begin{tabular}{c|c|cc|cc|c|cc|ccc}
        \hline
        \hline
        $N_\Sigma^\nu$ & $N_W^\omega$ & $N_\lambda^\omega$ & $N_\lambda^\nu$ & $N_M^\omega$ & $N_M^\nu$ & $N_G^\nu$ & $N_\Pi^\omega$ & $N_\Pi^\nu$ & $N_T^\omega$ & $N_T^{\nu^{(\prime)}}$ & $N_T^{\nu''}$ \\
        $2n$ & $3n+1$ & $2n+1$ & $2n$ & $n+1$ & $n$ & $6n$ & $3n+1$ & $12n+8$ & $2n+1$ & $2n$ & $7n+4$\\
        \hline
        \hline
    \end{tabular}
    \caption{Number of points in the frequency grids used in this work where $n\in\mathbb{N}$ is an additional parameter. The results leading to the phase diagrams (Figs.~\ref{fig:Stability_regions}, \ref{fig:Critical_regions} and \ref{fig:Stability_regions_without_W_inverse}) and runs with perturbative input (Figs.~\ref{fig:evals_dmu=0.0_weak}--\ref{fig:W_r_comparison_weak} and \ref{fig:GW_SBE_SOT_weak}--\ref{fig:W_r_comparison_weak_Im}) are computed with $n=14$. All the other results are computed with $n=18$.
    For the inversion of the Bethe--Salpeter equation~\eqref{eq:BSE_inverted}, we use increased boxes for $M_r$ and $T_r$, namely $N_T^\omega\times N_T^{\nu''}\times N_T^{\nu''}$.
    }
    \label{tab:Box-sizes}
\end{table}

Empirically, we see that unstable directions in the Jacobian $\mathcal{J}^{\omega\nu\nu'}_{\tilde{\omega}\tilde{\nu}\tilde{\nu}'}$ involving only SBE vertices are primarily generated at small frequencies while derivatives involving $M_r^{\omega\nu\nu'}$ [cf.\ Eq.~\eqref{eq:dM_dM}] may not decay fast for high frequencies \cite{Essl2026Stabilizing}. Thus, we save further numerical resources by reducing the size of frequency grids inside the Jacobian $\mathcal{J}^{\omega\nu\nu'}_{\tilde{\omega}\tilde{\nu}\tilde{\nu}'}$ to the size of the frequency grids used for $M_r^{\omega\nu\nu'}$ such that the Jacobian is a matrix of size $N_\mathrm{red}\times N_\mathrm{red}$ with 
\begin{align}
    N_\mathrm{red} = 1 + N^\nu_M + 4 N^\omega_M + 4 N^\omega_M N^\nu_M + 4 N^\omega_M (N^\nu_M)^2 < N.
    \label{eq:N_red}
\end{align}
Importantly, $N_\mathrm{red}$ includes every frequency used for $M_r^{\omega\nu\nu'}$ since the number of unstable directions that originate from a vertex divergence in $M_r$ depend on the size of the grid that is used (see Ref.~\cite{Essl2026Stabilizing} for more details). We use the stabilized iteration, Eq.~\eqref{eq:SOT}, for frequencies contained in the reduced Jacobian $\mathcal{J}^{\omega\nu\nu'}_{\tilde{\omega}\tilde{\nu}\tilde{\nu}'}$ and conventional damped iteration, Eq.~\eqref{eq:Damped-iteration}, for frequencies contained in the $N$-dimensional $\Psi$ but not in the $N_\mathrm{red}\times N_\mathrm{red}$ matrix $\mathcal{J}$. This is crucial to decrease the condition numbers of $\mathcal{U}$ and $\mathcal{P}$ and at the same time increase the numerical efficiency due to the significantly smaller size of the Jacobian that has to be diagonalized.

\subsection{Asymptotic corrections}
\label{sec:Asymptotics}

To minimize finite-size effects, we include several asymptotic corrections for one- and two-dimensional vertex functions.

To include the asymptotic tails of the Green's function $G^\nu$ in the expression of the Hartree term $\Sigma_\mathrm{H}$, we determine the first four asymptotic coefficients $c_n$ of $G^\nu$,
\begin{align}
    G^\nu \approx \sum_{n=1}^4 \frac{c_n}{\nu^n},
    \label{eq:G_asymptotics}
\end{align}
using linear least-squares fitting of the corresponding high-frequency tail that is inside the box.
Here, $c_1=-\ii$ is known exactly. The Hartree term $\Sigma_\mathrm{H}$, Eq.~\eqref{eq:Sigma_Hartree}, then includes exact Matsubara summations over terms like $c_n/\nu^n$:
\begin{align}
    \Sigma_\mathrm{H} &\approx U\left[\tfrac{1}{\beta}\sum_{\nu''\in\mathrm{box}}G^{\nu''}+\frac{1}{2}+\frac{\beta c_2}{4}+\frac{\beta^3 c_4}{48}-\tfrac{1}{\beta}\sum_{\nu''\in\mathrm{box}}\sum_{n=1}^4\frac{c_n}{(\nu'')^n}\right].
\end{align}
Here, the first summand corresponds to the limited summation over numerical data, the next three summands originate from exact Matsubara summations over the asymptotic corrections, Eq.~\eqref{eq:G_asymptotics}, and the last term subtracts the limited summation over the asymptotic correction.

Similarly, in the fermionic Dyson equation~\eqref{eq:G} for $G^\nu$, $\Sigma^\nu$ is approximated as $\Sigma^\nu \approx \sum_{n=1}^4 a_n/\nu^n$ if $\nu$ is outside of the respective frequency box with fitting coefficients $a_n$.

Summed bubbles $\tfrac{1}{\beta}\sum_{\nu}\Pi_r^{\omega\nu}$, Eq.~\eqref{eq:Pi_r}, which appear in the expression of the polarization $P_r^\omega$, Eq.~\eqref{eq:P_r}, are provided with similar corrections, i.e., we include exact Matsubara summations over terms like $c_mc_n/[\nu^m(\omega+\nu)^n]$ up to $m+n\leq 4$. Explicitly, we use
\begin{subequations}
    \begin{align}
        \nonumber\tfrac{1}{\beta}\sum_{\nu}\Pi^{\omega\nu}_\ph &\approx \tfrac{1}{\beta}\sum_{\nu\in\mathrm{box}}\Pi_\ph^{\omega\nu}
        - \left(\frac{\beta(c_1)^2}{4}+\frac{\beta^3c_1c_3}{24}+\frac{\beta^3(c_2)^2}{48}\right)\delta_{\omega}
        - \left(-\frac{\beta c_1c_3}{2\omega^2}+\frac{\beta(c_2)^2}{2\omega^2}\right)\delta_{\omega\neq 0}\\
        &\phantom{\approx} + \tfrac{1}{\beta}\sum_{\nu\in\mathrm{box}}\left(\frac{(c_1)^2}{\nu(\nu+\omega)}+\frac{c_1c_2}{\nu(\nu+\omega)^2}+\frac{c_2c_1}{\nu^2(\nu+\omega)}+\frac{c_1c_3}{\nu(\nu+\omega)^3}+\frac{c_3c_1}{\nu^3(\nu+\omega)}+\frac{(c_2)^2}{\nu^2(\nu+\omega)^2}\right),\displaybreak[2]\\
        \tfrac{1}{\beta}\sum_\nu\Pi^{\omega\nu}_\pp
        &\nonumber\approx\tfrac{1}{\beta}\sum_{\nu\in\mathrm{box}}\Pi_\pp^{\omega\nu}
        +\tfrac{1}{2} \left(-\frac{\beta(c_1)^2}{4}-\frac{\beta^3c_1c_3}{24}+\frac{\beta^3(c_2)^2}{48}\right)\delta_{\omega}
        +\tfrac{1}{2} \left(\frac{\beta c_1c_2}{2\omega}+\frac{\beta c_1c_3}{2\omega^2}+\frac{\beta(c_2)^2}{2\omega^2}\right)\delta_{\omega\neq 0}\\
        &\phantom{\approx} -\tfrac{1}{2} \tfrac{1}{\beta}\sum_{\nu\in\mathrm{box}}\left(-\frac{(c_1)^2}{\nu(\nu+\omega)}-\frac{c_1c_2}{\nu(\nu+\omega)^2}+\frac{c_2c_1}{\nu^2(\nu+\omega)}-\frac{c_1c_3}{\nu(\nu+\omega)^3}-\frac{c_3c_1}{\nu^3(\nu+\omega)}+\frac{(c_2)^2}{\nu^2(\nu+\omega)^2}\right).
    \end{align}
\end{subequations}
Here, the term including $\delta_{\omega\neq 0}$ refers to all nonzero bosonic Matsubara frequencies $\omega$. 

Furthermore, in the expressions for the self-energy $\tilde\Sigma$, \eqref{eq:Sigma_symmetrized}, the polarization $P_r$, and the $U$-irreducible vertices $T_r$, \eqref{eq:T_r_from_MBE}, we use the following asymptotic corrections for the bosonic propagators $W_r$ and the Hedin vertices:
\begin{align}
    \tilde W_r^\omega \approx \sum_{n=1}^4 \frac{A_{rn}}{\omega^n},
    \quad
    \tilde\lambda_r^{\omega\nu} \approx \frac{B_{r1}}{\nu(\nu-\omega)}+\frac{B_{r2}}{\nu(\nu+\omega)}.
\end{align}
From second-order perturbation theory, we conclude for the latter $B_{r1}=U^2/4$ and $B_{r2}=0$, which is well confirmed numerically by least-square regression over the selected high-frequency values even in the strong-coupling regime.

We observe that all these asymptotic corrections become crucial in the strong-coupling regime in order to make the calculations numerically feasible.

\subsection{Determination of $\mathcal{P}$ from Schur decomposition}
\label{sec:Schur}

Since the Jacobian $\mathcal{J}$ scales quadratically with the size of the state $\Psi$, its diagonalization and the determination of $\mathcal{P}$ are the bottleneck of our method. In particular, for large sizes, the accuracy of the stabilization matrix $\mathcal{P}$, Eq.~\eqref{eq:P-matrix}, strongly influences the convergence in the self-consistent schemes. 
Here, we present two algorithms that make the computation of $\mathcal{P}$ more robust and efficient.

Generally, the Jacobian as $N_\mathrm{red}\times N_\mathrm{red}$ matrix [cf.\ Eq.~\eqref{eq:N_red}] can be brought into Schur decomposition,
\begin{align}
    \mathcal{J} = \mathcal{Z}\mathcal{T}\mathcal{Z}^\dagger,
    \label{eq:Schur_decomposition}
\end{align}
where $\mathcal{Z}$ is a unitary matrix and $\mathcal{T}$ is an upper triangular matrix with the eigenvalues on its diagonal, i.e., $\mathcal{T}_{\alpha\alpha}=j_\alpha$.

For uniform mixing $p$, Eq.~\eqref{eq:p_max}, there is an inexpensive form of $\mathcal{P}$, Eq.~\eqref{eq:P-matrix}, in terms of the Schur decomposition. The eigenvalues $j_\alpha$ are sorted such that the first $n$ elements fulfill $\mathrm{Re}\, j_\alpha <1$ and the other $N_\mathrm{red}-n$ represent unstable eigenvalues. Writing the Schur decomposition, Eq.~\eqref{eq:Schur_decomposition}, in a block structure yields
\begin{align}
    \mathcal{J} = \left( \mathcal{Z}_1, \mathcal{Z}_2\right) \left(\begin{matrix}
        \mathcal{T}_{11} & \mathcal{T}_{12} \\
        0 & \mathcal{T}_{22}
    \end{matrix}\right)\left( \begin{matrix}
        [\mathcal{Z}^\dagger]_1 \\
        [\mathcal{Z}^\dagger]_2
    \end{matrix}\right),
\end{align}
where $\mathcal{Z}_1$ represents the first $n$ columns of $\mathcal{Z}$ ($\mathcal{Z}_2$ the last $N_\mathrm{red}-n$ columns), $[\mathcal{Z}^\dagger]_1$ the first $n$ rows of $\mathcal{Z}^\dagger$ ($[\mathcal{Z}^\dagger]_2$ the last $N_\mathrm{red}-n$ rows) and $\mathcal{T}_{11}$ is an $n\times n$ matrix including the stable eigenvalues $j_\alpha$ ($\mathcal{T}_{22}$ an $(N_\mathrm{red}-n)\times(N_\mathrm{red}-n)$ matrix including the unstable eigenvalues). $\mathcal{T}$ can be brought into block-diagonal structure by introducing $\mathcal{Q} = \left(\begin{matrix}
        \mathds{1} & \mathcal{X} \\ 0 & \mathds{1}
    \end{matrix}\right)$ with $\mathcal{X}$ fulfilling the Sylvester equation $\mathcal{T}_{11}\mathcal{X}-\mathcal{X}\mathcal{T}_{22}=-T_{12}$ such that
\begin{align}
    \mathcal{T} = \left(\begin{matrix}
        \mathcal{T}_{11} & \mathcal{T}_{12} \\
        0 & \mathcal{T}_{22}
    \end{matrix}\right) = \left(\begin{matrix}
        \mathds{1} & \mathcal{X} \\ 0 & \mathds{1}
    \end{matrix}\right) \left(\begin{matrix}
        \mathcal{T}_{11} & 0 \\ 0 & \mathcal{T}_{22}
    \end{matrix}\right) \left(\begin{matrix}
        \mathds{1} & -\mathcal{X} \\ 0 & \mathds{1}
    \end{matrix}\right) = \mathcal{Q}  \left(\begin{matrix}
        \mathcal{T}_{11} & 0 \\ 0 & \mathcal{T}_{22}
    \end{matrix}\right) \mathcal{Q}^{-1} \Rightarrow \mathcal{J} = \mathcal{ZQ}\left(\begin{matrix}
        \mathcal{T}_{11} & 0 \\ 0 & \mathcal{T}_{22}
    \end{matrix}\right)\mathcal{Q}^{-1}\mathcal{Z}^\dagger.
\end{align}
The $\mathcal{P}$ matrix then yields
\begin{align}
    \nonumber\mathcal{P} &= p \mathcal{ZQ} \left(\begin{matrix}
        \mathds{1}_{n\times n} & 0 \\
        0 & -\mathds{1}_{(N_\mathrm{red}-n)\times(N_\mathrm{red}-n)}
    \end{matrix}\right) \mathcal{Q}^{-1}\mathcal{Z}^\dagger
    = p \left(\mathcal{Z}_1,\mathcal{Z}_1\mathcal{X}+\mathcal{Z}_2\right)\left(\begin{matrix}
        \mathds{1}_{n\times n} & 0 \\
        0 & -\mathds{1}_{(N_\mathrm{red}-n)\times(N_\mathrm{red}-n)}
    \end{matrix}\right) \left(\begin{matrix}
        [\mathcal{Z}^\dagger]_1-\mathcal{X}[\mathcal{Z}^\dagger]_2 \\ [\mathcal{Z}^\dagger]_2
    \end{matrix}\right)\\
    &= p \left( \mathcal{Z}_1[\mathcal{Z}^\dagger]_1 - 2 \mathcal{Z}_1 \mathcal{X} [\mathcal{Z}^\dagger]_2 - \mathcal{Z}_2[\mathcal{Z}^\dagger]_2\right).
    \label{eq:P_uniform_Schur}
\end{align}
In this form, $\mathcal{P}$ commutes with the Jacobian $\mathcal{J}$ so the matrices share the same eigenvectors. Numerically, solving the Sylvester equation for $\mathcal{X}$ involving the upper triangular matrices $\mathcal{T}_{11}$ and $\mathcal{T}_{22}$ is much more beneficial than the inversion of the whole transformation matrix $\mathcal{U}$. Further, Eq.~\eqref{eq:P_uniform_Schur} yields a much lower condition number for  $\mathcal{P}$ than a direct computation of  $\mathcal{P}$ by $\mathcal{U}\mathcal{D}\mathcal{U}^{-1}$.

A straightforward modification of Eq.~\eqref{eq:P_uniform_Schur} for nonuniform mixing, Eq.~\eqref{eq:nonuniform_p}, stabilizes the fixed point locally, but $[\mathcal{P},\mathcal{J}]=0$ would be violated. Empirically, we noticed that the region in the space of solutions $\Psi$ is more restricted when $[\mathcal{P},\mathcal{J}]\neq 0$. We suppose that these restrictions originate from the fact that the eigendirections would not be completely flipped in a rigorous way. Therefore, we continue with a more involved strategy to set $\mathcal{P}$ with nonuniform mixing:
\begin{align}
    \mathcal{P}= \mathcal{Z}\mathcal{M}\mathcal{Z}^\dagger \quad\text{with}\quad \mathcal{M}_{\alpha\alpha} = p_\alpha = c\,\mathrm{Re}\,\frac{2}{1-j_\alpha} \quad\text{and}\quad \mathcal{M}_{\alpha>\beta} = 0.
\end{align}
The upper triangular elements $\mathcal{M}_{\alpha<\beta}$ are determined from the condition $[\mathcal{M},\mathcal{T}]=0$, which directly follows from $[\mathcal{P},\mathcal{J}]=0$. With $\mathcal{T}_{\alpha\alpha}=j_\alpha$, we conclude:
\begin{align}
    \nonumber 0 &= \sum_\gamma\left(\mathcal{M}_{\alpha\gamma}\mathcal{T}_{\gamma\beta}-\mathcal{T}_{\alpha\gamma}\mathcal{M}_{\gamma\beta}\right) = \left(p_\alpha-p_\beta\right)\mathcal{T}_{\alpha\beta}+(j_\beta-j_\alpha)\mathcal{M}_{\alpha\beta}+\sum_{\alpha<\gamma<\beta}\left(\mathcal{M}_{\alpha\gamma}\mathcal{T}_{\gamma\beta}-\mathcal{T}_{\alpha\gamma}\mathcal{M}_{\gamma\beta}\right)\\
    \Rightarrow\mathcal{M}_{\alpha\beta} &= \frac{p_\alpha-p_\beta}{j_\alpha-j_\beta}\mathcal{T}_{\alpha\beta}+\sum_{\alpha<\gamma<\beta}\frac{\mathcal{M}_{\alpha\gamma}\mathcal{T}_{\gamma\beta}-\mathcal{T}_{\alpha\gamma}\mathcal{M}_{\gamma\beta}}{j_\alpha-j_\beta}.
    \label{eq:M_alphabeta}
\end{align}
Here, the last sum only runs over restricted values of $\gamma$. Solving for $\mathcal{M}_{\alpha\beta}$ is done numerically. To avoid instabilities near degenerate eigenvalues $j_\alpha\simeq j_\beta$, we exploit the following identity:
\begin{align}
    \frac{p_\alpha-p_\beta}{j_\beta-j_\alpha} = \frac{2c}{(1-j_\alpha)(1-j_\beta)}+\ee^{-2\ii \phi_{\alpha\beta}}\frac{2c}{(1-j^*_\alpha)(1-j^*_\beta)},
\end{align} 
where $\phi_{\alpha\beta}=\mathrm{arg}(j_\alpha-j_\beta)$ is the phase of $j_\alpha-j_\beta$.
The stabilization matrix $\mathcal{P}$ is defined in such a way that it flips the sign of unstable directions in the Jacobian.

The Schur decomposition, Eq.~\eqref{eq:Schur_decomposition}, offers another representation of the Jacobian where $\mathcal{Z}$ transforms to the upper-triangular form of the Jacobian, the matrix $\mathcal{T}$ that contains the eigenvalues $j_\alpha$ on its diagonal. A numerically cheaper alternative to $\mathcal{P}=\mathcal{UDU}^{-1}$ is the expression $\mathcal{ZDZ}^\dagger$, which locally stabilizes the fixed point but does not commute with $\mathcal{J}$.

\subsection{Details on self-consistent schemes}
\label{sec:Weak-coupling_details}

For the runs in Sec.~\ref{sec:Stabilized_iteration_strong} and App.~\ref{sec:Strong_results_appendix} with exact input of the vertices being not iterated (cf.\ Fig.~\ref{fig:MBE-SOT-Scheme}), the self-consistent solutions at $U$ and $\delta\mu$ are obtained starting from the exact state at $U-\delta U$ with $\beta\delta U = 0.01$ and the same $\delta\mu$ for the iterated quantities, i.e., $\Psi_i=\Psi_\mathrm{phys}(U-\delta U,\delta\mu)$. The corresponding Jacobian is directly computed from the exact solution $\mathcal{J}[\Psi_\mathrm{phys}(U,\delta\mu)]$, from which we determine the stabilization matrix $\mathcal{P}$, Eq.~\eqref{eq:P-matrix} [cf.\ App.~\ref{sec:Schur}]. For MBE, we use uniform mixing $p(c)$, Eq.~\eqref{eq:p_max}, whereas, for SBE and TRILEX, we use nonuniform mixing $p_\alpha(c)$, Eq.~\eqref{eq:nonuniform_p}. Hereby, we set the additional parameter to $c=0.5$. Furthermore, we impose a minimal number of self-consistent iterations $n_\mathrm{min} = \min\left(15000,\lceil [\epsilon\,p(c=1)]^{-1}\rceil\right)$ depending on the necessary mixing parameter $p(c=1)$, Eq.~\eqref{eq:p_max}, and $\epsilon=\min_\alpha\left\vert 1-\mathrm{Re}\,j_\alpha\right\vert$, the minimal distance to a critical eigenvalue with $\mathrm{Re}\,j_\alpha=1$ [cf.\ Fig.~\ref{fig:Critical_regions}]. This way, we ensure that the iterative solution is driven far enough from the starting point. Successful convergence is gained when the relative difference of all iterated quantities in $\Psi_n\to\Psi_{n+1}$ is smaller than $10^{-4}$.  

Generally, there are two main reasons why convergence can become difficult. First, some eigenvalues can come very close to $\mathrm{Re}\, j_\alpha=1$, which intuitively corresponds to a flattened direction in the abstract space of solutions $\Psi$ and thus decelerates the convergence. To keep control of the computational expense, we set the upper limit for the self-consistent iterations to $n_\mathrm{max}=30000$, where some solutions are not converged yet.
Second, at vertex divergences, the divergent vertices $\lambda_c$ and $M_r$ are directly contained in the Jacobian $\mathcal{J}$ (cf.\ App.~\ref{sec:Jacobian-MBE}) and, thus, increase the eigenvalues significantly. Large eigenvalues $|j_\alpha|\gg 1$ again imply a very strong curvature in the abstract space of solutions $\Psi$ and therefore require a huge damping, i.e., small $p$, which again decelerates the convergence. The Jacobian evaluated at the fixed point $\Psi^*$ is only a local quantity. If the initial state $\Psi_i$ in the self-consistent scheme is not close enough to $\Psi^*$, it may not be in the basin of attraction of $\Psi^*$ and, therefore, the iteration may not converge to $\Psi^*$ even though it is locally stable.

\begin{figure*}
	\centering
    \includegraphics[width=0.85\textwidth]{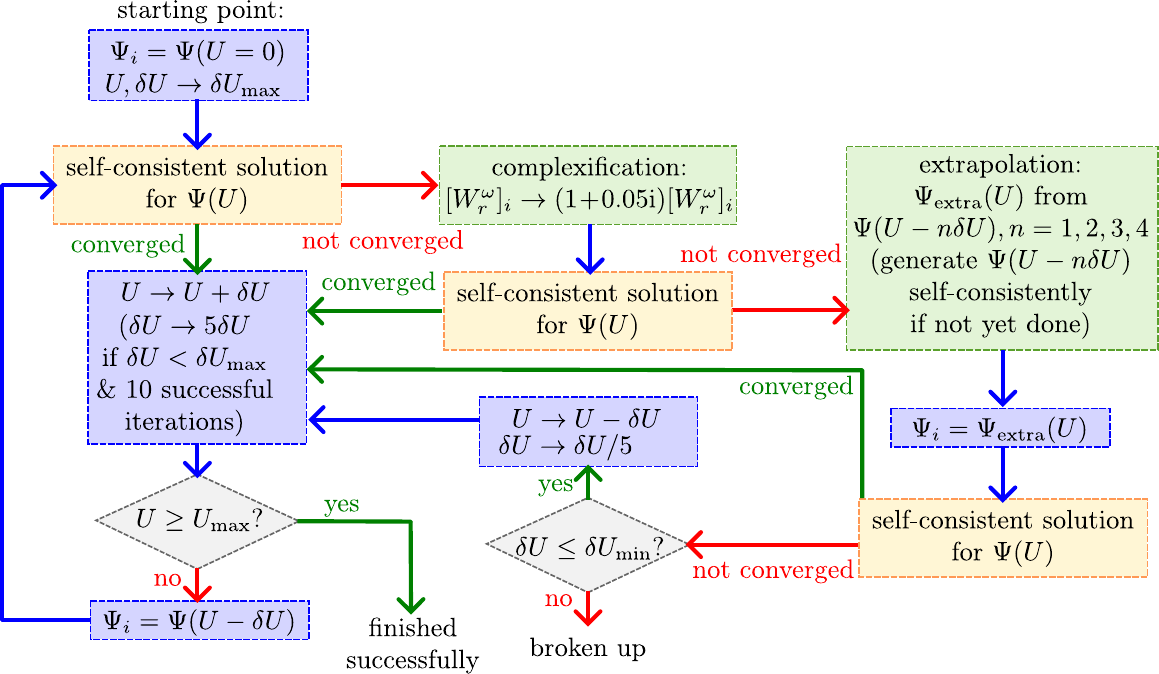}
	\caption{Schematic overview on the algorithm producing self-consistent solutions with perturbative input. Self-consistent solutions are performed with increasing values of $U$. Depending on the success of convergence, the initial state $\Psi_i$ is constructed from previous converged results, states with complexified bosonic propagators $W_r$, or extrapolated states. Each self-consistent loop (yellow boxes) consists of two consecutive runs: one with $\mathcal{J}[\Psi_i,U]$ and the second one with $\mathcal{J}[\Psi(U),U]$ where $\Psi(U)$ is the converged result from the first run.
    }
	\label{fig:Perturbative_algorithm}
\end{figure*}

\begin{comment}
\MGC{We need to evaluate if we want to show the stability regions of the perturbative schemes.} For a more general picture, we show the maximal mixing parameters $p_\mathrm{max}(U,\delta\mu)$ and unstable regions from the Jacobians of the weak approaches in Fig.~\ref{fig:Stability_regions_weak} similar to Fig.~\ref{fig:Stability_regions} for the nonperturbative approaches. Overall, the unstable regions are smaller in the weak approaches. For GW [cf.\ Fig.~\ref{fig:Stability_regions_weak}(a)], we cannot even find an unstable region. As already seen from the eigenvalues [cf.\ Fig.~\ref{fig:evals_dmu=0.0_weak}], weak MBE [cf.\ Fig.~\ref{fig:Stability_regions_weak}(c)] provides significant changes compared to its nonperturbative version [cf.\ Fig.~\ref{fig:Stability_regions}(c)] since vertex divergences are absent and thus cannot generate instabilities. Furthermore, we observe a similar stable region at large imbalances, i.e., $\delta\mu \gtrsim 3/4\, U$ since here the weak-coupling approach is more and more satisfied.
\end{comment}

For the runs in Sec.~\ref{sec:Weak-coupling_results} and App.~\ref{sec:Weak_coupling_appendix} with perturbative input of the vertices being not iterated (cf.\ Fig.~\ref{fig:MBE-SOT-Scheme}), consecutive self-consistent solutions for increasing $U$ are required, which involves a more sophisticated algorithm, illustrated in Fig.~\ref{fig:Perturbative_algorithm}. Note that for practical calculations where the physical fixed point is not known {\sl a priori}, a similar algorithm is needed. 

We track the self-consistent solution $\Psi(U)$ as the interaction strength $U$ is swept from the noninteracting point $U=0$ upward to a maximal value $U_\mathrm{max}$. The self-consistent solution (yellow boxes in Fig.~\ref{fig:Perturbative_algorithm}) involves two successions using stabilized iteration: a first self-consistent loop starts with $\Psi_i$ and the associated Jacobian $\mathcal{J}[\Psi_i(U)]$ and a second pass re-evaluates the Jacobian at the resulting (more accurate) solution $\mathcal{J}[\Psi_\mathrm{converged}(U)]$ and re-converges from there. After successful convergence, $U$ is increased by $\delta U$ and the next self-consistent computation starts with the converged result from the previous step, i.e., $\Psi_i=\Psi(U-\delta U)$.

When the self-consistent solution fails, we use three strategies to overcome this failure. The restriction of the perturbative approximation can lead to unphysical results, even when following the path of the physical fixed point in parameter space, i.e., there might exist no physical fixed point anymore. 
In particular, the exact solution of the HA only includes real bosonic propagators $W_c$ and $W_m$, Eq.~\eqref{eq:HA_W_r}. To promote convergence and extend the space of solutions, in a first trial, we add a finite imaginary part to the initial $W_r$ (green box in the center of Fig.~\ref{fig:Perturbative_algorithm}). After another failure, we extrapolate the initial value from extrapolation of previous states (right green box in Fig.~\ref{fig:Perturbative_algorithm}). If that fails again, our third strategy is to decrease the step size $\delta U$ (blue box in the center). Only if $\delta U$ goes below the minimal threshold $\delta U_\mathrm{min}=10^{-4}$, the algorithm is broken up unsuccessfully.

The runs with perturbative input are performed with a minimal and maximal number of self-consistent iterations being $n_\mathrm{min}=\mathrm{min}(15000,\lceil [\epsilon p(c=1)]^{-1}\rceil)$ and $n_\mathrm{max}=240000$. We use uniform mixing $p_\alpha$, Eq.~\eqref{eq:p_max}, for all diagrammatic approaches.

\section{Additional results}

\subsection{Stability regions}
\label{sec:app_stability_regions}

\begin{figure*}
	\centering
    \includegraphics[width=\textwidth]{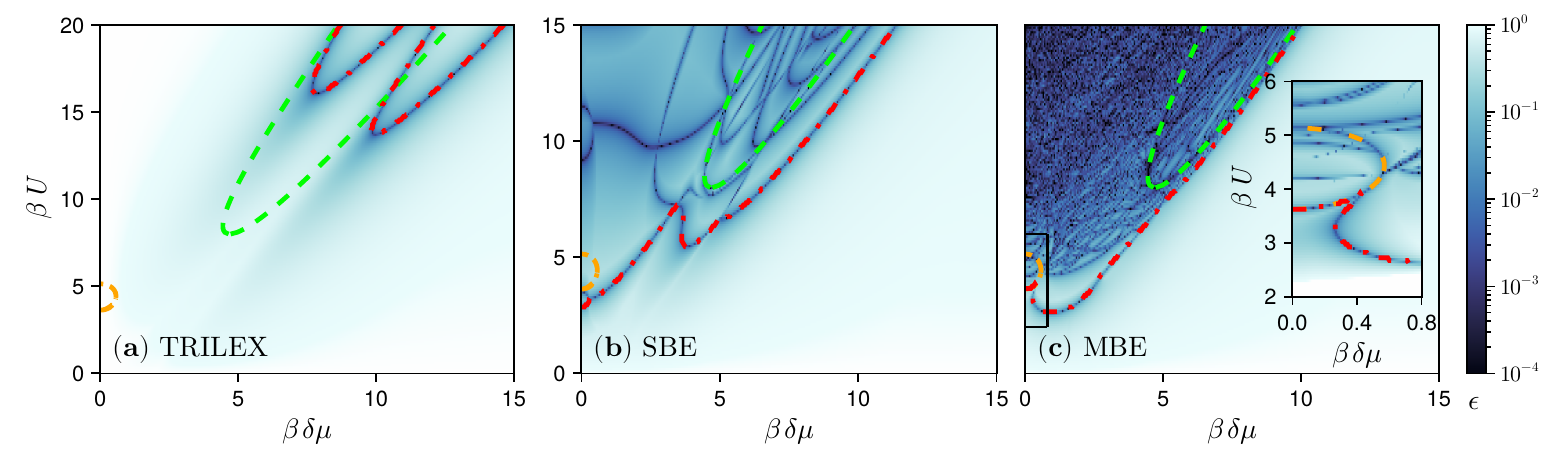}
	\caption{Minimal distance of an eigenvalue from criticality, i.e., $\epsilon = \min_\alpha |1-\mathrm{Re}\,j_\alpha|$, for the respective diagrammatic schemes (a) TRILEX, (b) SBE, (c) MBE (cf.\ Fig.~\ref{fig:MBE-SOT-Scheme}). The red dashed lines mark where the first crossing  $\mathrm{Re}\,j_\alpha = 1$ occurs, i.e., in the regions above/within the red lines, the conventional damped iteration does not converge to the physical solution. The orange and green dashed lines mark the first 2PI divergence and the SBE divergence. The inset in panel (c) magnifies the region around the first vertex divergence.
    }
	\label{fig:Critical_regions}
\end{figure*}

Another parameter to examine the convergence of self-consistent schemes is $\epsilon=\min_\alpha |1-\mathrm{Re}\,j_\alpha|$, the minimal distance of an eigenvalue to the critical values $\mathrm{Re}\,j_\alpha=1$. Intuitively, the parameter $\epsilon$ measures how ``flat'' the direction that is closest to being unstable is. Therefore, convergence is extremely slow when $\epsilon$ is small since the Jacobian hardly drives the solution to the fixed point. In Fig.~\ref{fig:Critical_regions}, we plot $\epsilon$ similar to $p(c=1)$ in Fig.~\ref{fig:Stability_regions}. The parameter $\epsilon$ is insightful as it points out problematic regions, where some eigenvalues are extremely flat around $\mathrm{Re}\,j_\alpha\simeq 1$, e.g., for SBE at half-filling near $\beta U \simeq 9.3$ [cf.\ Figs.~\ref{fig:SBE_SOT_dmu=0.0}(d) and \ref{fig:Critical_regions}(b)], which is not detected by $p(c=1)$ [cf.\ \ref{fig:Stability_regions}(b)].

\begin{figure*}
	\centering
    \includegraphics[width=\linewidth]{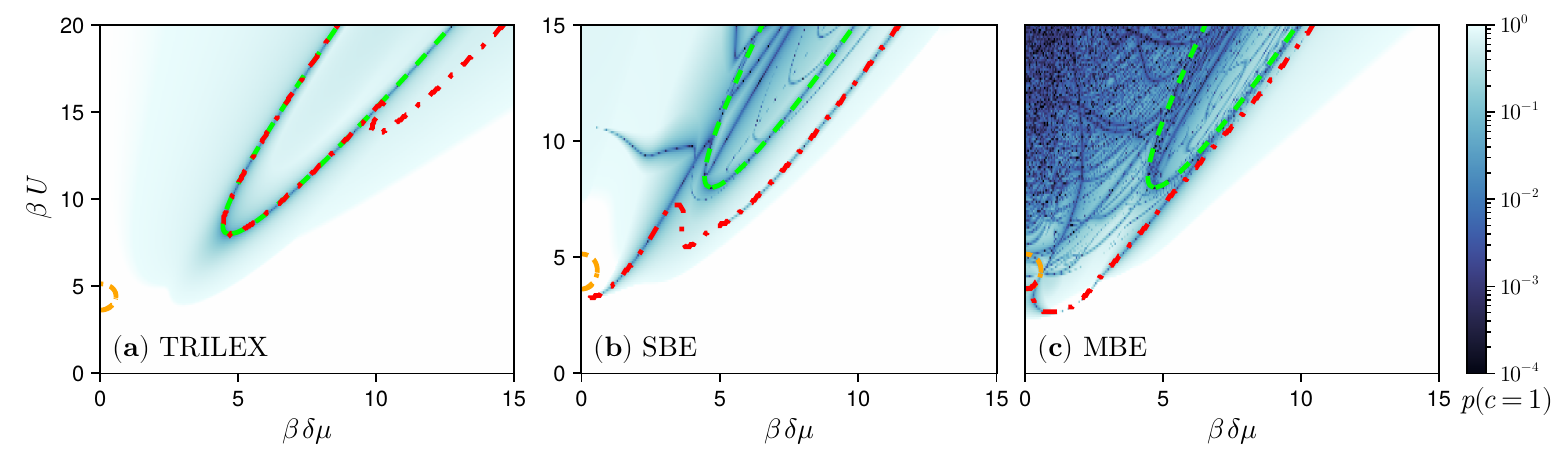}
	\caption{Similar plots as in Fig.~\ref{fig:Stability_regions}. Only for the bosonic propagators $W^\omega_r=U_r+U_rP^\omega_rW_r$ is used instead of $W^\omega_r=(U_r^{-1}-P^\omega_r)^{-1}$. The unstable regions marked by dashed red lines are larger compared to the ones shown in the main text.
    }
    \label{fig:Stability_regions_without_W_inverse}
\end{figure*}

The iterative maps in Eq.~\eqref{eq:f[Psi]} can be changed by keeping the physical solution as a fixed point of the iterative system. 
As an illustrative example, we can use the recursive form of the bosonic Dyson equation,
\begin{align}
    f_{W_r}^\omega[\Sigma,W_r,\lambda_r]= U_r + U_r P^\omega_r W^\omega_r,
    \label{eq:W_r_recursive}
\end{align}
which is often used instead of the inverted form in Eq.~\eqref{eq:f_W}. This changes the Jacobian block of $f_{W_r}^\omega$ to
\begin{align}
    \frac{\delta f_{W_r}^\omega}{\delta W^{\tilde\omega}_{\tilde r}} = \delta_{r\tilde r}\delta_{\omega\tilde\omega}U_r P_r^\omega,\quad
    \frac{\delta f_{W_r}^\omega}{\delta \lambda^{\tilde\omega\tilde\nu}_{\tilde r}} =\delta_{r\tilde r}\delta_{\omega\tilde\omega}U_r\tfrac{1}{\beta}\Pi^{\omega\tilde{\nu}}_rW^\omega_r,\quad
    \frac{\delta f_{W_r}^\omega}{\delta M^{\tilde\omega\tilde\nu\tilde\nu'}_{\tilde r}} = 0.
    \label{eq:dW_dW_recursive}
\end{align}

Figure~\ref{fig:Stability_regions_without_W_inverse} shows the stability region of the HA analogous to Fig.~\ref{fig:Stability_regions}, when the recursive bosonic Dyson equation~\eqref{eq:W_r_recursive} is used.
We find that the stability region changes significantly for the TRILEX scheme, where a larger unstable regions is noticeable compared to that in Fig.~\ref{fig:Stability_regions}(a). As $\lambda_c$ appears linearly in the recursive equation~\eqref{eq:W_r_recursive} through $P_r$, Eq.~\eqref{eq:P_r}, the SBE divergence seems to have a more direct influence on the stability region.
We find a similar behavior for the ZPM. 
At the same time, for SBE and MBE [cf.\ Fig.~\ref{fig:Stability_regions_without_W_inverse}(b)--(c)], we do not find a qualitative change in the stability regions compared to before. 

\subsection{Channels in the Schwinger--Dyson equation}
\label{sec:change_it_maps}

Another possibility of changing the iterative maps without changing the fixed point are the different versions of the Schwinger--Dyson equation~\eqref{eq:Sigma}. Here, we explicitly show that the symmetrized version with respect to the $c$ and $m$ channel, Eq.~\eqref{eq:Sigma_symmetrized}, turns out to be the most efficient numerically. From the exact formulas of $\tilde W_r=W_r-U_r$, Eq.~\eqref{eq:HA_W_r}, and $\tilde \lambda_r=\lambda_r-1$, Eq.~\eqref{eq:HA_lambda}, for the HA, we deduce:
\begin{subequations}
\begin{align}
    W_c^\omega\lambda_c^{\omega\nu}+W_m^\omega\lambda_m^{\omega\nu}&= (\tilde W_c^\omega+\tilde W_m^\omega)\left[1+U\frac{1-2n}{y(\nu)+Un}-U^2\frac{(1-n)n}{(y(\nu)+Un)(y(\nu+\omega)+Un)}\right],\displaybreak[2]\\
    \text{where~}\tilde W_c^\omega+\tilde W_m^\omega&=U^2\beta\delta_\omega 2n(1-n).
\end{align}
\end{subequations}
Consequently, the whole summand in Eq.~\eqref{eq:Sigma_symmetrized} is proportional to $\delta_\omega$ and the sum is trivial.

\subsection{Strong-coupling results at $\beta\delta\mu=1.0$}
\label{sec:Strong_results_appendix}

\begin{figure*}
	\centering
    \includegraphics[width=\textwidth]{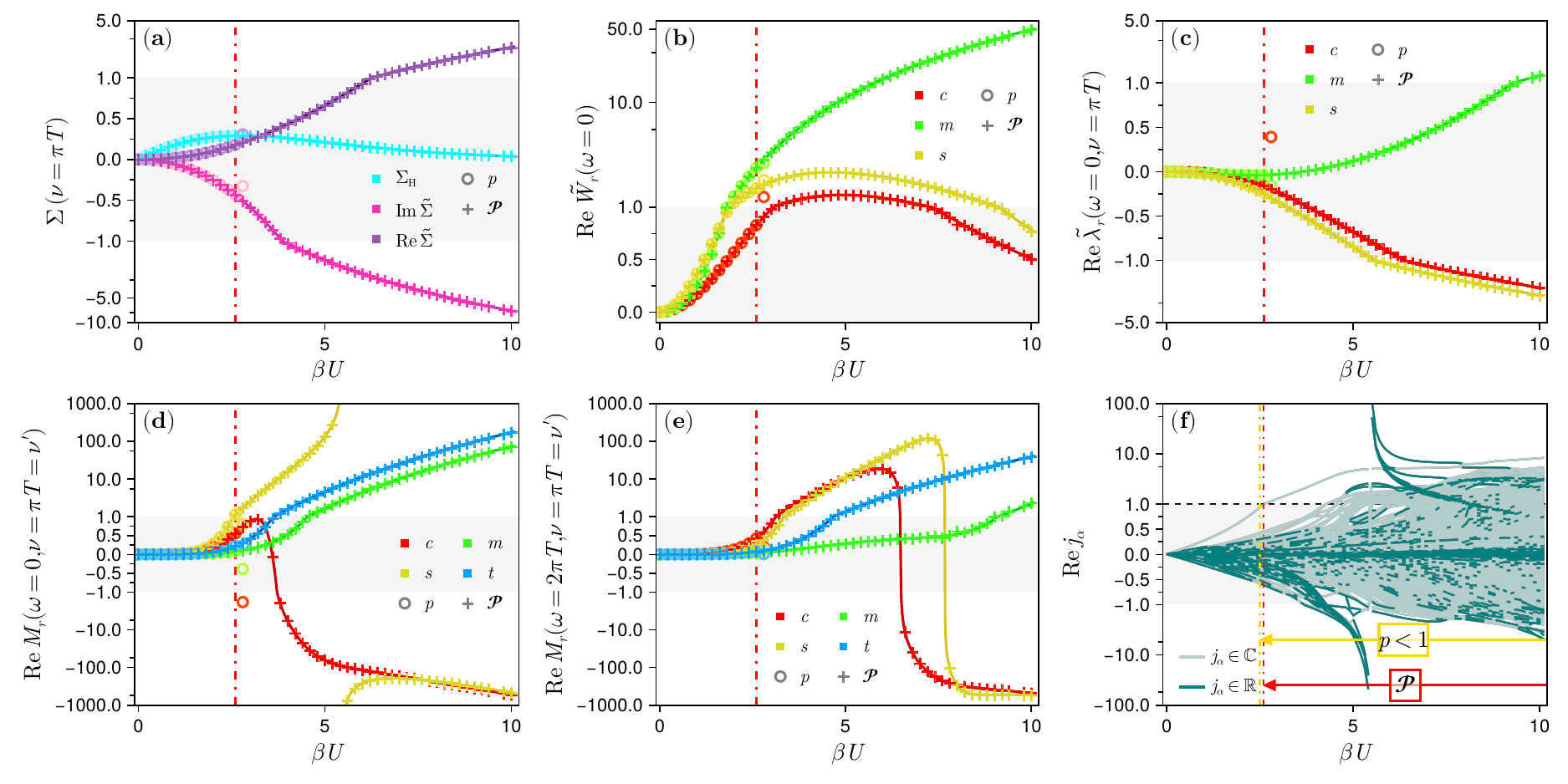}
	\caption{Converged quantities at specific Matsubara frequencies in dependence on $U$ from MBE at half-filling. Details about colors and markers are provided in Fig.~\ref{fig:GW_SOT_dmu=12.0}(d). (a) Self-energies $\Sigma$ at $\nu=\pi T$, (b) bosonic propagators $\tilde W_r$ at $\omega=0$, (c) Hedin vertices $\tilde\lambda_r$ at $\omega=0$ and $\nu=\pi T$, (d) MBE vertices $M_r$ at $\omega=0$ and $\nu=\pi T=\nu'$, (e) MBE vertices $M_r$ at $\omega=2\pi T$ and $\nu=\pi T=\nu'$, (f) real parts of the eigenvalues $j_\alpha$.
    }
	\label{fig:MBE_SOT_dmu=1.0}
\end{figure*}

Figure~\ref{fig:MBE_SOT_dmu=1.0} is analogous to Fig.~\ref{fig:MBE_SOT_dmu=0.0} in the main text, only the data are obtained out of half-filling with $\beta\delta\mu=1.0$. 
The significant change is that the first instability at $\beta U\simeq 2.6$ (red dashed line) occurs clearly before the first vertex divergence in the $s$ channel [yellow line in Fig.~\ref{fig:MBE_SOT_dmu=1.0}(d)], which was anticipated from Figs.~\ref{fig:Vertex_divergences}(a)--(b) and \ref{fig:Stability_regions}(c).

\begin{comment}
\begin{figure*}
	\centering
	\includegraphics[width=0.32\textwidth]{260202_HA_evals_U_N_freqs_N_p_0.5_SDE_GW_R_prev_etainv_deltamu_1.0.png}
    \hfill
	\includegraphics[width=0.32\textwidth]{260202_HA_evals_U_N_freqs_N_p_0.5_SDE_R_prev_etainv_deltamu_1.0.png}
	\hfill
    \includegraphics[width=0.32\textwidth]{260202_HA_evals_U_N_freqs_N_p_0.5_SDE_MBE_R_prev_etainv_deltamu_1.0.png}
	\caption{Real parts of the eigenvalues $\mathrm{Re}\,j_\alpha$ of the Jacobian, Eqs.~\eqref{eq:J_GW}--\eqref{eq:Jacobian-MBE}, for the different \textit{strong} diagrammatic schemes (a) GW$^3$, (b) SBE, (c) MBE (cf.\ Fig.~\ref{fig:MBE-SOT-Scheme}). The cuts are generated in dependence of $U$ at $\beta\delta\mu=1.0$.}
	\label{fig:evals_dmu=1.0}
\end{figure*}
\end{comment}

\begin{comment}
\begin{figure}
	\centering
    \includegraphics[width=0.48\textwidth]{260720_HA_SU2_U_N_freqs_9.0_p_0.5_SDE_R_sot_prev_etainv_deltamu_1.0.pdf}
	\caption{Results for the converged vertex functions from SBE at $\delta \mu = 1.0$ at specific Matsubara frequencies in dependence on $U$. The colors and markers are defined in Fig.~\ref{fig:MBE_SOT_dmu=1.0}(f). (a) One-particle functions at $\nu=\pi T$, (b) bosonic propagators $\tilde W_r$ at $\omega=0$, (c) Hedin vertices $\tilde\lambda_r$ at $\omega=0$ and $\nu=\pi T$, (d)--(f) same for the imaginary parts of the vertices and real parts of the self-energy and Green's function.}
	\label{fig:SBE_SOT_dmu=1.0}
\end{figure}
\end{comment}

\subsection{Additional results for the ZPM}
\label{sec:ZP_additional}

In this section, we further investigate the apparent similarity between the HA and the ZPM. For this we show the eigenvalues of the Jacobian for the three diagrammatic approaches (TRILEX, SBE, MBE), Eqs.~\eqref{eq:J_GW}--\eqref{eq:Jacobian-MBE}, evaluated at the exact physical fixed point of the ZPM, and compare them to those of the HA.  
Due to the small size of the Jacobian for the ZPM ($5\times5$ for TRILEX, $9\times9$ for SBE, and $13\times13$ for MBE), this model represents a good opportunity to investigate further the kinds of instabilities that occur in the different approaches. 

\begin{figure*}
    \centering
    \includegraphics[width=1\linewidth]{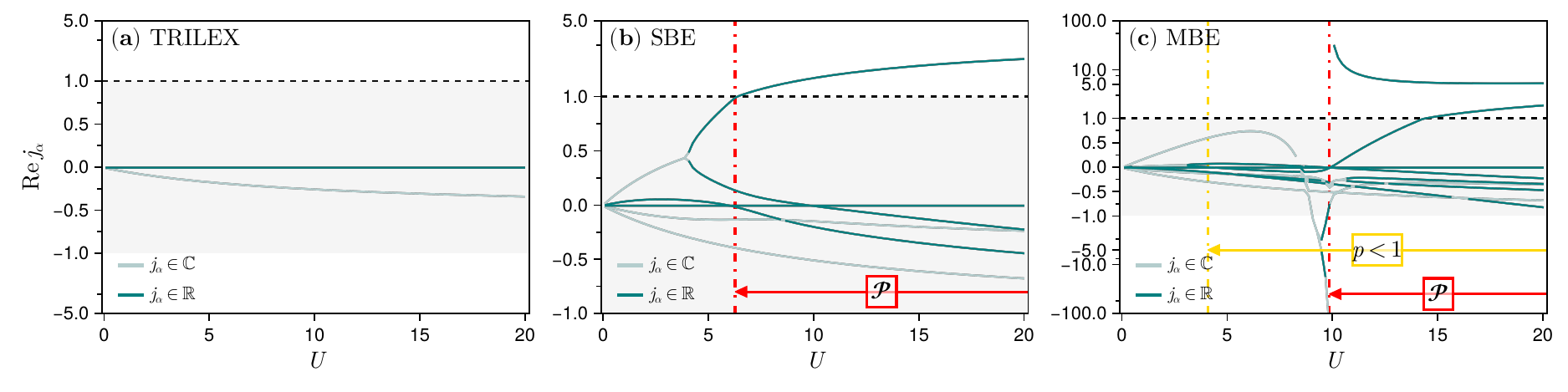}
    \caption{Real parts of the eigenvalues $j_\alpha$ of the Jacobian, Eqs.~\eqref{eq:J_GW}--\eqref{eq:Jacobian-MBE}, from the different  diagrammatic schemes (a) TRILEX, (b) SBE, (c) MBE for the ZPM. The lines are generated in dependence of $U$ at half-filling. Dark teal lines represent purely real $j_\alpha$ and light teal ones complex-valued $j_\alpha$. Dashed vertical lines and arrows mark regions with finite mixing (yellow) and instabilities $\mathrm{Re}\,j_\alpha > 1$ (red).
    }
    \label{fig:ZP_evals_dmu=0.0}
\end{figure*}

Starting with a path at particle-hole symmetry, i.e., $\mathrm{Re}\,\delta\mu=0$, and increasing $U$, in Fig.~\ref{fig:ZP_evals_dmu=0.0}, we show the real part of the eigenvalues $j_\alpha$ of the Jacobian for (a) TRILEX, (b) SBE, and (c) MBE.
We find that, as expected, TRILEX does not show an instability at particle-hole symmetry [Fig.~\ref{fig:ZP_evals_dmu=0.0}(a)]. 
However, for SBE [Fig.~\ref{fig:ZP_evals_dmu=0.0}(b)], we find one purely real eigenvalue (dark teal line) that causes an instability. 
In Fig.~\ref{fig:ZP_evals_dmu=0.0}(c), we find that, for MBE, two purely real eigenvalues cause instabilities: one eigenvalue that has an odd pole and corresponds to the 2PI divergence in $M_c$ at $U=(\mathrm{Im}\,\delta\mu\;)^2$ and one eigenvalue that crosses 1 at $U\simeq 14.0$.

Comparing these results with Figs.~\ref{fig:SBE_SOT_dmu=0.0}(d) and \ref{fig:MBE_SOT_dmu=0.0}(f) for the HA in the main text, we find that, even though the complicated frequency structure of the HA leads to many more eigenvalues, the overall structure is remarkably similar. The only difference  is that the complicated frequency structure of the HA leads to more unstable eigenvalues from which some also become complex at higher $U$ values.

\begin{figure*}
    \centering
    \includegraphics[width=1\linewidth]{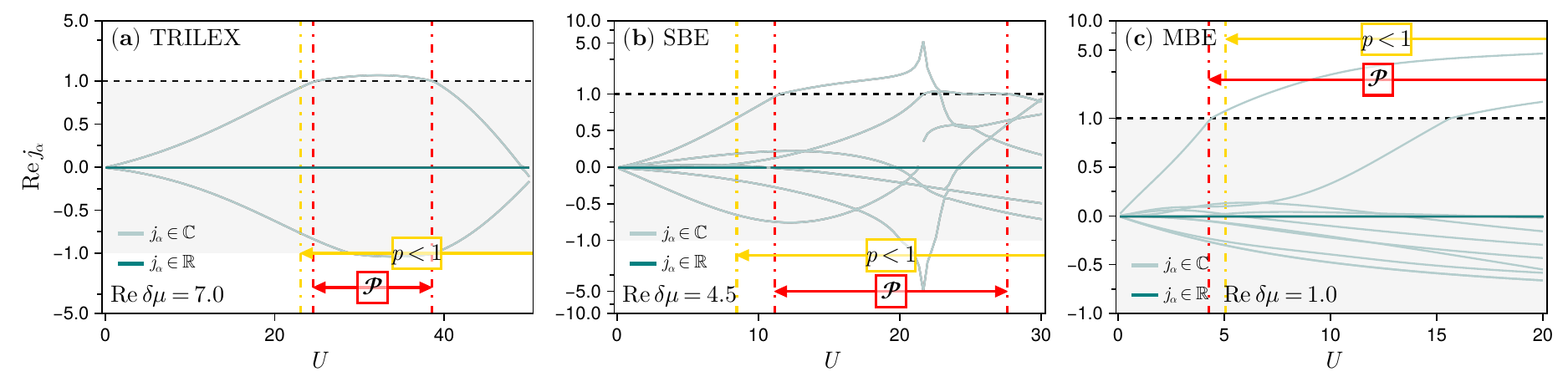}
    \caption{Real parts of the eigenvalues $j_\alpha$ of the Jacobian, Eqs.~\eqref{eq:J_GW}--\eqref{eq:Jacobian-MBE}, from the different  diagrammatic schemes (a) TRILEX, (b) SBE, (c) MBE for the ZPM. The lines are generated in dependence of $U$ at different values of $\delta\mu$. Dark teal lines represent purely real $j_\alpha$ and light teal ones complex-valued $j_\alpha$. Dashed vertical lines and arrows mark regions with finite mixing (yellow) and instabilities $\mathrm{Re}\,j_\alpha > 1$ (red).
    }
    \label{fig:ZP_evals_dmu!=0.0}
\end{figure*}

Now turning our attention to selected paths out of particle-hole symmetry, $\mathrm{Re}\,\delta\mu\neq 0$, we find, in Fig.~\ref{fig:ZP_evals_dmu!=0.0}(a), that TRILEX exhibits an instability, which disappears again when going to higher $U$ values and is caused by a complex eigenvalue. Again, this behavior is very similar to Fig.~\ref{fig:GW_SOT_dmu=12.0}(c), which shows the eigenvalues of the Jacobian for the HA out of half-filling. The main difference that the instability is cause by a complex eigenvalue in the ZPM and by a real eigenvalue in the HA can probably be explained by the fact that $W_r$, with the exception of $W_s(\omega\neq0)$, must be real in the HA while they are all complex out of particle-hole symmetry in the ZPM.

In Fig.~\ref{fig:ZP_evals_dmu!=0.0}(b), we find that the SBE approach exhibits a first instability caused by a complex eigenvalue at $U\simeq 11.0$. Note that the doping is chosen such that the isolated point of the SBE divergence in $\lambda_c$ is in the parameter path. This SBE divergence at $U\simeq 21.6$ is also visible in the eigenvalues of the Jacobian. A comparison to Fig.~\ref{fig:SBE_SOT_dmu=6.0}(d) for the HA, again, reveals a striking similarity.

Finally, we investigate the eigenvalues of the Jacobian for the MBE approach in Fig.~\ref{fig:ZP_evals_dmu!=0.0}(c). There, we find two instable complex eigenvalues, which is again very similar to Fig.~\ref{fig:MBE_SOT_dmu=1.0}(f) for the HA. The main difference is that the divergence in $M_s$ occurring out of half-filling in the HA is absent in the ZPM.

\subsection{Additional weak-coupling results}
\label{sec:Weak_coupling_appendix}

\begin{figure}
	\centering
	\includegraphics[width=\linewidth]{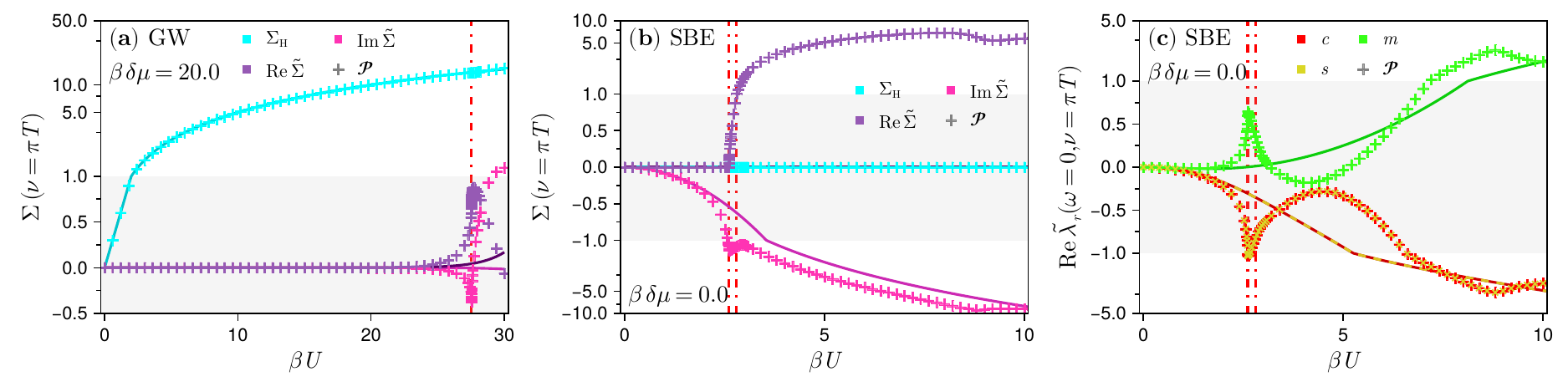}
	\caption{Converged quantities at specific Matsubara frequencies in dependence on $U$ from GW and SBE with perturbative input. Details about colors and markers are provided in Fig.~\ref{fig:GW_SOT_dmu=12.0}(d). Self-energies $\Sigma$ at $\nu=\pi T$ from (a) GW at $\beta\delta\mu=20.0$ and (b) SBE at half-filling, (c) Hedin vertices $\tilde\lambda_r$ at $\omega=0$ and $\nu=\pi T$ from SBE at half-filling.
    }
	\label{fig:GW_SBE_SOT_weak}
\end{figure}

\begin{figure*}
	\centering
    \includegraphics[width=\linewidth]{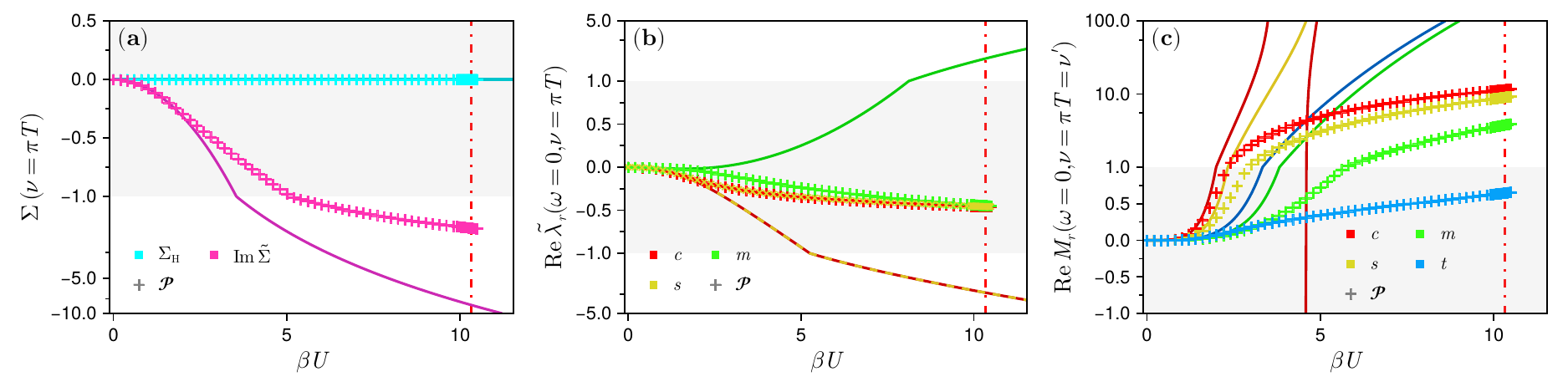}
	\caption{Converged quantities at specific Matsubara frequencies in dependence on $U$ from perturbative MBE at half-filling. Details about colors and markers are provided in Fig.~\ref{fig:GW_SOT_dmu=12.0}(d). (a) Self-energies $\Sigma$ at $\nu=\pi T$, (b) Hedin vertices $\tilde\lambda_r$ at $\omega=0$ and $\nu=\pi T$, (c) MBE vertices $M_r$ at $\omega=0$ and $\nu=\pi T=\nu'$.
    }
	\label{fig:MBE_SOT_dmu=0.0_weak}
\end{figure*}

\begin{figure*}
	\centering
    \includegraphics[width=\textwidth]{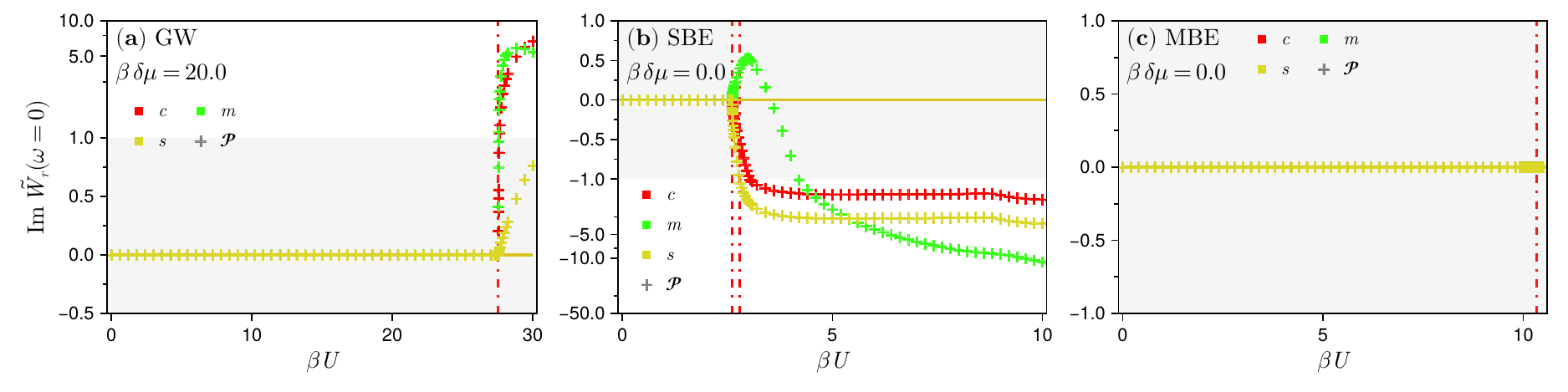}
	\caption{Converged imaginary parts of the bosonic propagators $W_r$ at $\omega=0$ in dependence on $U$ for the different diagrammatic schemes (a) GW, (b) SBE, (c) MBE with perturbative input (cf.\ Fig.~\ref{fig:MBE-SOT-Scheme}). The data are presented in dependence on $U$ at different values of $\delta\mu$. Details about colors and markers are provided in Fig.~\ref{fig:GW_SOT_dmu=12.0}(d). 
    }
	\label{fig:W_r_comparison_weak_Im}
\end{figure*}

Figures~\ref{fig:GW_SBE_SOT_weak} and \ref{fig:MBE_SOT_dmu=0.0_weak} show more converged vertex functions at half-filling, obtained from the different diagrammatic schemes with perturbative input. They complement Figs.~\ref{fig:evals_dmu=0.0_weak}--\ref{fig:W_r_comparison_weak} from the main text. In Fig.~\ref{fig:GW_SBE_SOT_weak}(b), we observe that also the converged results for $\mathrm{Re}\,\tilde\Sigma$ (purple pluses) become finite behind the instability (red dashed lines), which contradicts the exact result, Eq.~\eqref{eq:HA_G}, at half-filling. Interestingly, the converged solutions of the Hedin vertices $\tilde\lambda$ [pluses in Fig.~\ref{fig:GW_SBE_SOT_weak}(c)] oscillate around the exact solution (lines), but they come with a finite imaginary part (not shown explicitly) opposed to the exact solution, Eq.~\eqref{eq:HA_lambda}. Furthermore, Fig.~\ref{fig:W_r_comparison_weak_Im} shows the imaginary part of the bosonic propagators $W_r$ at $\omega=0$, which vanishes in the exact result [cf.\ Eq.~\ref{eq:HA_W_r}], but is finite in the converged results behind the first instability.

\twocolumngrid

\bibliography{main.bib}

\end{document}